\documentclass[journal,table]{IEEEtran} 

\ifCLASSINFOpdf
\usepackage[pdftex]{graphicx}
\else
\fi

\usepackage{multirow}
\usepackage{balance}
\usepackage{cite}
\usepackage{graphicx}
\usepackage{float}
\usepackage{amsmath}
\usepackage[utf8]{inputenc}
\usepackage[final]{pdfpages}
\usepackage{pdfpages}
\usepackage{lipsum}
\usepackage{textcase}
\usepackage{xurl}
\usepackage{amsmath,esint} 
\usepackage{epstopdf}
\usepackage{array}
\usepackage{xcolor}
\usepackage{subcaption} 
\usepackage[normalem]{ulem}

\usepackage{bm}

\DeclareRobustCommand{\uvec}[1]{{%
  \ifcsname uvec#1\endcsname
     \csname uvec#1\endcsname
   \else
    \bm{\hat{\mathbf{#1}}}%
   \fi
}}

\newcolumntype{P}[1]{>{\centering\arraybackslash}p{#1}}
\newcolumntype{M}[1]{>{\centering\arraybackslash}m{#1}}

\begin{document}


\title{A Survey on Detection, Classification, and
Tracking of UAVs using Radar and Communications Systems}

\author{\IEEEauthorblockN{Wahab Khawaja\IEEEauthorrefmark{1},~Martins Ezuma\IEEEauthorrefmark{2},~\IEEEmembership{Member, IEEE},~Vasilii Semkin\IEEEauthorrefmark{3},~Fatih Erden\IEEEauthorrefmark{2},~Ozgur Ozdemir\IEEEauthorrefmark{2},~\IEEEmembership{Member, IEEE},  and~Ismail Guvenc\IEEEauthorrefmark{2},~\IEEEmembership{Fellow, IEEE}
}

\IEEEauthorblockA{\IEEEauthorrefmark{1}Department of Electrical and Computer Engineering Department, Aarhus University, Aarhus, Denmark, 8200}

\IEEEauthorblockA{\IEEEauthorrefmark{2} Electrical and Computer Engineering Department, North Carolina State University, Raleigh, NC 27606, USA}

\IEEEauthorblockA{\IEEEauthorrefmark{3}VTT Technical Research Centre of Finland, Tietotie 3, 02150 Espoo, Finland}

Email: wahab.ali@must.edu.pk, ezuma3000@gmail.com, vasilii.semkin@vtt.fi, erdenfatih@gmail.com, \{oozdemi, iguvenc\}@ncsu.edu
}
\maketitle

\begin{abstract}
The use of unmanned aerial vehicles~(UAVs) for a variety of commercial, civilian, and defense applications has increased many folds in recent years. While UAVs are expected to transform future air operations, there are instances where they can be used for malicious purposes. In this context, the detection, classification, and tracking~(DCT) of UAVs~(DCT-U) for safety and surveillance of national air space is a challenging task when compared to DCT of manned aerial vehicles. In this survey, we discuss the threats and challenges from malicious UAVs and we subsequently study three radio frequency~(RF)-based systems for DCT-U. These RF-based systems include radars, communication systems, and RF analyzers. Radar systems are further divided into conventional and modern radar systems, while communication systems can be used for joint communications and sensing~(JC\&S) in active mode and act as a source of illumination to passive radars for DCT-U. The limitations of the three RF-based systems are also provided. The survey briefly discusses non-RF systems for DCT-U and their limitations. Future directions based on the lessons learned are provided at the end of the survey.
\end{abstract}


\IEEEpeerreviewmaketitle

\section{Introduction}
Radar systems are widely used for the detection, classification, and tracking~(DCT) of aerial vehicles. Radar technology was first introduced in 1935~\cite{radar_1945}. The initial radar systems were bulky with large mechanically rotated antennas, were ground-based, and they had large power requirements, limited processing capabilities, limited accuracy, high cost. They were also highly vulnerable to electronic countermeasures~(ECM). Radars have seen decades of improvements in overcoming many challenges. Modern radar systems employ cutting-edge electronics, such as Gallium Nitride, along with compact antennas, phased arrays, and highly efficient signal processing techniques. These advancements are pivotal in achieving remarkable capabilities including extended detection ranges, rapid response times, streamlined signal processing, exceptional accuracy, low probability of false alarm~(PFA), unambiguous detection and tracking of aerial vehicles even in cluttered environments, simultaneous tracking of multiple aerial vehicles, seamless integration with diverse sensors (airborne, ground, and sea-based), and operational adaptability across various terrains~\cite{radartypes1,radartypes2}. Also, the extensive training data of different terrains and potential aerial vehicles aided with artificial intelligence~(AI) classification algorithms have helped in the real-time classification of different types of aerial vehicles in complex environments. Furthermore, the $5$G and beyond are envisioned to use joint sensing and communications~\cite{joint_radar_comm} that rely on radar technology. 

\begin{figure}[!t] 
    \centering
		\includegraphics[width=\columnwidth]{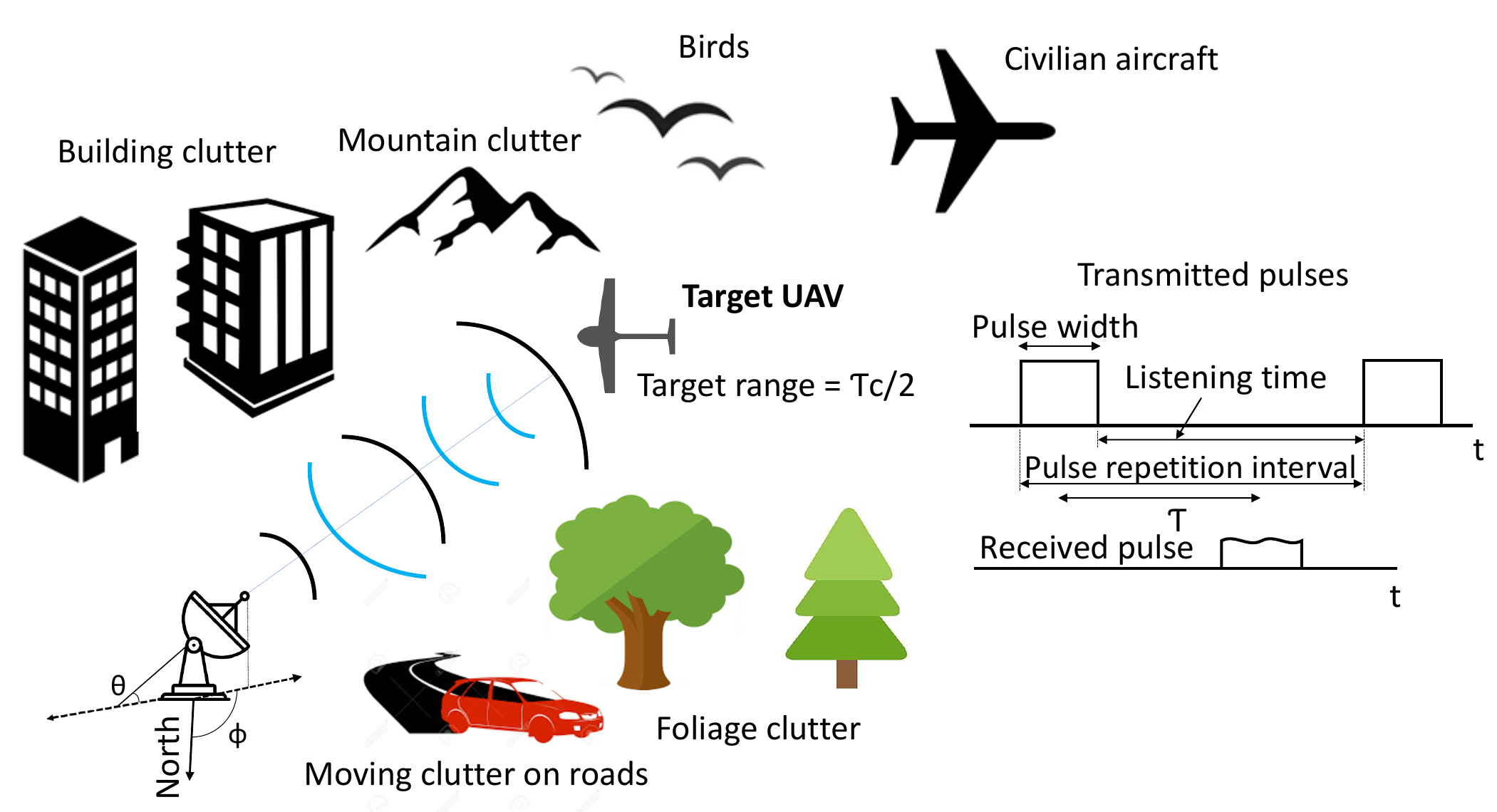}
	   \caption{Detection of a UAV using a conventional pulse radar. Different types of clutter are also shown. Here $\tau$ denotes time delay and $c$ is the speed of light. } \label{Fig:radar_basic1}
\end{figure}


The unmanned aerial vehicle~(UAV) technology is currently one of the fastest-growing technologies in the world. According to \cite{stats}, the overall UAV global market share~(in the military, law enforcement, government, commercial, and consumer domains) is estimated at \$$27.4$ billion in $2021$ and is expected to reach \$$58.4$ billion by $2026$. However, there are many incidents around the world in the recent decade when UAVs are used for malicious purposes, e.g.,~\cite{NSA,SA}. Malicious UAV detection, tracking, and classification (e.g., distinguishing from hobbyist UAVs and birds) are challenging to both conventional and modern radar systems at desired ranges. This is mainly due to the small size, complex shapes, and non-metallic construction material of UAVs, and their ability to fly close to terrain. Moreover, their simple design and ease of manufacturing from off-the-shelf components, and diverse types have made countermeasures against UAVs challenging. 

   \begin{table*}[t]
	\begin{center}
     \footnotesize
		\caption{Comparison of the scope and topics covered in this survey with existing popular related surveys. The compared topics are provided in Fig.~\ref{Fig:Topics_covered}. In the table, L, M, and H are used to represent the extent of the coverage of topics, respectively, where L represents no/negligible coverage, M represents basic coverage, and H represents high coverage.} \label{Table:relevant_surveys}
\begin{tabular}{@{}|P{ 0.8cm}|P{8.3cm}|P{0.32cm}|P{0.32cm}|P{0.32cm}|P{0.32cm}|P{0.32cm}|P{0.32cm}|P{0.32cm}|P{0.32cm}|P{0.32cm}|@{}}
 \hline
\textbf{Ref.}&\textbf{Scope}&\textbf{S1}&\textbf{S2}&\textbf{S3}&\textbf{S4}&\textbf{S5}&\textbf{S6}&\textbf{S7}&\textbf{S8}&\textbf{S9}\\
\hline
\cite{wahab_uav_threats}&The future design, threats, and challenges from UAVs, and radar and non-radar methods for the detection, tracking, and classification of UAVs available in the literature are covered &\cellcolor{green!50} H&\cellcolor{green!26}M&\cellcolor{green!26}M&\cellcolor{green!26}M&\cellcolor{green!8}L&\cellcolor{green!26}M&\cellcolor{green!50}H&\cellcolor{green!8}L&\cellcolor{green!26}M\\
\hline
\cite{multipleUAVs_difficulty}&In this survey, different detection schemes for UAVs using radar systems are provided. The main focus of the paper is to highlight the requirement for new radar systems that can detect small, slow, and low-flying UAVs&\cellcolor{green!26}M&\cellcolor{green!26}M&\cellcolor{green!50}H&\cellcolor{green!8}L&\cellcolor{green!8}L&\cellcolor{green!8}L&\cellcolor{green!26}M&\cellcolor{green!8}L&\cellcolor{green!8}L \\
\hline
\cite{new_review1}&In this review paper, essential metrics for UAV classification using radar data, limitations of radar systems for automatic UAV classification, and future directions for UAV classification using radar are provided&\cellcolor{green!26}M&\cellcolor{green!50}H&\cellcolor{green!8}L&\cellcolor{green!8}L&\cellcolor{green!8}L&\cellcolor{green!8}L&\cellcolor{green!26}M&\cellcolor{green!8}L&\cellcolor{green!8}L \\
\hline
\cite{new_survey1}&This survey provides UAV detection using radar, acoustic, EO sensors, RF analysis of signals emitted by UAVs, and multi-sensor fusion. The performance of different sensors is also compared&\cellcolor{green!26}M&\cellcolor{green!50}H&\cellcolor{green!8}L&\cellcolor{green!50}H&\cellcolor{green!26}M&\cellcolor{green!50}H&\cellcolor{green!50}H&\cellcolor{green!8}L&\cellcolor{green!50}H \\
\hline
\cite{new_survey2}&This survey provides UAV detection and classification using data from radar, EO, acoustic, and RF analysis of UAV-transmitted signals and hybrid sensors. Different machine learning~(ML) and deep learning algorithms applied to the data are also discussed&\cellcolor{green!8}L&\cellcolor{green!50}H&\cellcolor{green!8}L&\cellcolor{green!50}H&\cellcolor{green!8}L&\cellcolor{green!50}H&\cellcolor{green!26}M&\cellcolor{green!8}L&\cellcolor{green!26}M \\
\hline
\cite{UAV_Survey_Farshad}&The survey covers different cyber and physical threats from UAVs. Popular radar and non-radar methods for the detection, tracking, and interdiction of UAVs are also covered&\cellcolor{green!50}H&\cellcolor{green!26}M&\cellcolor{green!26}M&\cellcolor{green!26}M&\cellcolor{green!8}L&\cellcolor{green!26}M&\cellcolor{green!26}M&\cellcolor{green!8}L&\cellcolor{green!26}M \\
\hline
\cite{literature2}&This survey focuses on different methods for the detection and classification of UAVs and birds&\cellcolor{green!26}M&\cellcolor{green!26}M&\cellcolor{green!50}H&\cellcolor{green!26}M&\cellcolor{green!8}L&\cellcolor{green!8}L&\cellcolor{green!50}H&\cellcolor{green!8}L&\cellcolor{green!8}L \\
\hline
\cite{UAV_technol2}&In this survey state-of-the-art methods for the detection of UAVs are provided. Moreover, different interdiction methods for countering non-cooperative UAVs are discussed&\cellcolor{green!50}H&\cellcolor{green!26}M&\cellcolor{green!26}M&\cellcolor{green!26}M&\cellcolor{green!26}M&\cellcolor{green!26}M&\cellcolor{green!50}H&\cellcolor{green!8}L&\cellcolor{green!26}M \\
\hline
\cite{literature5}&In this survey detection, tracking, classification, and disabling of malicious UAVs using different sensors placed at the ground and aerial platforms is discussed. A discussion of the current and future market growth of UAVs is also provided&\cellcolor{green!50}H&\cellcolor{green!26}M&\cellcolor{green!26}M&\cellcolor{green!26}M&\cellcolor{green!8}L&\cellcolor{green!26}M&\cellcolor{green!50}H&\cellcolor{green!8}L&\cellcolor{green!50}H \\
\hline
\cite{survey_new1}&State-of-the-art research for the DCT-U using different sensors and ML algorithms is provided in this survey&\cellcolor{green!26}M&\cellcolor{green!26}M&\cellcolor{green!26}M&\cellcolor{green!26}M&\cellcolor{green!8}L&\cellcolor{green!26}M&\cellcolor{green!26}M&\cellcolor{green!8}L&\cellcolor{green!26}M \\
\hline
\cite{survey_new2}&In this survey, different detection and defense techniques against UAVs using RF-based systems implemented using SDR are provided&\cellcolor{green!26}M&\cellcolor{green!26}M&\cellcolor{green!26}M&\cellcolor{green!26}M&\cellcolor{green!8}L&\cellcolor{green!50}H&\cellcolor{green!26}M&\cellcolor{green!8}L&\cellcolor{green!26}M \\
\hline
\cite{comm_jcas_survey1}&This survey provides joint communication and sensing challenges and solutions employing mobile networks &\cellcolor{green!8}L&\cellcolor{green!50}H&\cellcolor{green!26}M&\cellcolor{green!8}L&\cellcolor{green!50}H&\cellcolor{green!8}L&\cellcolor{green!26}M&\cellcolor{green!26}M&\cellcolor{green!8}L \\ \hline
\cite{comm_jcas_survey2}&In this survey different signal processing techniques for JC\&S are discussed&\cellcolor{green!8}L&\cellcolor{green!50}H&\cellcolor{green!26}M&\cellcolor{green!8}L&\cellcolor{green!50}H&\cellcolor{green!8}L&\cellcolor{green!8}L&\cellcolor{green!8}L&\cellcolor{green!8}L \\
\hline
\cite{review_lte}&In this survey, use of LTE as a source of illumination for passive radars and corresponding applications are provided&\cellcolor{green!8}L&\cellcolor{green!8}L&\cellcolor{green!8}L&\cellcolor{green!8}L&\cellcolor{green!50}H&\cellcolor{green!8}L&\cellcolor{green!8}L&\cellcolor{green!8}L&\cellcolor{green!8}L \\
\hline
\cite{review_dvbt}&Detection and tracking of UAVs using passive DVB-T radar is discussed in this survey&\cellcolor{green!26}M&\cellcolor{green!8}L&\cellcolor{green!8}L&\cellcolor{green!8}L&\cellcolor{green!50}H&\cellcolor{green!8}L&\cellcolor{green!8}L&\cellcolor{green!8}L&\cellcolor{green!8}L \\
\hline
\cite{survey_SOO}&In this survey detection and classification of UAVs using RF-based techniques and ML are discussed&\cellcolor{green!26}M&\cellcolor{green!8}L&\cellcolor{green!8}L&\cellcolor{green!8}L&\cellcolor{green!8}L&\cellcolor{green!50}H&\cellcolor{green!8}L&\cellcolor{green!8}L&\cellcolor{green!8}L \\
\hline
Our survey&In this work, we have provided a comprehensive survey on the capabilities of UAVs, threats and challenges from malicious UAVs, DCT-U using radar, communication systems, RF analyzers, and non-RF systems. The limitations of radar, communication systems, RF analyzers, and non-RF systems for DCT-U are also provided&\cellcolor{green!50}H&\cellcolor{green!50}H&\cellcolor{green!50}H&\cellcolor{green!50}H&\cellcolor{green!50}H&\cellcolor{green!50}H&\cellcolor{green!50}H&\cellcolor{green!50}H&\cellcolor{green!50}H \\
\hline
\end{tabular}
		\end{center}
			\end{table*}

\begin{figure*}[!t] 
    \centering
		\includegraphics[width=.95\textwidth]{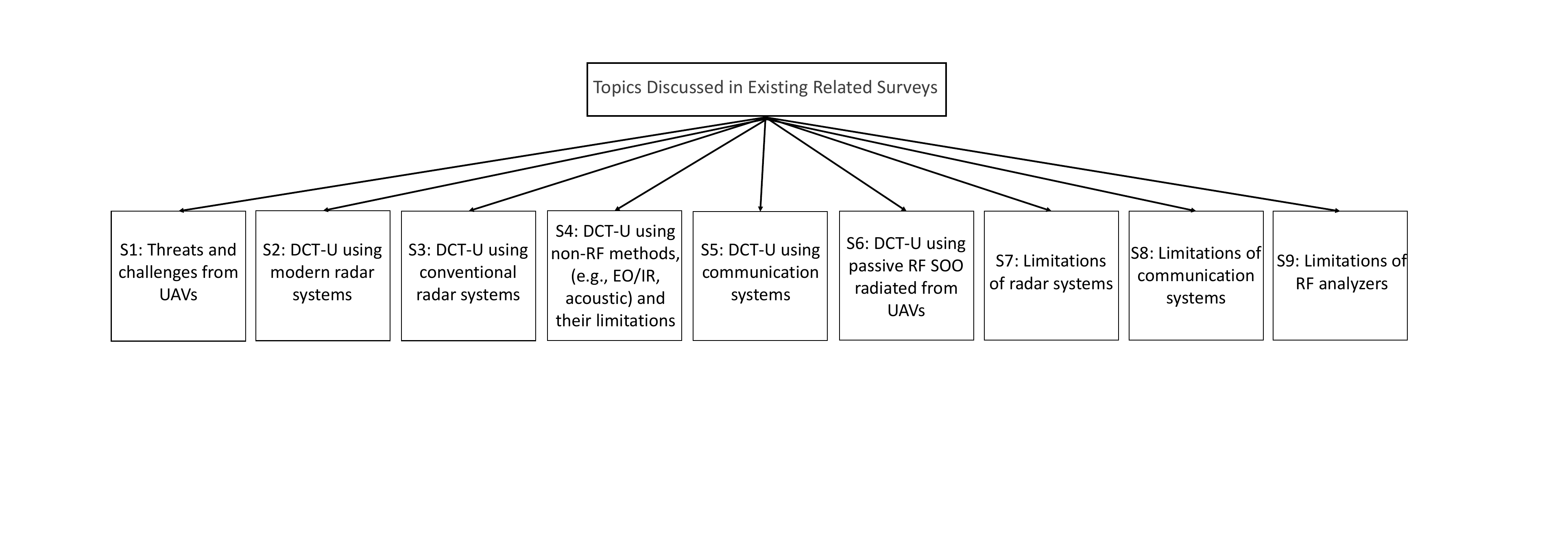}
	   \caption{Topics discussed in existing related surveys for comparison with our survey.}  \label{Fig:Topics_covered}
\end{figure*}

The UAVs used by amateur users may also introduce major threats if their users do not follow regulatory rules~\cite{UAV_amateur02,UAV_amateur1,UAV_amateur2}. With all these, it is critically important to find ways to counter malicious UAVs as well as unintended threats from amateur UAVs~\cite{wahab_uav_threats,UAV_amateur2}. Fig.~\ref{Fig:radar_basic1} shows a basic UAV detection scenario using a pulse radar. The different types of clutter that can affect the detection and tracking of the UAV are also shown. There are numerous research efforts carried out in academia and industry to counter the threats and challenges of malicious UAVs. According to a NATO Review Report~\cite{C_UAS}, the global share of UAV countering technologies is expected at $\$6.6$~billion by $2024$.


There are different methods available in the literature for the detection, tracking, and classification of malicious UAVs. Popular methods include radar systems, electro-optical/infra-red~(EO/IR) imaging, radio frequency~(RF) analysis, and sound/noise analysis of UAVs~\cite{wahab_uav_threats}. Radar systems are dominantly used compared to other techniques~\cite{modern_detect, drone_detect_techs,wahab_uav_threats}. However, the probability of miss detection for UAVs using radar systems is high mainly due to the small radar cross section~(RCS), RCS similarity to RCS of birds, and high maneuverability close to the terrain. Moreover, as UAVs are used for different applications and recreational purposes without a centralized traffic management and monitoring system, it is sometimes difficult to learn about the intent of the UAV flight. 

Alternatively, or as a complementary approach, communication systems that have generally good coverage in densely populated urban areas can be used for DCT-U. Joint communications and sensing~(JC\&S) systems generally based on mobile communication networks can be used for communications and target detection in active mode efficiently utilizing the available spectrum~\cite{comm_jcas_survey1}. Moreover, the transmissions from popular communication systems, e.g., long-term evolution~(LTE) and frequency modulation~(FM) broadcasts can be analyzed passively to detect, track, and classify UAVs~\cite{Comm_gen}. DCT-U can also be performed by passively analyzing communication signals of opportunity~(SOO) emitted from UAVs using RF analyzers~\cite{RF_analysis}. 

\subsection{Existing Surveys and Reviews in the Literature}
There are studies available in the literature that provide a review of efforts for DCT-U. Table~\ref{Table:relevant_surveys} provides a comparison of our survey with related existing surveys. The scope and topics covered in this survey compared to existing surveys are highlighted in Table~\ref{Table:relevant_surveys}. In \cite{wahab_uav_threats}, threats and countermeasures against malicious UAVs were covered. The countermeasures included radar and non-radar methods. In \cite{multipleUAVs_difficulty}, a review was provided for the detection and classification of small UAVs using radars. For the efficient detection of small UAVs, portable, highly mobile, inexpensive, and highly efficient radar systems were suggested. The technical requirements and implementation details were also provided in the paper. In \cite{UAV_Survey_Farshad}, a survey on cyber and physical threats from UAVs was covered. Different methods were also reviewed in the survey for DCT-U. In \cite{literature2}, recent advances in the detection and classification of rotary wing UAVs and birds using radar systems in complex scenarios were discussed. The study covered echo modeling, fretting characteristics recognition, the fusion of distributed multi-view features for UAV detection, and classification using deep learning.

\begin{table*}[htbp]
	\begin{center}		
    \footnotesize
		\caption{Abbreviations used in this paper.}\label{Table:Table_acronym}
        	\begin{tabular}{|P{1.8cm}|P{6.5cm}|P{1.8cm}|P{6.5cm}|}
			\hline			\textbf{Acronym}&\textbf{Text}&\textbf{Acronym}&\textbf{Text}\\
            \hline
      AAM & advanced air mobility       
 &   MIMO&multiple-input-multiple-output\\
            \hline
ADS-B& automatic dependent surveillance-broadcast protocol
 &  ML&machine learning \\
            \hline     
AESA & active electronically scanned array
 & MMPAR&multifunction multibeam phased array radar  \\
            \hline 
 AI & artificial intelligence
 & mmWave & millimeter wave
  \\
            \hline    
AM & amplitude modulation
 & MUSIC&multiple signal classification
     \\
            \hline    
BS& base station
 & NLOS & non-LOS 
  \\
            \hline
CFAR& constant false alarm rate
 &  OFDM & orthogonal frequency-division multiplexing    \\
            \hline
CIR& channel impulse response
 & OTFS & orthogonal time frequency space
 \\
            \hline     
CNN& convolutional neural network
 & PCL & passive coherent location
  \\
            \hline     
CSI& channel state information
 & 
    PDF &  probability density function\\
            \hline     
CW& continuous wave
 & PESA & passive electronically scanned array
   \\
            \hline     
DAB& digital audio broadcast&PFA & probability of false alarm
\\
\hline
DARPA&Defense Advanced Research Projects Agency
 & PN&pseudorandom   \\
            \hline    
DCT-U& detection, classification, and tracking of UAVs
 & PRF&pulse repetition frequency  \\
            \hline    
DVB-S& digital video broadcast satellite
 &  PRI&pulse repetition interval  \\
            \hline    
DVB-T& digital video broadcast terrestrial
 &  PW&pulse width  \\
            \hline  
ECCM& electronic countercountermeasures
 &  RATR&radar automatic target recognition \\
            \hline   
ECM& electronic countermeasures
 & RCS&radar cross section \\
            \hline   
EO/IR&electro-optical/infrared
 &  RF&radio frequency    \\
            \hline   
FAA &Federal Aviation Administration
 &  RX&receiver  \\
            \hline  
FFT&fast Fourier transform & SAR&synthetic aperture radar\\
\hline        
FM&frequency modulation
 &   SDR&software-defined radio  \\
            \hline   
FMCW&frequency-modulated continuous wave
 &   SNR&signal-to-noise ratio \\
            \hline   
FSR&forward scattering radar
 &  SOO&signals of opportunity \\
            \hline   
GBM&grading boosting method
 &    SVM&support vector machine \\
            \hline   
GS&ground station
 &  TX&transmitter \\
            \hline   
GSM & global system for mobile communication
 &   UAV&unmanned aerial vehicle  \\
            \hline   
HRRP&high range resolution profile
 &   UHF&ultra high frequency  \\
            \hline   
IoT&Internet of things
 &    USRP& universal software radio peripheral \\
            \hline  
JC\&S&joint communications and sensing
 &  UTM&unmanned aircraft system traffic management \\
           \hline    
LOS&line-of-sight
 &   UWB&ultra-wideband \\
            \hline   
LTE&long-term evolution
 &  VHF&very high frequency \\
            \hline
		  \end{tabular}
		\end{center}
 \end{table*}

Counter-UAV methods and systems were studied in \cite{UAV_technol2}, which included discussions on techniques such as radar, data fusion, passive RF, acoustic, vision, and jamming. The challenges in countering UAVs and future trends were also covered in the survey. In \cite{literature3}, different approaches using radar systems for the detection and localization of UAVs in academia and industry were provided. The hardware and software methods and commonly used localization algorithms were also discussed in the survey. In \cite{literature5}, detailed information was provided on counter-UAV systems. The counter-UAV systems were divided into ground and aerial platforms. The ground platform-based counter-UAV system consisted of mobile, static, and human-packable sensors, whereas, the aerial platform consisted of low and high-altitude sensors. 

Several surveys have explored the use of communication systems for DCT of various targets, including UAVs. JC\&S approaches were investigated in previous surveys~\cite{comm_jcas_survey1, comm_jcas_survey2}. Additionally, in \cite{review_lte,review_dvbt}, comprehensive reviews were presented for the utilization of communication systems as a means of illuminating UAVs for passive DCT. In \cite{survey_SOO}, a review study on  detection, classification, and tracking of UAVs~(DCT-U) based on RF SOO emitted by UAVs was provided. Table~\ref{Table:relevant_surveys} provides a comparison of the topics covered and the scope of the existing literature in survey papers, contrasting it with the contents of this survey. The numerical values, ranging from 0 to 2 in Table~\ref{Table:relevant_surveys}, signify the extent of coverage for each respective topic, and letters A to H represent the topics covered.


\subsection{Contributions and Organization of This Survey}
Compared to the existing literature, this work presents a comprehensive survey of the capabilities, potential threats from UAVs, and challenges in DCT-U. This survey focuses mainly on three RF-based methods for DCT-U. These three RF-based methods include radar systems, communication systems, and RF analyzers. The DCT-U using radar systems is further categorized into conventional and modern radar systems. Communication systems supporting both active and passive sensing for DCT-U are also discussed. Additionally, RF analyzers that passively monitor RF SOO radiated from UAVs for DCT-U are provided. Furthermore, the advantages and limitations of the three RF-based methods for DCT-U are provided. Non-RF methods for DCT-U based on acoustic, EO/IR, lidar, and sensor fusion are also briefly covered in this survey. Finally, future directions for advancing DCT-U capabilities are provided. To the best of the knowledge of the authors, there is no detailed survey available in the literature that provides comprehensive details on the threats from UAVs and challenges in the DCT-U and different methods~(RF and non-RF) to counter malicious UAVs and it is our goal to fill this gap. Table~\ref{Table:relevant_surveys} provides a comparison of our survey with already available related surveys. 

The rest of the paper is organized as follows. Section~\ref{Section:Radar_basics} discusses the basics of radar systems for DCT-U. The existing and future aerial threats and challenging features of UAVs that make it difficult to detect and track them are provided in Section~\ref{Section:future_threats}. Section~\ref{Section:radar_comm_det} reviews different types of conventional and modern radar systems for DCT-U, Section~\ref{Section:Comm_Sys} covers communication systems used for DCT-U, Section~\ref{Section:SOO} discusses different RF analyzers that passively capture RF SOO radiated from UAVs for DCT-U, the future directions based on the lessons learned are provided in Section~\ref{Section:Future_directions}, and finally, Section~\ref{Section:conclusions} concludes the paper. The abbreviations used in the paper are provided in Table~\ref{Table:Table_acronym}.


\section{Radio waves and Radar Basics}   \label{Section:Radar_basics}
In this section, the basics of radio wave propagation and radar systems including the interaction of radio waves with aerial vehicles and surroundings, radar parameters, and associated signal processing techniques are discussed. 

\begin{figure*}[!t] 
    \centering
	\begin{subfigure}{\columnwidth}
    \centering
	\includegraphics[width=0.82\textwidth]{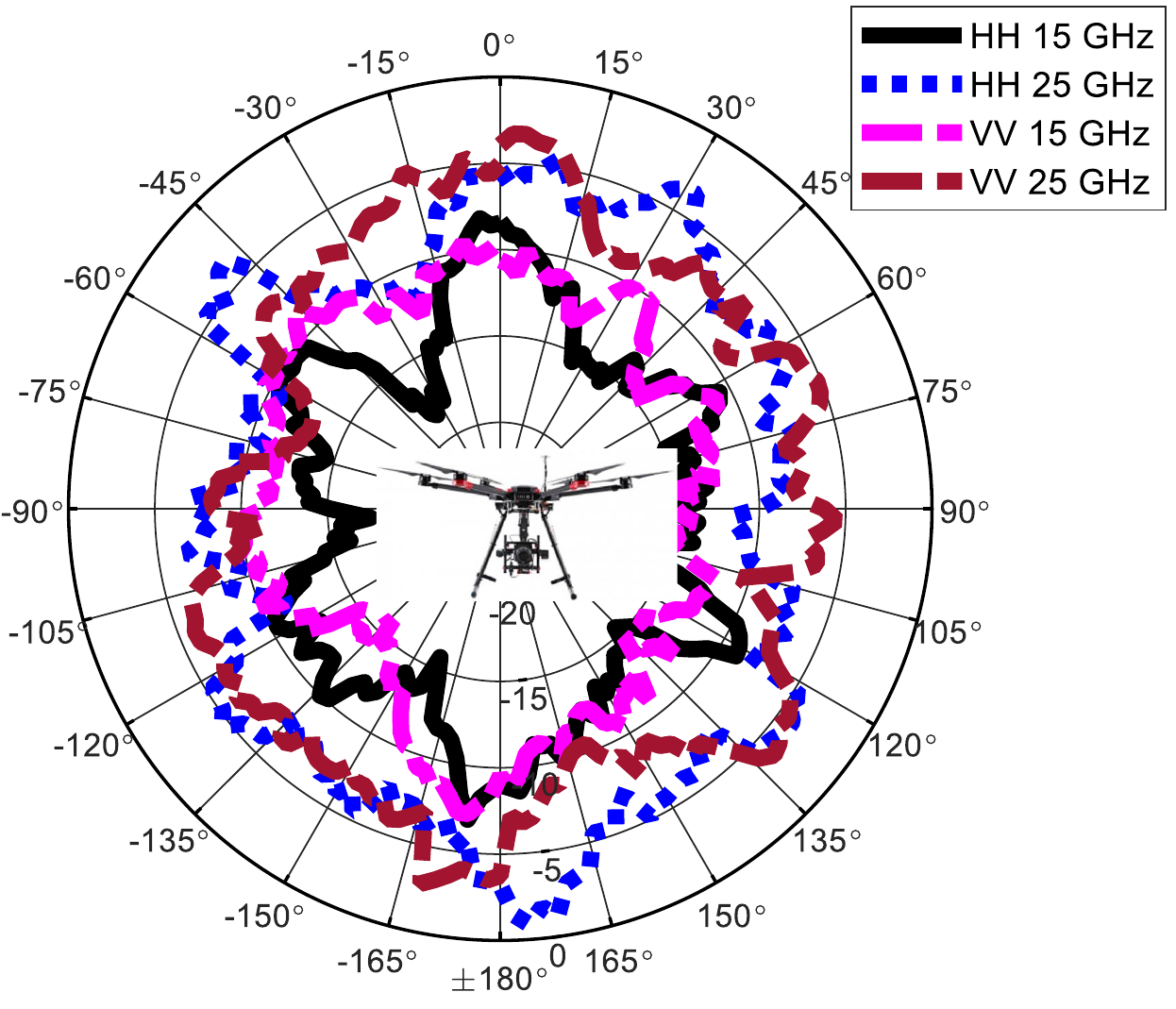}
	  \caption{}  
    \end{subfigure}    
    \begin{subfigure}{\columnwidth}
    \centering
	\includegraphics[width=0.82\textwidth]{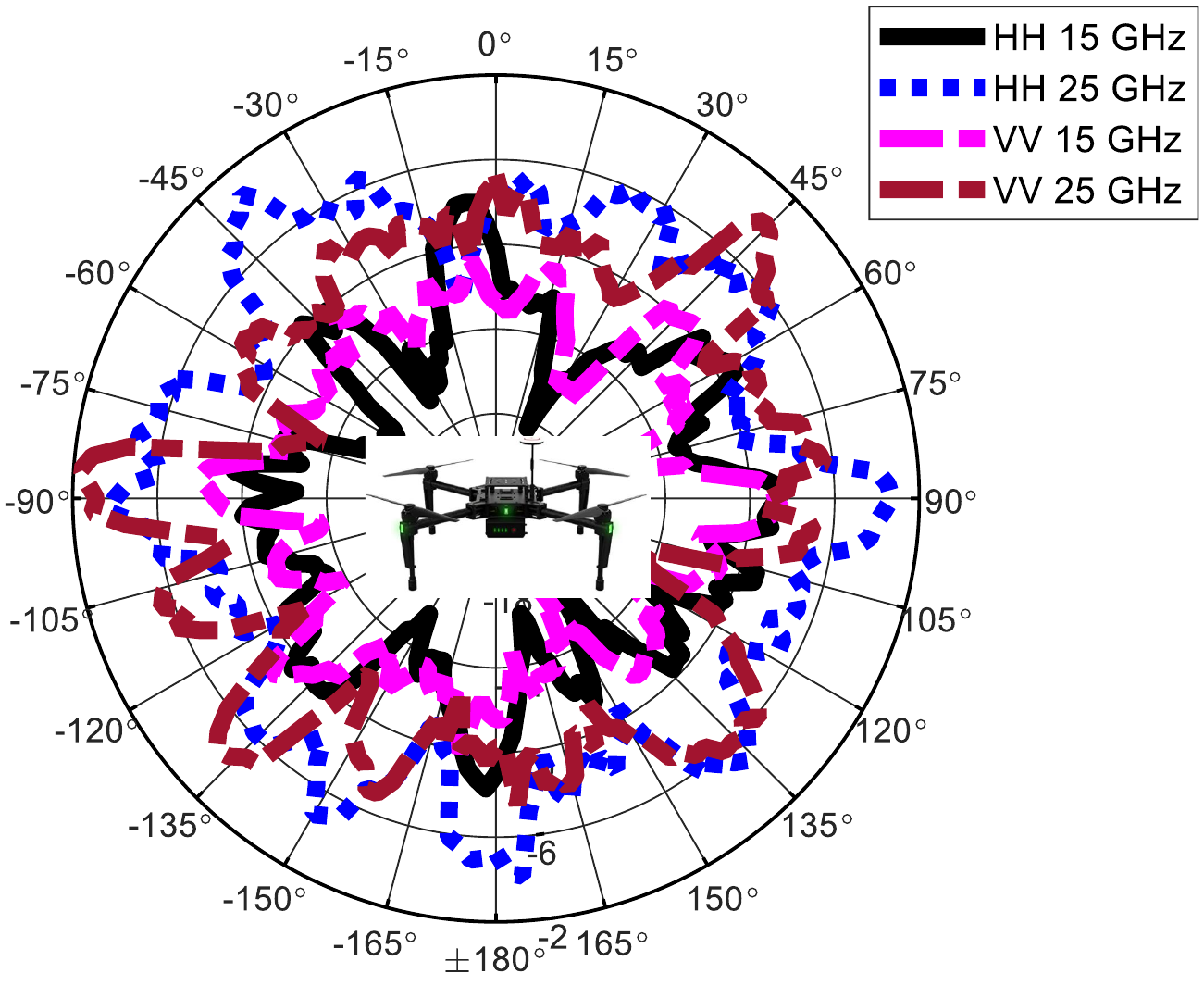}
	  \caption{}  
    \end{subfigure}
    \begin{subfigure}{\columnwidth}
    \centering
	\includegraphics[width=0.86\textwidth]{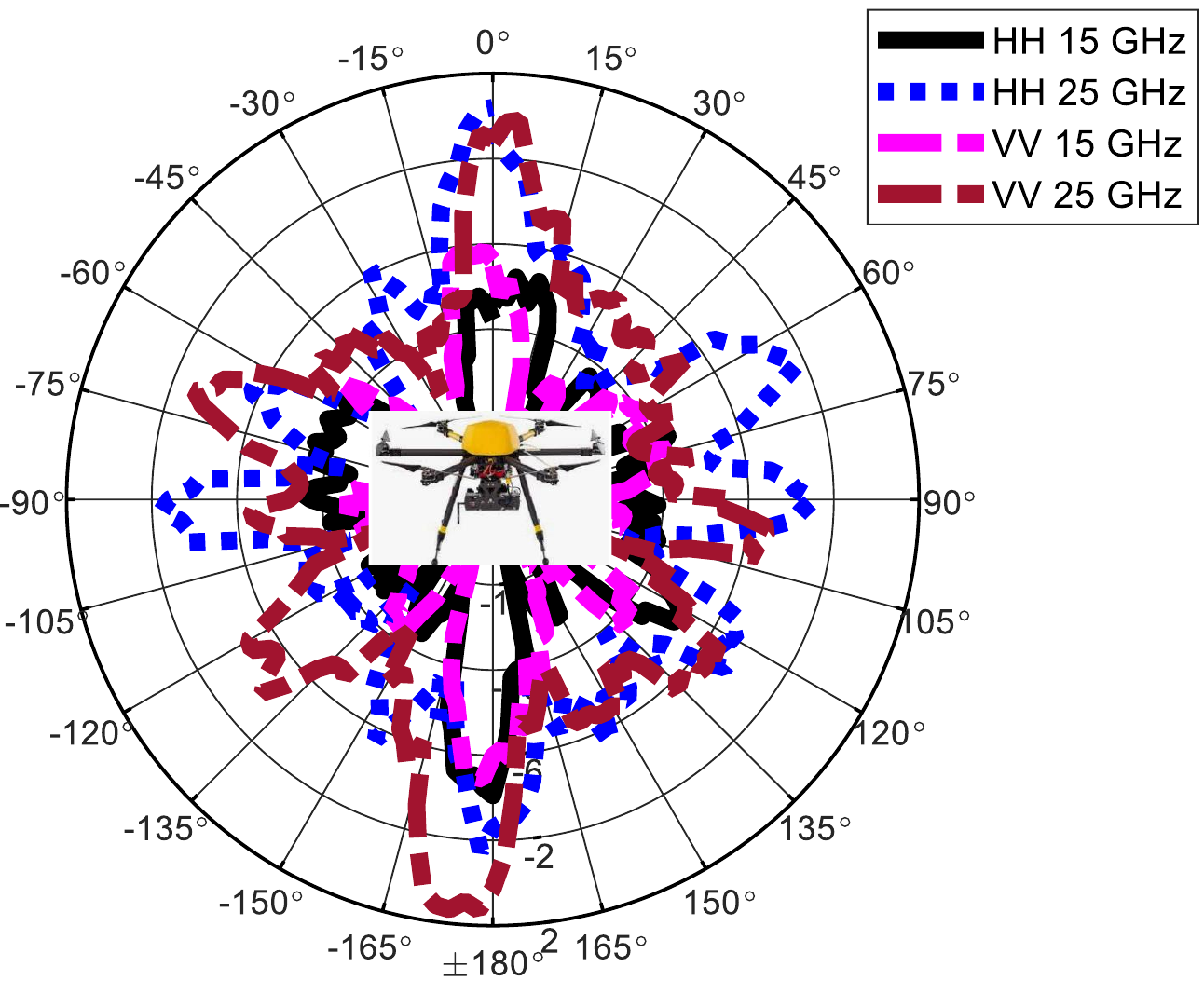}
	  \caption{}  
    \end{subfigure}
    \begin{subfigure}{\columnwidth}
    \centering
	\includegraphics[width=0.86\textwidth]{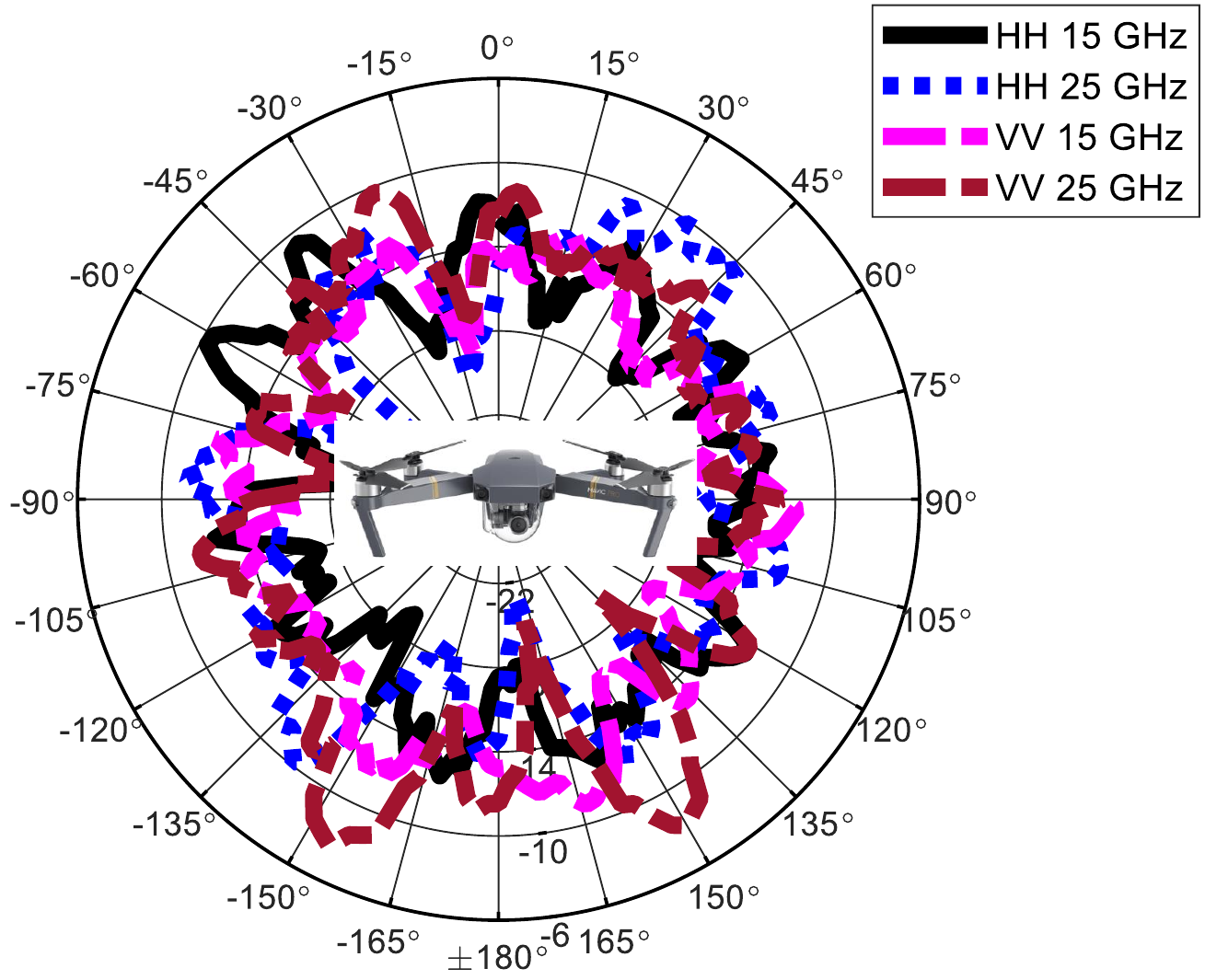}
	  \caption{}  
    \end{subfigure}
    \caption{The RCS of four different UAVs measured at 15~GHz and 25~GHz using vertical-vertical and horizontal-horizontal polarization~(regenerated from \cite{class_new9}). The measured RCS (in dBsm) is plotted against the azimuth angle for the range [0$^\circ$~360$^\circ$]. The four UAVs are (a) DJI Matrice 600 Pro, (b) DJI Matrice 100, (c) Trimble
zx5, and (d) DJI Mavic Pro 1.} \label{Fig:RCS_anechoic_martins}
\end{figure*}

\subsection{Radio Waves Interaction with the Aerial Vehicles and Surroundings} \label{Section:EM_surroundings}
A simple radar system is based on capturing the reflected radio waves from a desired object. The detection of an aerial vehicle using radio waves depends on the physical and motion characteristics of the target and surrounding environment. The physical characteristics of an aerial vehicle are estimated using the RCS of the aerial vehicle by the radar. The RCS of an aerial vehicle $\sigma$ is given as 
\begin{equation}
\sigma = \lim_{R \to \infty} 4\pi R^2 \frac{|E_{\rm s}|^2}{|E_{\rm i}|^2}, 
\end{equation}
where $R$ is the slant range between radar and aerial vehicle, while $E_{\rm i}$ and $E_{\rm s}$ are the incident and scattered electric fields from an aerial vehicle, respectively. The RCS of an aerial vehicle depends on the frequency, polarization, angle of illumination, geometry, and electrical properties of the material of the aerial vehicle. The RCS can also be used for classification. In \cite{class_new9}, RCS of six different types of UAVs measured at $15$~GHz and $25$~GHz were used for classification. The RCS of four UAVs measured at $15$~GHz and $25$~GHz and at respective polarization pairs of vertical-vertical and horizontal-horizontal are provided in Fig.~\ref{Fig:RCS_anechoic_martins}.

The motion characteristics of an aerial vehicle include velocity, pitch, yaw, roll angles, and rate of climb. The major motion characteristic is velocity. The velocity of an aerial vehicle is generally obtained by measuring the Doppler shift in frequency of the received radio signal given as $f_{\rm d} = \frac{2v\cos\alpha}{\lambda}$, where $v$ is the velocity of the aerial vehicle, $\alpha$ is the angle between the radar's line of sight towards the aerial vehicle and direction of travel of the aerial vehicle, and $\lambda$ is the wavelength. In addition to the main Doppler shift~$f_{\rm d}$, there are additional Doppler shifts due to the motion/rotation of sub-parts of the aerial vehicle. Such additional Doppler shifts at the micro level are categorized as micro-Doppler, which is a helpful feature often used in the classification of an aerial vehicle~\cite{micro_doppler}.

\begin{figure*}[!t]
	\centering
	\includegraphics[width=1.4\columnwidth]{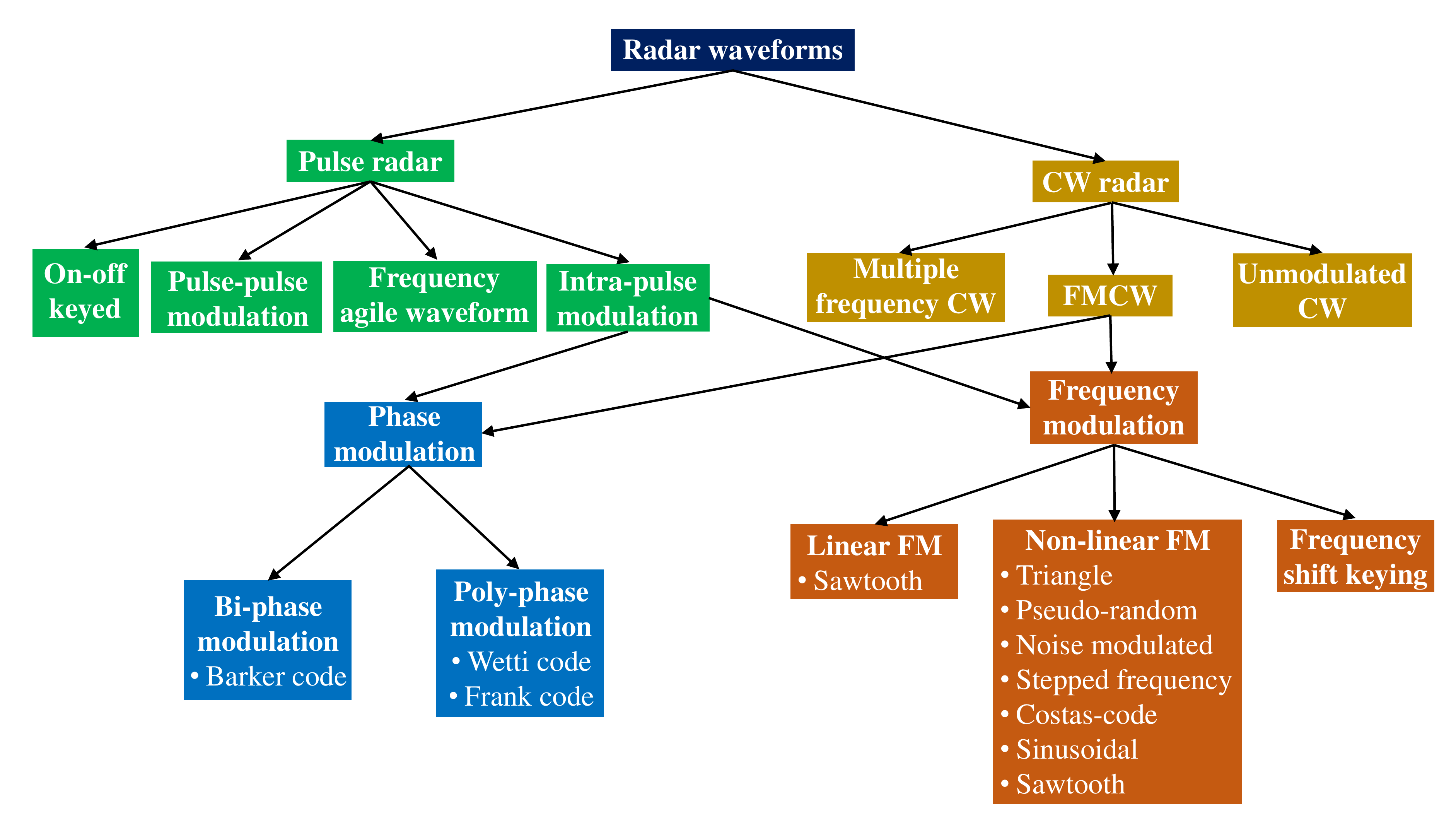}
	\caption{Different types of radar waveforms~(regenerated from \cite{new_fig2})}.\label{Fig:Waveforms_radar}
\end{figure*}

The collection of radio wave reflections~(specular and diffuse) from objects that are not of interest such as buildings and other man-made structures, birds, trees, and hills constitute radar clutter~\cite{radar_clutter_basics}. The radar clutter can be static or dynamic. A major source of static clutter is the ground/sea surface reflection~\cite{GRC}. Dynamic clutter can be from moving vehicles/objects in the environment, e.g., moving cars, flying planes, the rotation of wind turbines, and the movement of foliage. The clutter is related to the angular and range resolution of a radar system. If the angular resolution is small (i.e., large beamwidth) large clutter is observed and vice versa. If the range resolution is high, the clutter can be differentiated from the aerial vehicle, and the signal-to-clutter ratio increases. The clutter cross-section can be thousands of times larger than the cross-section of the aerial vehicle observed by radar.
 
The propagation of radio waves also depends on the terrain and atmospheric effects. The propagation effects~(reflection, diffraction, and scattering) on radio waves will be different in different terrains. Therefore, a radar system working in a rural terrain will require calibration before it can operate in an urban terrain. The radio waves also suffer propagation losses due to atmospheric absorption in addition to free space loss~\cite{atmos_loss}. Atmospheric conditions, e.g., rain, hail, snow, and upper atmospheric conditions can directly affect the detection by radar~\cite{atmospheric_new}. The rain, hail, snow, fog, and other precipitation conditions can result in echoes from these particles that superimpose the target echoes. The atmospheric effects, e.g., wind in a forest area or strong air currents at sea can also affect the propagation characteristics of radio waves~\cite{wind_foliage,wind_sea}. 

\subsection{Radar Parameters}  \label{Section:TX_RX}
Different radar parameters can affect DCT-U. Major radar parameters are transmit power, sounding signal, pulse width~(PW) and duty cycle, pulse repetition interval~(PRI), frequency and bandwidth, modulation and coding, intrapulse modulation, and antenna parameters. The transmit power of radars can vary from a few milliwatts to megawatts. The transmit power is mainly dependent on the application and is constrained by the platform. The high transmit power allows a greater range for radar systems by overcoming free space attenuation. To detect UAVs with small RCS at long ranges, large transmit power is often required.

Different signal waveforms used by radar systems are provided in Fig.~\ref{Fig:Waveforms_radar}. Mainly the waveforms are divided into pulse and continuous wave~(CW) waveforms. Pulse and CW waveforms are further divided into other waveform types. Each waveform in Fig.~\ref{Fig:Waveforms_radar} has specific use and advantage. Overall, CW radar systems are simpler compared to pulse radar systems. However, pulse radar systems have a high dynamic range due to the isolation of transmitter~(TX) and receiver~(RX). The high dynamic range allows long-range detection capability. Furthermore, continuous high-energy transmissions are not possible from a single TX/RX antenna~(monostatic radar) and it can damage the RX if not stopped by a Duplexer. Therefore, a listening time is required for pulse radars. The listening time is constrained by the duty cycle, PW, and PRI. 

\begin{figure}[!t]
	\centering
	\includegraphics[width=0.96\columnwidth]{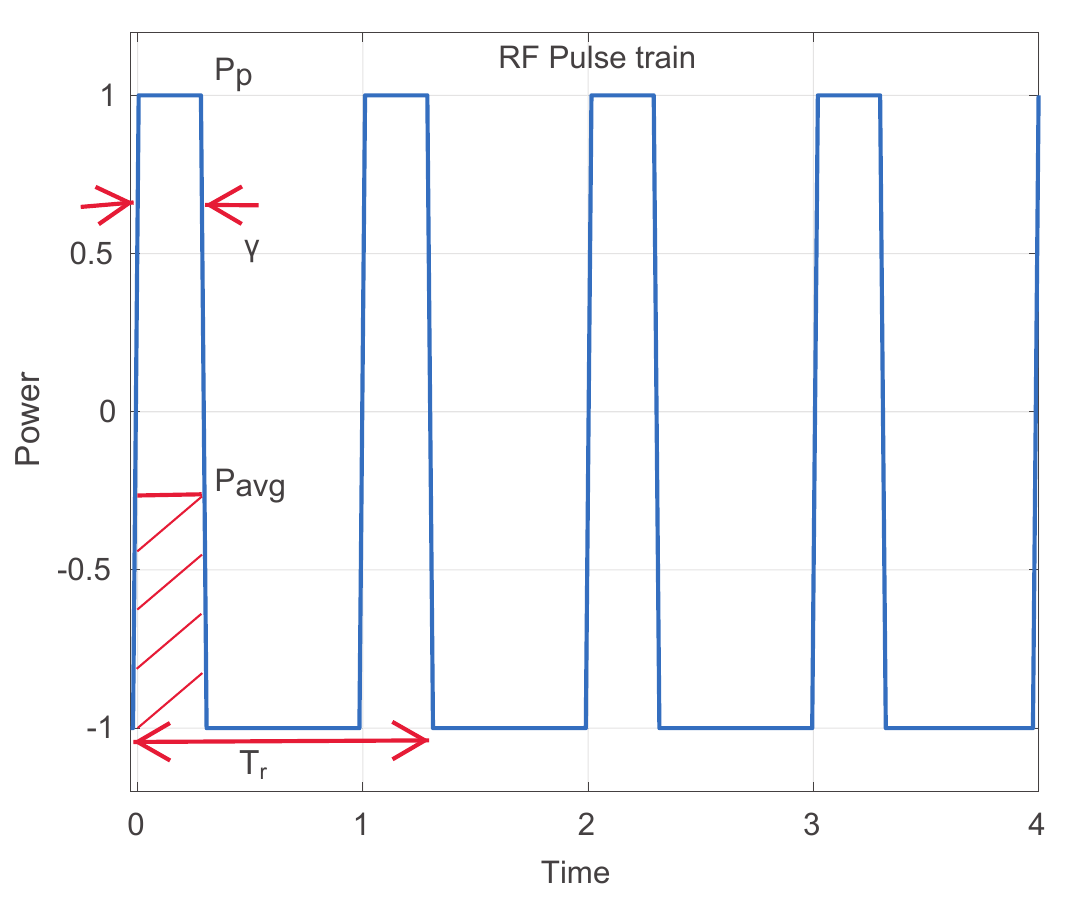}
	\caption{An RF pulse train showing the PW, PRI, and pulse peak and average power. }\label{Fig:pulsetrain}
\end{figure}

The peak power, average power, PW, PRI, pulse repetition frequency~(PRF), and duty cycle for a radar pulse train shown in Fig.~\ref{Fig:pulsetrain} are related to each other as follows~\cite{pulse_parameters}:
\begin{equation}
    \rm{duty~cycle}  = P_{\rm avg}/P_{\rm p} = \rm{\gamma}/\rm{T_{\rm r}} = \rm{\gamma} \times \rm{f_{\rm r}}, \label{Eq:dutycycle}
\end{equation}
where $P_{\rm avg}$ is the average power, $P_{\rm p}$ is the peak power, $\gamma$ is the PW, $T_{\rm r}$ is the PRI, and $f_{\rm r}$ is the PRF. The parameters of short and long pulse radar test waveforms are provided in Table~\ref{Table:shortwaveforms}. The design of both short and long pulse radar test waveforms in Table~\ref{Table:shortwaveforms} are based on the compliance measurement procedure outlined by the FCC for unlicensed devices. Noticeably, the short pulse waveforms have a higher number of pulses per burst compared to long pulse waveforms, whereas, the PW of short pulse waveforms is significantly smaller than the long pulse waveforms as expected. The minimum percentage of successful detection and PRI is higher for long-pulse waveforms compared to short-pulse waveforms.

\begin{table*}[htbp]
	\begin{center}
     \footnotesize
		\caption{Short and long pulse radar test waveforms~(adapted from }\cite{ITUwaveforms}). \label{Table:shortwaveforms}
\begin{tabular}{|p{1.3cm}|p{1cm}|p{1cm}|p{1.5cm}|p{3.4cm}|p{2cm}|p{2cm}|}
 \hline
&\textbf{Type of Radar}&\textbf{PW ($\mu$s)} &\textbf{PRI ($\mu$s)}&\textbf{Number of pulses} &\textbf{Minimum
percentage of successful detection}&\textbf{Minimum number of trials}\\ \cline{2-7}
\multirow{5}{*}{\textbf{Short pulse}}&0&1&1428&18&See note~1 in \cite{ITUwaveforms}&See note~1 in \cite{ITUwaveforms}\\ \cline{2-7}
&1&1&Test~A, and Test~B, in \cite{ITUwaveforms}&$\rm{Roundup}\Big( \big(\frac{1}{360}\big).\big(\frac{19.10^6}{PRI}\big)\Big)$&60\%&30\\ \cline{2-7}
&2&1-5&150-230&23-29&60\%&30\\ \cline{2-7}
&3&6-10&200-500&16-18&60\%&30\\ \cline{2-7}
&4&11-20&200-500&12-16&60\%&30\\ \hline
\multirow{1}{*}{\textbf{Long pulse}}&5&50-100&1000-2000&1-3&80\%&30\\
\hline
\end{tabular}
		\end{center}
			\end{table*}


   \begin{table*}[t]
	\begin{center}
     \footnotesize
		\caption{Different frequency bands for radars and their applications.} \label{Table:Freq_target}
\begin{tabular}{@{}|P{ 1.5cm}|P{2.2cm}|P{10.8cm}|@{}}
\hline
 \textbf{Radar band}& \textbf{Frequency (GHz)}& \textbf{Popular applications}\\
\hline
Millimeter&40-100&UAV detection, tracking, and navigation, airborne radar, spaceborne radar, SAR\\
\hline
Ka&26.5–40&Airborne close range targeting, airport surveillance, traffic speed detection, SAR\\
\hline
K&18–26.5&Small UAV detection, airborne close range targeting, traffic speed detection\\
\hline
Ku&12.5–18&High resolution mapping, satellite altimetry, air-traffic control, air-borne radar\\
\hline
X&8–12.5&Short range tracking,  guidance, UAV detection and tracking, marine radar, air-traffic control\\
\hline
C&4–8&Long range tracking, weather monitoring, SAR\\
\hline
S&2–4&Moderate range surveillance, air-traffic control, weather monitoring, surface ship radar \\
\hline
L&1–2&Long range surveillance, small UAV detection, atmospheric studies, air-traffic control, SAR\\
\hline
UHF&0.3–1&Very long range early warning against aerial threats, anti-stealth\\
\hline
VHF&0.03 to 0.3&Very long range early warning against aerial threats, anti-stealth\\
\hline
\end{tabular}
		\end{center}
			\end{table*}
   
The majority of the radar systems operate between $400$~MHz to $36$~GHz frequency range. The use of a particular frequency band depends on 1) the role of the radar, i.e., search, tracking, or guidance; 2) the type of aerial vehicles expected to detect and track; 3) the terrain; 4) the platform over which radar is mounted; and 5) range of the radar. The long-range search radar uses low frequencies generally VHF and UHF bands. The low frequencies allow long-range detection capabilities. The close-range tracking and guidance radar systems require higher resolution and narrow beams compared to search radars and therefore use higher frequencies compared to search radars. The different frequency bands used by radar systems and corresponding applications are provided in Table~\ref{Table:Freq_target}. Major applications of high-frequency bands are close-range targeting, tracking, and high-resolution imaging, e.g., for small RCS UAVs, whereas, low-frequency bands are mainly used for long-range target detection. 

The bandwidth of a radar is an important parameter and determines the range resolution of the target. The bandwidth of a pulse radar is equal to the reciprocal of the transmitted PW. The range resolution of a radar is dependent on the bandwidth or PW. The greater the bandwidth, the better the range resolution. The bandwidth and the signal power~(or range) are exchangeable due to the dependence of both on the PW. Ultra-wideband~(UWB) radars are popular due to their large bandwidth~(high range resolution)~\cite{UWB_highresol}. The multipath components from a UWB radar can be resolved in the order of centimeters, therefore, fine details of the aerial vehicle or image can be obtained. The large bandwidth is also helpful in distinguishing an aerial vehicle from the clutter.

Different types of modulations are used by different types of radar systems~\cite{modulation1}. Modulation types for the CW and pulse radar are shown in Fig.~\ref{Fig:Waveforms_radar}. In addition to basic pulse modulations, pulse radar can use phase and frequency intrapulse modulations. Similarly, frequency-modulated continuous wave~(FMCW) radar can have different types of phase and frequency intrapulse modulations as shown in Fig.~\ref{Fig:Waveforms_radar}. The major benefit of intra-pulse modulations is higher range resolution without compromising the signal-to-noise ratio~(SNR)~\cite{pulsecompressionradar}. The phase intrapulse modulations are further divided into bi-phase and poly-phase modulations whereas the frequency intrapulse modulations are further divided into linear and non-linear FM and frequency shift keying. The bi-phase and poly-phase intrapulse modulations use different codes shown in Fig.~\ref{Fig:Waveforms_radar}. Sequences of numbers of different lengths are used in these codes. Other radar coding schemes are provided in~\cite{coding1,coding2}.

Directional antennas are mainly used by radar systems for the detection and tracking of aerial vehicles. The high gain~(and narrow beamwidth in the azimuth and elevation planes) from a directional antenna helps to achieve long-range~\big(see (\ref{Eq:radar_range})\big), fine clutter rejection and high angular resolution. In addition, the position of the narrow beamwidth directional beam can be used to roughly estimate the position of the aerial vehicle. Popular radar antennas were simple parabolic reflectors and Cassegrain feed parabolic reflectors that were mechanically rotated. Nowadays, electronically scanned phased arrays are used that require minimum mechanical assembly for rotation. The electronically scanned phased arrays are further divided into passive and active types~\cite{radar_clutter_basics}. Active electronically scanned array~(AESA) is commonly used in modern radar systems. Analog, digital, or hybrid beamforming can be used in AESA~\cite{beamforming}. 

\begin{table*}[t]
	\begin{center}
     \footnotesize
		\caption{Common features and specifications of different types of UAVs. In the table, $L$, $W$, and $H$ represent the length, width, and height, respectively. The provided data is taken from different internet sources.} \label{Table:UAV_types}
\begin{tabular}{@{}|P{ 3.1cm}|P{2.5cm}|P{2.5cm}|P{2.5cm}|P{2.5cm}|@{}}
\hline
\multicolumn{1}{|c|}{}&\multicolumn{4}{|c|}{\textbf{Specifications of different types of UAVs}}\\		
\hline
 \textbf{UAV type}& \textbf{Single/multi-rotor}& \textbf{Flat/tilt wing}& \textbf{Hot air balloons}&\textbf{Satellites}\\
\hline
\textbf{Max. Size ($L$$\times$$W$$\times$$H$)}& 1m$\times$1m$\times$0.63m& 14m$\times$1.7m$\times$4m&17m$\times$17m$\times$24.5m&73m$\times$109m$\times$20m\\
\hline
\textbf{Airframe material}& Carbon fiber& Carbon fiber&Nylon, polyester&Aluminum alloys\\
\hline
\textbf{Max. flight Endurance}& 3 hours& 42 hours&475 hours&15 years\\
\hline
\textbf{Max. payload capacity}& 8 kg& 100 kg&600 kg&29,000 kg\\
\hline
\textbf{Max. flight ceiling}& 6 km& 18 km&21 km&35786 km\\
\hline
\textbf{Max. speed}& 29 m/s& 130 m/s&4.47 m/s&3138.9 m/s\\
\hline
\textbf{Propulsion System}& Propeller&Propeller&NA&Chemical thrusters\\
\hline
\textbf{Operating frequencies}& 900 MHz - 5.8 GHz& 900 MHz - 5.8 GHz&123.3 - 123.5 MHz&L, C, X, Ku, Ka band\\
\hline
\textbf{Navigation}& Internal/external& Internal/external&NA&Internal/external\\
\hline
\textbf{Heat signature}&Small&Very small&Large&Small\\
\hline
\textbf{RCS}& Small& Small/medium&Large&Large\\
\hline
\end{tabular}
		\end{center}
			\end{table*}
   
\subsection{Popular Radar Signal Processing Techniques} \label{Section:Signal_processing}
The majority of the signal processing techniques are similar for different types of radar systems. The main signal processing takes place at the radar RX. The common signal processing units at the radar RX include an analog-to-digital converter, low-noise amplifier, frequency mixer and filter, matched filter, and threshold detector. The maximum detection range of a commonly used radar is given as 
\begin{equation}
   R_{\rm max} = \bigg({\frac{P_{\rm T}G^2\lambda^2\sigma}{P_{\rm R, min}(4\pi)^3L}}\bigg)^{\frac{1}{4}}, \label{Eq:radar_range}
\end{equation}
where $R_{\rm max}$ is the maximum range for the radar, $P_{\rm T}$ is the transmit power, $G$ is the antenna gain, $P_{\rm R, min}$ is the minimum received power, and $L$ represent the total losses. The detection of multiple targets using a commonly used pulse radar is challenging due to range ambiguity. Radar-based signal processing techniques, for example, provided in~\cite{multiple_targets1, multiple_targets2, multiple_targets3} can be used for the detection and tracking of multiple targets simultaneously.

Generally, a target is tracked using its position and velocity information. The track equation for radar is given as
\begin{equation}
   S/N = \frac{P_{\rm T}G^2\lambda^2\sigma}{(4\pi)^3 R^4 k T_{\rm s}B_{\rm n}L}, \label{Eq:track_eq}
\end{equation}
where $S/N$ is the SNR, $k$ is the Boltzman constant, $T_{\rm s}$ is the system noise temperature, and $B_{\rm n}$ is the noise bandwidth of the RX. The average noise power is represented by $kT_{\rm s}B_{\rm n}$. There are different tracking algorithms available in the literature. In \cite{track_new8} FMCW radar and MUSIC algorithm were used for multiple UAV tracking. In particular, two different UAVs were tracked simultaneously in an open area using the MUSIC algorithm at low altitudes. In \cite{track_new10}, high-resolution radar and Hough transform were used for tracking small UAVs. Hough transform provided a steady and continuous tracking of UAV using a linearly distributed micro-Doppler signature. The Hough transform helped to resolve the problem of scatterer clustering. Moreover, there are different methods available in the literature for the localization of a moving object based on target tracks~\cite{local1,local2,local3}. Kalman and particle filters are also popular for the localization and tracking of mobile nodes with radars~\cite{localization_kalman_particle}.    

In a pulse radar system, the coherent integration of received pulses is used to increase the SNR. The coherent integration requires both the in-phase and quadrature components of the received signal. Furthermore, compressed sensing can be used for sparse radar signals and a reduced number of sample scenarios. For example in ~\cite{compressed2,compressed1}, compressed sensing was used in the reconstruction of the sparse target scene that helped in the detection of the target in complex scenarios. Other radar estimation problems, e.g., finding the high-resolution direction of arrival of signals from multiple sources at a single snapshot were obtained through compressed sensing techniques in~\cite{compress_sensing2}.

Doppler processing is essential for estimating the velocity of an aerial vehicle in all the radar systems. The frequency shift of the received signal from the center frequency is used to obtain the Doppler shift and subsequently the velocity estimate of the aerial vehicle. Micro-Doppler processing can also be used to obtain the micro-Doppler signature of a moving aerial vehicle. In older radar systems, RCS and the velocity of an aerial vehicle were used for classification. However, nowadays, modern classification algorithms use additional target features for classification, e.g., micro-Doppler signature. The classification is also aided by AI algorithms. Deep learning AI algorithms help to accurately identify the type of the aerial vehicle~\cite{cognitive_deeplearning,UAV_ML_new2}.

\section{Analysis of Aerial Threats and Challenges} \label{Section:future_threats}
\begin{figure*}[!t]
	\centering
	\includegraphics[width=\textwidth]{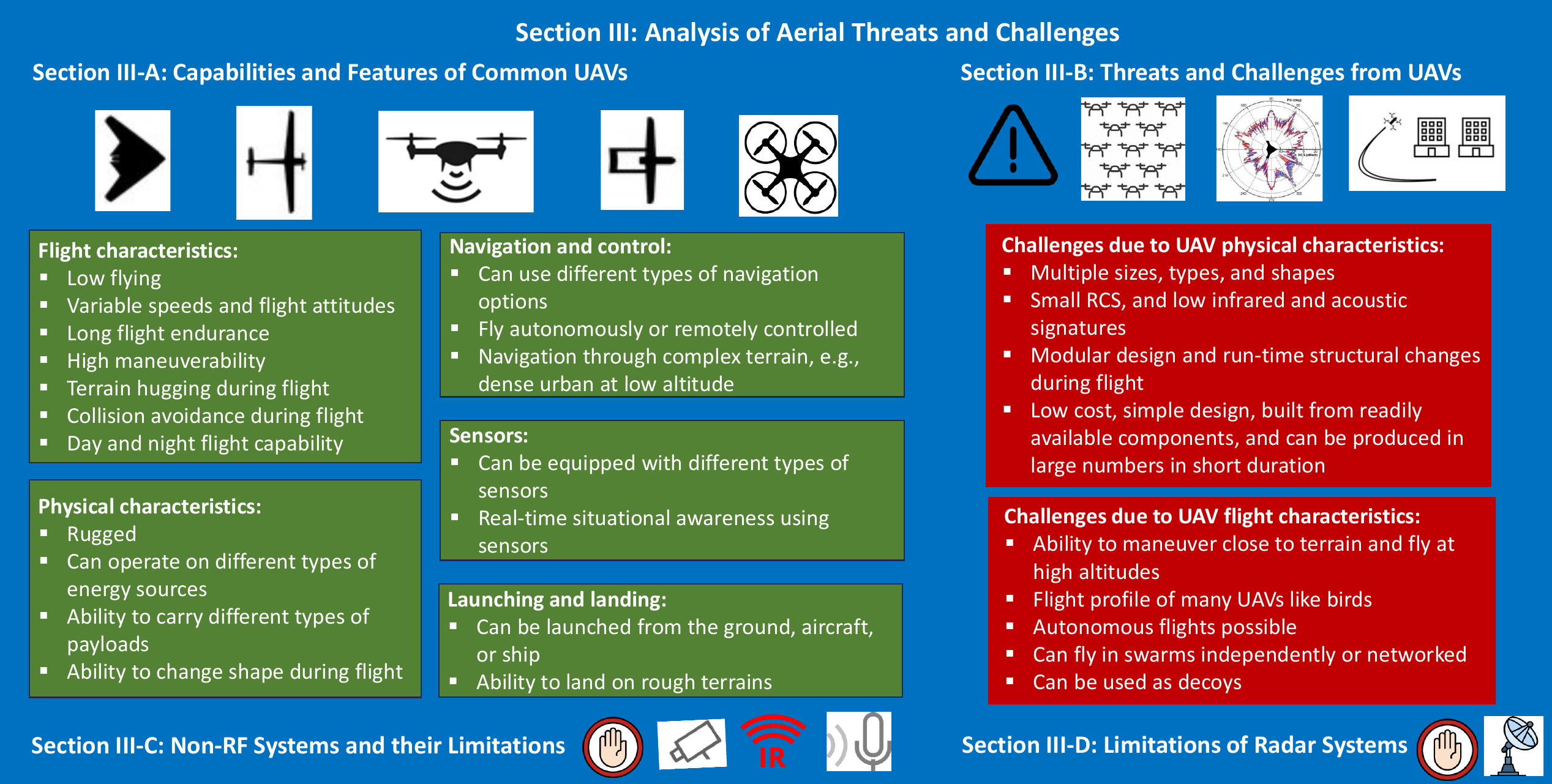}
	\caption{Capabilities and features of common UAVs and challenges due to these capabilities. Topics covered in Section~\ref{Section:future_threats} are also provided in this figure. The figure additionally outlines the structure of Section~\ref{Section:future_threats}.}  \label{Fig:capab_challenges}
\end{figure*}

In this section, the capabilities of UAVs, current and future aerial threats and challenges from UAVs, and limitations of radar systems in DCT-U are provided. A brief discussion of non-RF systems and their limitations in the DCT-U are also provided. Fig.~\ref{Fig:capab_challenges} highlights the capabilities and features of common UAVs and challenges in their detection. Fig.~\ref{Fig:capab_challenges} also provides a layout of the topics covered in this section.

\subsection{Capabilities and Features of Common UAVs} 

There are different types of UAVs available with different features and capabilities. The current and expected future capabilities and applications of UAVs were provided in~\cite{new_UAVchallenges}. UAV characteristics were also discussed, including flight altitude, payload, speed, range, and flight time. Table~\ref{Table:UAV_types} shows the major types and specifications of typical UAVs. Table~\ref{Table:UAV_types} provides maximum values of size, flight endurance, payload capacity, flight ceiling, and speed for UAVs sorted from the internet. Table~\ref{Table:UAV_types} shows that the UAVs generally have small size, weight, power consumption, heat signature, and RCS compared to manned aerial vehicles.
Moreover, fixed-wing and rotary-wing UAVs offer medium flight endurance, payload capacity, flight ceiling, and speed. The cost of UAVs is also significantly small. They can be easily produced in large numbers using simple designs and readily available off-the-shelf components. The simple design allows various sizes of UAVs. For example, a UAV can fit into the palm of a hand and have a wing span of more than $43$~m. We have summarized the major capabilities and features of UAVs in Fig.~\ref{Fig:capab_challenges}.

\begin{table*}[t]
	\begin{center}
     \footnotesize
		\caption{RCS of different types of UAVs measured using different types of radar systems. Key performance metrics and performance results of these systems are also provided.} \label{Table:RCS_radarsystems}
\begin{tabular}{@{}|P{ 2.2cm}|P{2.1cm}|P{1.7cm}|P{2.2cm}|P{3.0cm}|P{2.9cm}|P{0.6cm}|@{}}
 \hline
\textbf{UAV type}&\textbf{RCS}&\textbf{Center frequency}&\textbf{Sounding signal}&\textbf{Key performance metrics}&\textbf{Performance results}&\textbf{Ref.}\\
\hline
DJI Matrice 600 Pro, DJI Matrice 100, Trimble zx5, DJI Mavic Pro 1, DJI Inspire 1 Pro, DJI Phantom 4 Pro& Lognormal, generalized extreme
value, and gamma distributions&15 GHz, 25 GHz&CW &Far-field distance, polarized components, antenna gains, transmit power, SNR, and background subtraction &Classification accuracy based on RCS varied from $35.53$\% at $0$~dB SNR to $100$\% at $10$~dB SNR for different polarization and frequency &\cite{RCS_clutter1}\\
\hline
DJI Phantom 4 Pro& 0.01~m$^2$&25~GHz&FMCW & Probability of detection and false alarm, signal to clutter ratio, range resolution, detection range, and micro-Doppler signatures&UAV with RCS of $0.01$~m$^2$ was detected at $150$~m distance for a transmit power of $30$~dBm&\cite{RCS_clutter2}\\
\hline
DJI Phantom 3& 0.01~m$^2$&Ku band&Pulse based phased array radar & Detection range, range and velocity resolution, and track while scan capability &Detection range up to $10$~km, and capable of tracking $100$ targets simultaneously&\cite{clutter3}\\
\hline
Mavic Pro& 0.03~m$^2$&2.4~GHz&CW, linear frequency modulated &Detection range, range and velocity resolution, probability of detection, and waveform type and duration&The effective detection range was up to $1020$~m, range resolution was $3.75$~m. The radar was capable of detecting targets at speeds less than $1.5$~m/sec and velocity resolution was $0.21$~m/sec&\cite{clutter4}\\
\hline
Iris+, X8, and High one& See Table~II in \cite{clutter5}&3~GHz, 9.7~GHz, 15~GHz, and 24.3~GHz&Pulse Ku-band
short range battlefield radar&Maximum detection range, detection and false alarm probability, and SNR&UAVs were detected at a distance of $1$~km, with estimated SNR varied from $6$~dB to $26$~dB, and detection probability varied from $0.234$ to $0.997$ for different types of UAVs&\cite{clutter5}\\
\hline
Phantom 3, fixed-wing buzzard& Maximum value 0.09~m$^2$&X-band&FMCW&Detection range, clutter and its rejection capability, and Doppler effect&Reliable detection was made at $600$~m for fixed wing and $700$~m for quadcopter when oriented to maximize RCS&\cite{clutter6}\\
\hline
UAV& 0.08~m$^2$&S-band&Pulses&Detection range and detection probability, and clutter management&UAV detection range was $2,340$~m with a detection probability of $0.5$ for a RCS median value of $0.08$~m$^2$. The detection ranges for various bird species ranged from $707$~m to $1,537$~m &\cite{clutter11}\\
\hline
\end{tabular}
		\end{center}
			\end{table*}

\subsection{Threats and Challenges from UAVs}
UAVs operated by amateur users who do not follow regulations pose a significant threat to the surroundings. Moreover, UAVs can be used by non-state actors for malicious purposes. Threats from UAVs were discussed in \cite{wahab_uav_threats,new_uavthreats} and highlighted various vulnerabilities. Countermeasures against the UAV threats were also provided. The threat analysis of UAVs in different scenarios was also provided in \cite{new_uavthreats2}. A realistic threat scenario using hacking was simulated to provide awareness. Examples of major current and future threats from malicious UAVs can be listed as follows: 1) use by non-state actors during conflicts; 2) conventional,
biological, and chemical threats carried by UAVs; 3) threats to sensitive infrastructure, e.g., dams, bridges, important buildings, power generation plants, chemical, and nuclear facilities; 4) threats to important persons, vehicles, and locations; 5) threats to crowded areas; 6) malicious activities e.g., starting a fire, and identity theft; 7) smuggling of contraband articles using UAVs; 8) threats to the aviation industry; 9) planting improvised explosive devices, and mines on ground and at sea; 10) hacking UAVs and flying them for malicious purposes; 11) eavesdropping; 12) ECM using UAVs; 13) use as decoys; 14) different types of UAVs flying in a swarm; 15) sea surface hugging UAVs, and 16) threats from high-altitude flying platforms, e.g. hot-air balloons that can fly at high altitudes and carry large payloads.

There are many challenges to the accurate and timely DCT-U. In~\cite{new_UAVchallenges,wahab_uav_threats,new_uavthreats2}, different challenges in UAV detection and overcoming UAV threats and vulnerabilities were discussed. Challenges in the detection and classification of UAVs were also covered in the survey~\cite{new_uavchallenges2}. The major such challenges due to the unique capabilities of UAVs are summarized in Fig.~\ref{Fig:capab_challenges} and given as follows:
\begin{itemize}
    \item Multiple types and shapes of UAVs are difficult to classify.
    \item UAVs flying in a discontinuous flight path are difficult to detect and track. For example, multiple landings during the flight of a UAV at random intervals can pose challenges to detection and tracking. 
    \item Hybrid UAVs that can fly and move over land and water can achieve long operational ranges.
    \item Modular design and changes during the flight of UAVs, e.g., the continuous disintegration of UAVs into multiple autonomous UAVs can make detection, tracking, and classification challenging.  
    \item Smart structural modifications of UAVs using AI and 3D printing during flight can make DCT-U difficult. 
    \item UAVs can use AI to make autonomous run-time decisions based on the environment, clutter, and RF emissions in the environment and find the optimum route. AI-controlled autonomous UAVs have better situational awareness and are difficult to detect and track. Moreover, trajectory control of UAVs using multiple sources, e.g., onboard smartphones, and navigation using telecommunication coverage can make the UAV resilient to different RF countermeasures.
    \item Another major challenge is the classification of small, low-flying UAVs from birds and the identification of malicious UAVs from hobbyist UAVs.
    \item UAVs flying in a swarm are difficult to detect, track, and classify~\cite{multipleUAVs_difficulty,swarm_detect,UAV_swarms_new} especially when they fly at different altitudes and at different velocities in a specific trajectory are difficult to detect, track, and classify.
    \item Decoy UAVs that exist either physically or created virtually using ECM techniques for deception in a swarm can result in the wastage of resources on their detection and tracking.
    \item UAVs can be produced in large numbers due to simple design and off-the-shelf readily available components. On the flip side, effectively countering UAVs necessitates highly sophisticated systems. However, it's important to note that these advanced counter-UAV systems come with a considerable financial burden, making them expensive to acquire and manage.
    \item In a complex environment, e.g., in a dense urban area with high-rise buildings and dynamic clutter~(from civilian planes, cars, pedestrians, and birds), the detection and tracking of small UAVs becomes challenging. 
	\end{itemize}

\subsection{Non-RF Systems and their Limitations} \label{Section:Nonradar_nonRF}
The popular non-RF systems for DCT-U are EO/IR and acoustic. A brief introduction and limitations of these non-RF systems for DCT-U are as follows: 
\begin{itemize}
\item Passive imaging sensors such as EO/IR can be used for DCT-U with high precision. High-resolution imaging provided by EO sensors helps to capture detailed images of UAVs for accurate DCT-U. IR sensors provide DCT-U in day and night. In \cite{EO_new,new_EO1}, the performance of different algorithms for real-time detection of UAVs using EO sensors was provided. A thermal infrared camera is used for the nighttime detection of UAVs in \cite{IR_new}. However, in an EO/IR sensor, the complexity of the target's background and thermal image saturation affects the detection and classification performance of the EO/IR sensors for UAVs. The range of the EO/IR is also limited by the horizon. The atmospheric effects, e.g., haze, fog, rain, and snow can affect the performance of EO/IR sensors. Detection and tracking of a UAV were carried out using a camera and a radar in~\cite{new_radar_camera}. The accuracy of the radar measurements and images was enhanced using a Kalman filter and convolutional neural network~(CNN), respectively.
\item Acoustic signals from a flying UAV can be detected using an acoustic sensor, e.g., a microphone. The specific acoustic signal patterns can be analyzed for DCT-U. Acoustic sensors can better detect UAVs than EO/IR sensors in challenging environments, e.g., dense foliage. In ~\cite{survey_new1,acoustic_track_new}, acoustic signals captured by a microphone were used for tracking and classifying UAVs. A major limitation of acoustic methods is their poor performance in noisy environments and their range is limited. The acoustic methods work in the passive mode and a single acoustic sensor cannot provide precise localization of the UAV. Furthermore, the acoustic methods work poorly against high-altitude UAVs. 
\item Sensor fusion of different sensors can be used for DCT-U. For example, EO/IR and acoustic sensors can be used in a network. Multiple heterogeneous sensors can use sensor fusion to overcome the weaknesses of the individual sensors and provide combined strength~\cite{sensor_fusion1}. However, there are limitations due to networking and delays arising from the central data processing of different sensors.   
\end{itemize}

 \begin{table*}[t]
	\begin{center}
     \footnotesize
		\caption{Comparison of different sensors over a range of performance indicators (adapted from ~\cite{comparison}).} \label{Table:comparison_radar}
\begin{tabular}{@{}|P{ 4.0cm}|P{ 1.8cm}|P{ 1.8cm}|P{1.3cm}|P{1.3cm}|P{1.3cm}|P{1.3cm}|P{1.3cm}|@{}}
\hline
&\textbf{Long range radar with micro-Doppler}&\textbf{Short range high frequency radar with micro-Doppler}&\textbf{Airborne short range radar}&\textbf{RF triangulation}&\textbf{Clustered audio sensors}&\textbf{EO/IR sensor}&\textbf{Airborne EO/IR sensor}\\
\hline
\textbf{Range}&Extreme&Medium&Medium&Good&Short&Medium&Medium\\
\hline
\textbf{Dynamic environment coverage}&Bad&Bad&Good&Good&Good&Bad&Good\\
\hline
\textbf{Detection in urban environment}&Medium&Medium&Medium&Medium&Medium&Bad&Good\\
\hline
\textbf{Detection in darkness}&Good&Good&Good&Good&Good&Good&Good\\
\hline
\textbf{Distance to target}&Good&Good&Good&Good&Bad&Bad&Bad\\
\hline
\textbf{Speed of target}&Good&Good&Good&Good&Bad&Bad&Good\\
\hline
\textbf{Mobility}&Bad&Medium&Good&Bad&Medium&Good&Good \\
\hline
\textbf{Detection in fog}&Good&Good&Good&Good&Good&Bad&Bad\\
\hline
\textbf{Detection in rain}&Good&Good&Medium&Good&Medium&Medium&Medium\\
\hline
\textbf{Weight}&High&Medium&Low&High&Medium&Low&Low\\
\hline
\textbf{Rogue UAV detection}&Yes&Yes&Yes&No&Yes&Yes&Yes\\
\hline
\textbf{UAV swarm detection}&Medium&Medium&Good&Good&Good&Good&Good\\
\hline
\textbf{Processing power}&High&High&High&High&Low&Medium&Medium\\
\hline
\textbf{Power consumption}&High&Low&Low&Medium&Medium&Low&Low\\
\hline
\textbf{Low speed UAV detection}&Good&Good&Good&Good&Good&Good&Good\\
\hline
\textbf{Low cost}&Bad&Medium&Medium&Medium&Medium&Good&Good\\
\hline
\textbf{Maintenance}&Good&Good&Bad&Good&Medium&Medium&Bad\\
\hline
\textbf{Installation ease}&Bad&Medium&Good&Medium&Medium&Medium&Good\\
\hline
\end{tabular}
		\end{center}
			\end{table*}

\subsection{Limitations of Radar Systems}  \label{Section:Limitations_radars}
DCT-U of typical UAVs is challenging due to their unique capabilities discussed earlier. While radar is the most popular sensor for DCT-U, there are some common limitations of radar systems listed as follows:
\begin{itemize}
    \item In general, radar systems are inefficient because a large amount of energy is transmitted, whereas, only a fraction of the transmitted energy is received back from clutter and from a potential aerial vehicle. Also, the RF energy emissions from radar systems expose them to detection. 
    \item The commonly available small UAVs have small RCS. In addition to the small RCS of UAVs, if radiation-absorbent material and specific shapes are used, the RCS of the UAVs can be further reduced. Small RCS UAVs are difficult to detect by common radar systems at long ranges and cluttered surroundings. The RCS of different types of UAVs reported in the literature at different frequencies are overviewed in Table~\ref{Table:RCS_radarsystems}. From Table~\ref{Table:RCS_radarsystems}, it can be observed that the majority of RCS measurements were carried out at frequency bands~$>2~$GHz and the RCS of the majority of commercially available UAVs is less than $0.1$~m$^2$.
    \item Extreme flight maneuvers are possible with UAVs due to the absence of a pilot. For example, UAVs can fly very close to the terrain to avoid detection by radar systems.
   \item The range and Doppler ambiguities in popular pulse-Doppler radars put bounds on the range and velocity measurements of the target. The range and Doppler ambiguities become notable when detecting multiple UAVs at different ranges and velocities, e.g., in a swarm.  
   \item The DCT-U of multiple small UAVs in a swarm requires high angular and range resolution radar. However, the scan duration and complexity of the radar system increase with the increase of the range and angular resolution. 
    \item The detection of small and low-flying UAVs in highly cluttered environments requires adaptive adjustments of radar thresholds in order to keep the PFA low.
    \item The typical radar systems have limitations in tracking very high altitude, highly maneuverable, and high-speed UAVs. 
    \item The radar systems are vulnerable to ECM applied by UAVs and require electronic counter-countermeasures~(ECCM) against ECM. Popular offensive ECM that can be applied by UAVs include jamming hostile radio signals, spoofing the radio signals with false data~\cite{jamming}, or transmitting multiple copies of the echo signals as decoys~\cite{ECCM,ECCM2}.
    \item Noise, RF interference, and clutter affect the performance of a radar system.
    \end{itemize}

A comparison of radar and non-RF systems over different performance indicators~\cite{comparison} is provided in Table~\ref{Table:comparison_radar}. From Table~\ref{Table:comparison_radar}, it can be observed that short-range airborne radar has the overall best performance when detecting UAVs. The short-range high-frequency radar with micro-Doppler also has good performance in detecting UAVs compared to other sensors. Furthermore, some of the limitations of the EO/IR sensor can be overcome by mounting it on the airborne platform.


\section{DCT-U using Conventional and Modern Radar Systems}  \label{Section:radar_comm_det}
 In this section, the literature on the use of conventional and modern radar systems for DCT-U is covered. This section also outlines how modern radar systems address the limitations of typical radar systems for DCT-U.
			
\subsection{Conventional Radar Systems for DCT-U} \label{Section:Radar_types}
Conventional radar systems can be categorized as radar systems that have simple hardware, and basic signal processing and are typically used. For example, a monostatic real aperture radar with basic signal processing is shown earlier in the survey in Fig.~\ref{Fig:radar_basic1}. Pulse Doppler, CW, and FMCW radars can also be categorized as conventional radars and are often used for DCT-U. In \cite{pulse_doppler_uav1}, a pulse Doppler radar and CNN were used for the detection and localization of UAVs with minimum constant false alarm rate~(CFAR). A two-head CNN was used, one for the patch of range-Doppler map classification and the other for the offset regression between the patch center and the target. Subsequently, a non-maximum suppression was used to keep CFAR low for small UAVs. 

A conventional pulse Doppler radar and Hough transform were used for the detection and tracking of small UAVs in \cite{track_new10}. The linear distributed features of micro-Doppler and Hough transform helped to detect and identify small UAVs. In \cite{UAV_CW}, a CW radar was used for the detection of small UAVs. Two stages were used for the detection of UAVs using CW radar. First was the extraction of UAV parameters using the Gaussian mixture model and then trajectory extraction using multi-frame information. In \cite{fmcw1}, an X-band radar was used for the detection and tracking of small UAVs with RCS of $.01$~m$^2$. The FMCW radar was able to detect small quadcopter UAVs at a distance of $20$~m to $500$~m. The radar was able to differentiate between two UAVs flying at a distance of $20$~m apart. FMCW radar was also used for the detection of UAVs in~\cite{UAV_FMCW_new,UAV_FMCW_new35,SVM_classify1}. 

Other commonly used radars for DCT-U include marine and staring radars. In \cite{clutter6} a marine broadband FMCW radar operating at X-band was used for the detection of two different types of UAVs. Measurements were carried out to detect rotary wing and fixed wing UAVs using the marine radar at a distance of $400$~m. Similarly, in \cite{marine_radar_new}, low-cost marine radar was used for the detection of UAVs. The marine radar was mounted on a UAV and provided $360^{\circ}$ coverage over a large area. In \cite{UAV_Lband_staring} a staring radar operating at L-band was used to differentiate between UAVs and birds. The measurements were taken in an urban area and the measurement data was used as training data for classification algorithms. In \cite{UAV_Lband}, Aveillant Gamekeeper staring radar was used for surveillance of non-cooperative UAVs. The radar operated at L-band and used extended dwell that helped in small UAV detection over a range of several kilometers. Moreover, ML classifiers were used to distinguish between UAVs and birds using the $3$D radar data. In \cite{new_surveillance_radar}, a surveillance radar was used for detection and classification using deep learning based on range-Doppler images, which were processed using the YOLO framework.

\begin{figure}[!t]
	\centering	\includegraphics[width=\columnwidth]{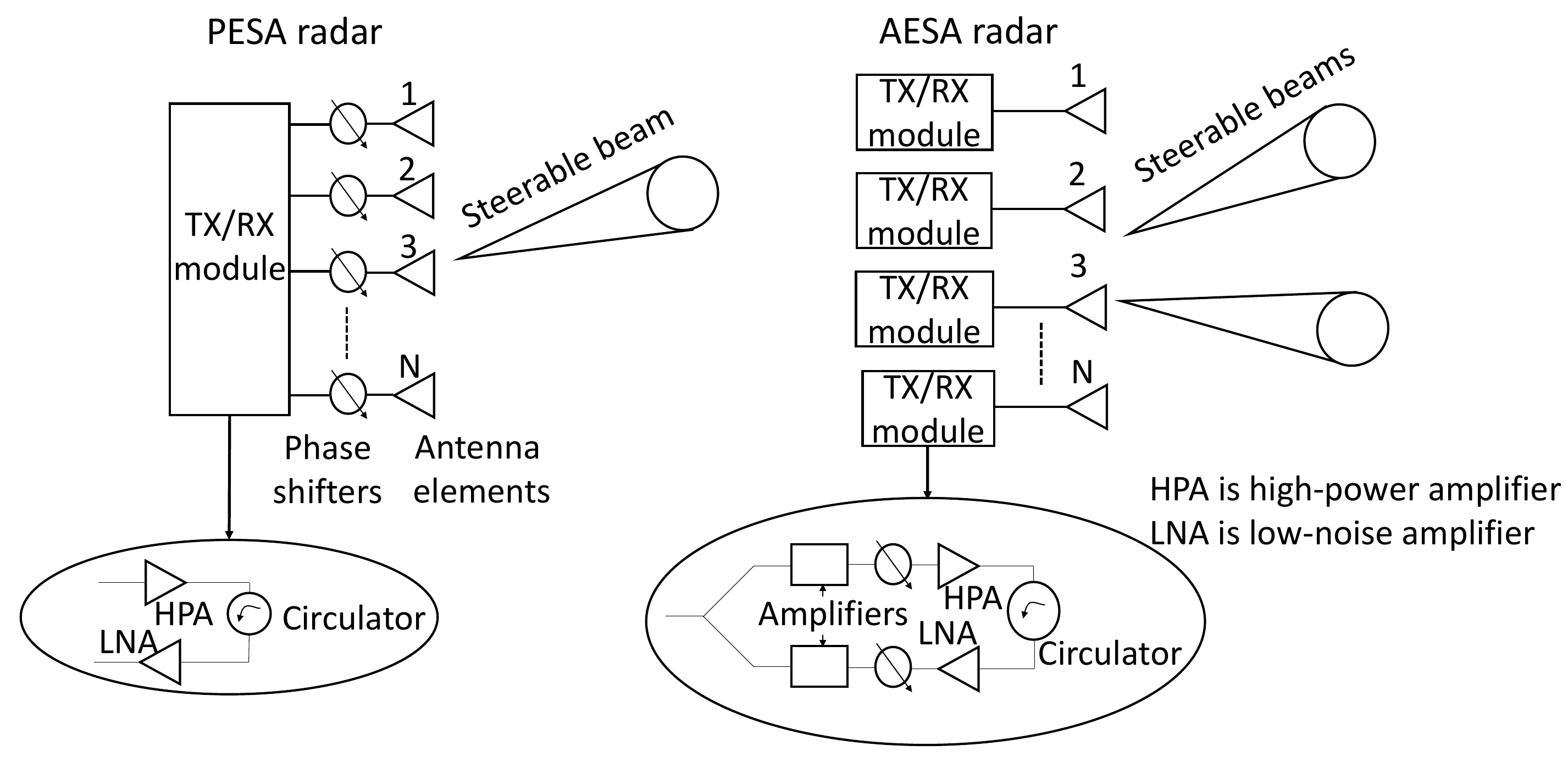}
	\caption{PESA and AESA radars are shown. AESA radars have multiple TX/RX modules that can work independently and act as individual radars~(regenerated from \cite{AESA1,new_fig02}}). \label{Fig:AESA_PESA}
\end{figure}

\subsection{Detection and Tracking of UAVs Using Modern Radar Systems}
Many modern radar systems use millimeter wave~(mmWave) frequencies for UAV detection and tracking, as given in~\cite{new_mmwaveradar, new_mmwaveradar2,Wahab_AESA}. The mmWave frequencies help realize concise electronically steered phased arrays, which are used in the majority of modern radar systems. Passive electronically scanned array~(PESA) and AESA are two main types of electronically steered phased-array-based radar systems shown in Fig.~\ref{Fig:AESA_PESA}. AESA is widely used nowadays to perform search and track tasks simultaneously~\cite{AESA1,AESA2}. Multiple AESA beams are used to perform different tasks simultaneously, e.g., communications, passive listening, and imaging, in addition to search and track tasks. The range equation for AESA radar~\cite{AESA_rangeeq} is modified compared to the conventional radar equation provided in (\ref{Eq:radar_range}). In particular, the contribution of individual TX/RX modules is considered in the AESA range equation given as
\begin{equation}
  R = \bigg({\frac{N^3p\pi^2\lambda^2\sigma T_{\rm d}}{(4\pi)^3kT_{\rm s}D_{\rm x}(n')L_{\rm t}L_{\rm a}}}\bigg)^\frac{1}{4}, \label{Eq:AESA}
\end{equation}
where $N$ is the number of TX and RX modules, $p$ is the mean power of each TX/RX module, the antenna gain at the TX and RX is $G = \pi N$ for the broadside direction with $N$ TX/RX modules, $T_{\rm d}$ is the dwell time, $D_{\rm x}(n')$ is the effective detectability factor~(see \cite{AESA_rangeeq}), $L_{\rm t}$ is the transmission line loss, and $L_{\rm a}$ is the atmospheric loss.

\begin{figure}[t]
\captionsetup[subfigure]{justification=centering}
     \begin{subfigure}[b]{0.5\textwidth}
         \centering
         \includegraphics[width=\textwidth]{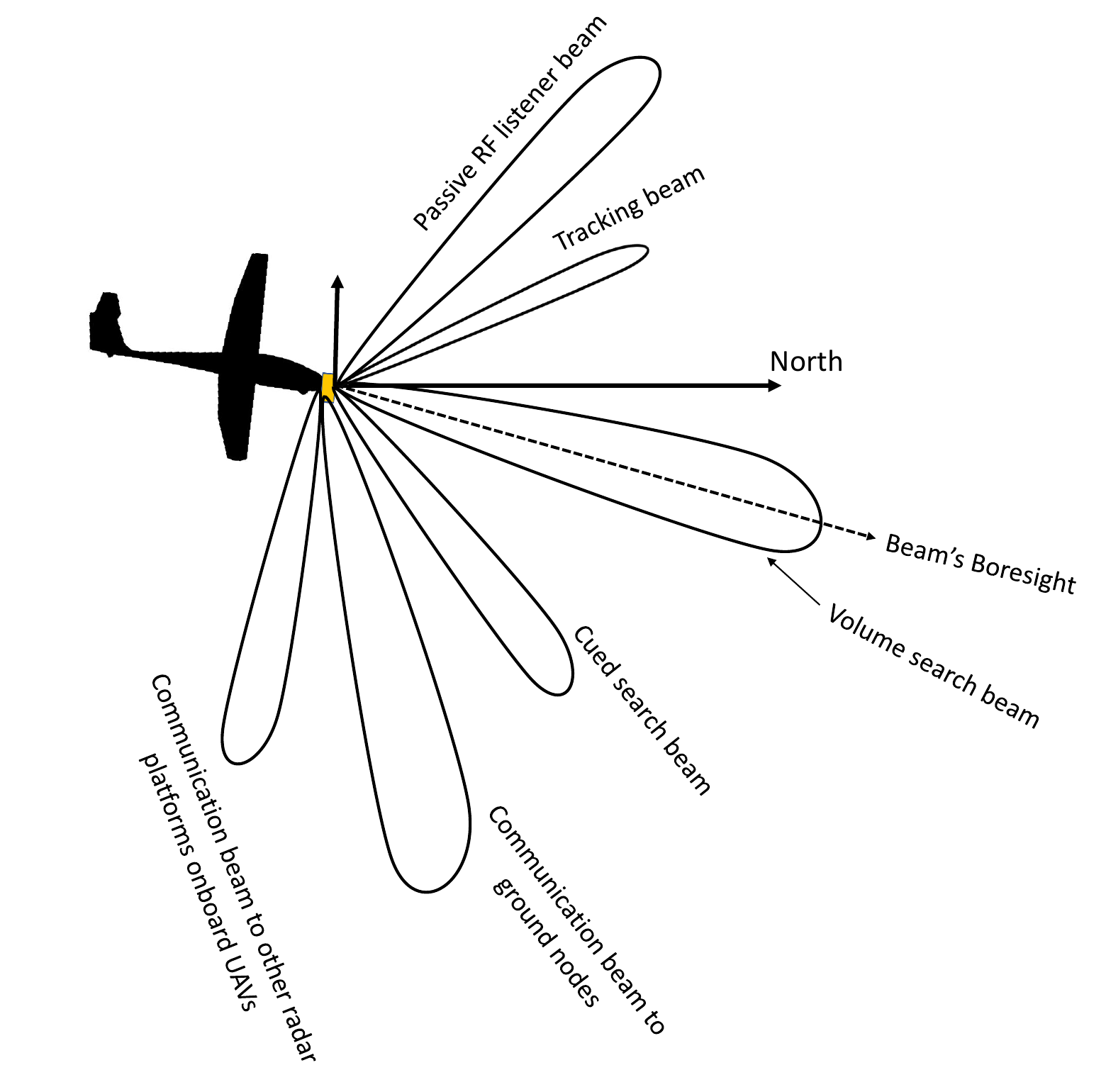}
         \caption{{}}
    \end{subfigure}
     \hfill
    \begin{subfigure}[b]{0.5\textwidth}
         \centering
         \includegraphics[width=\textwidth]{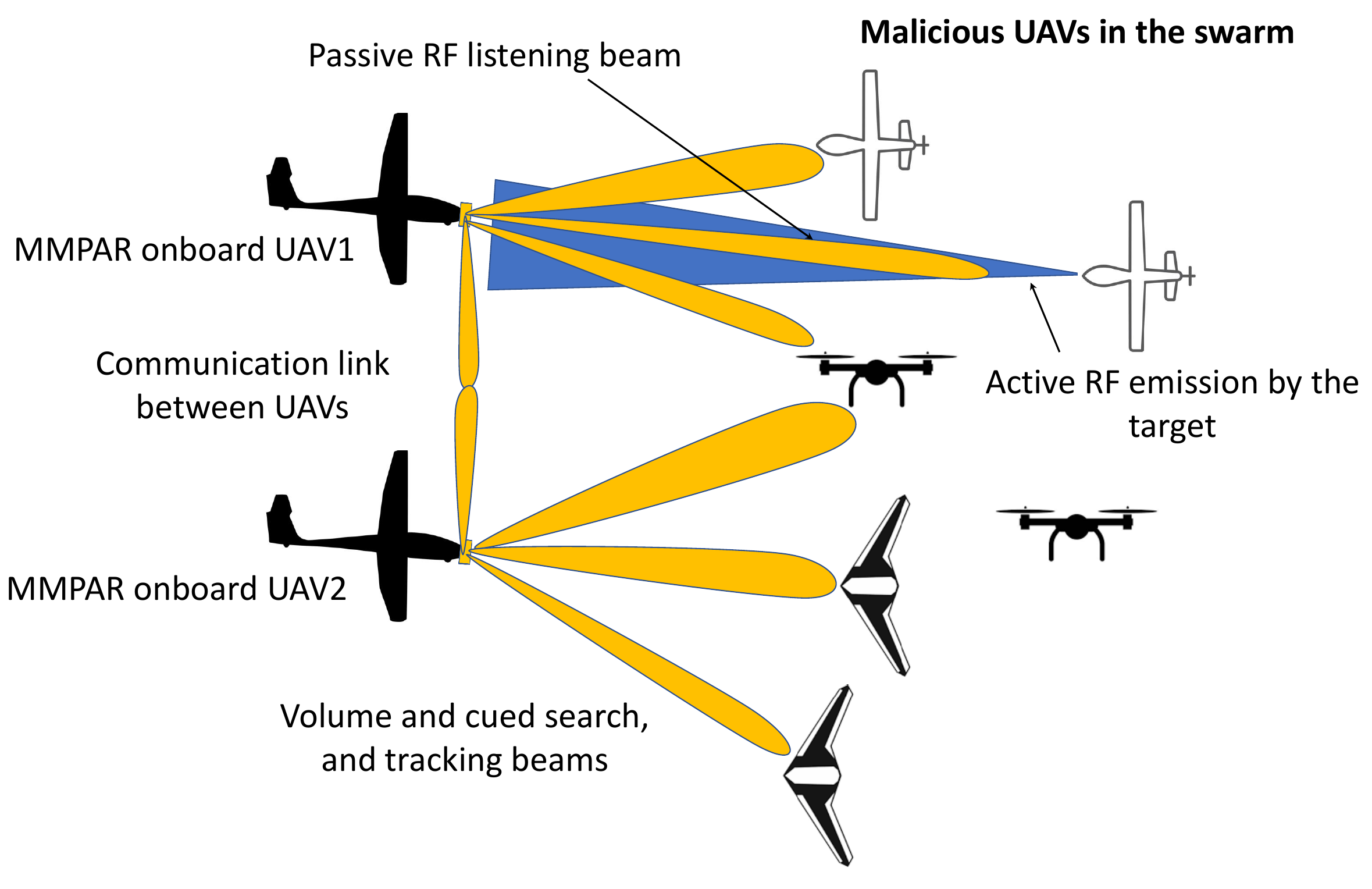}
         \caption{{}}
    \end{subfigure}
         \caption{MMPAR radar onboard UAV for DCT-U swarm~(regenerated from~\cite{Wahab_AESA}); (a) MMPAR onboard a UAV using different types of beams simultaneously, (b) a scenario where two MMPAR onboard two UAVs detect and track multiple UAVs simultaneously using volume and cued search, and tracking beams. Also, the passive RF listening beam is used for the detection of active RF transmissions from a malicious UAV in the swarm. The communication beams between the two UAVs provide a continuous communication link. }    \label{Fig:MMPAR_fig}
\end{figure}

In \cite{Wahab_AESA}, a multifunction multibeam phased array radar~(MMPAR) placed onboard a UAV was used to detect and track multiple UAVs in a swarm simultaneously. The MMPAR in \cite{Wahab_AESA} worked autonomously using reinforcement learning. The multifunction beams used by MMPAR and malicious UAV swarm detection and tracking scenarios using MMPAR onboard UAVs are shown in Fig.~\ref{Fig:MMPAR_fig}(a) and Fig.~\ref{Fig:MMPAR_fig}(b), respectively. The multiple radar beams are used for volume search, cued search, tracking, communications, and passive RF listening. The overall radar resources are shared among different radar beams at different scanning instances. In addition to AESA radar, other modern radar systems for DCT-U are provided in the following subsections.

    \subsubsection{Bistatic and Multistatic Radars} Bistatic and multistatic radar systems can collect reflections from a UAV at different locations and increase the detection probability for a small RCS target. In \cite{UAV_passive_bistatic}, a multibeam and multiband software-defined bistatic passive radar was used for the detection of small RCS UAVs. In \cite{track_new2}, a multistatic radar system was used for the detection and tracking of a micro-UAV. Micro-Doppler signatures were used to differentiate between UAV returns and ground clutter. The measurements from multiple bistatic pairs were fused together for accurate tracking of the UAV. 
    \subsubsection{Cognitive Radars} Cognitive radar systems can autonomously detect and track UAVs by selecting optimum radar parameters based on the scenario~\cite{cognitive_radar_survey}. In \cite{cognitive_radar2}, a cognitive radar system for the detection and classification of micro-UAVs was proposed. A binarized deep belief network and micro-Doppler signatures were used for the detection and classification of UAVs. Similar to cognitive radar systems, adaptive radar systems can also be used for the detection and tracking of UAVs. In \cite{monopulse_uav1}, a phased array C-band radar operating at multiple channels was used. The adaptive architecture helped to resolve a single UAV from multiple UAVs in the range-Doppler and angular domains while optimally utilizing the resources. In~\cite{cognitive_ECCM}, a cognitive radar overcomes an adversary's ECM by deviating from optimal behavior. This misleads the adversary and makes the ECM ineffective while maintaining its own performance optimally.
    \subsubsection{Software-defined Radio-based Radars} Software-defined radios~(SDRs) offer many advantages, e.g., flexibility during run-time changes, system upgrades without significant hardware changes, and cost-effective solutions. SDRs can be used to implement different types of radar systems for the detection and tracking of UAVs~\cite{survey_new2}. In \cite{SDR_UAV1}, an SDR-based FMCW radar was used for the detection and localization of UAVs. The detection and localization of UAVs were carried out using off-the-shelf commonly available SDRs. Similarly, in \cite{survey_new2}, low-cost SDRs were used for the detection, tracking, jamming, and spoofing of unauthorized UAVs. Real-time measurements were carried out to prove the effectiveness of the proposed radar system.

           \begin{table*}[htbp]
	\begin{center}
     \footnotesize
		\caption{Tracking of UAVs by radar systems using different tracking algorithms. The performance metrics and results for tracking are also provided.} \label{Table:Tracking_algos}
\begin{tabular}{@{}|P{1.8cm}|P{1.8cm}|P{1.8cm}|P{2.5cm}|P{3.2cm}|P{3.1cm}|P{0.65cm}|@{}}
 \hline
\textbf{UAV type}&\textbf{Radar/sensor type} &\textbf{Tracking features}&\textbf{Tracking algorithm}&\textbf{Key performance metrics}&\textbf{Performance results}&\textbf{Ref.}\\
\hline
F450 and phantom 3&FMCW radar, acoustic and optical sensors&Data from three sensors&Kalman filter, single source detection algorithm, multiple drone detection algorithm, MUSIC algorithm&Detection range, azimuth coverage, simultaneous UAV tracking, and effect of background structures in urban environment&NA&\cite{track_new8}\\
\hline
Miniature rotor UAV&Ku-band radar&Micro-Doppler features&Hough transform&Detection and tracking accuracy, Doppler bandwidth, and signal to clutter ratio&The conventional method had a root mean square error~(RMSE) of $1.53$~m, whereas the tracking accuracy increased with Hough Transform and RMSE reduced to $0.53$~m&\cite{track_new10}\\
\hline
Micro-UAV&Multistatic radar NetRAD&Micro-Doppler, 2D aerial vehicle state, bi-static range&Extended Kalman filter&Track loss rate, tracking latency, and tracking initiation, maintenance, stability and smoothness&Accurate tracking of UAV was achieved with RMSE values between $0.5$~m to $1.5$~m&\cite{track_new2}\\
\hline
Motoar Sky MS-670&FMCW Interferometric radar&Radial velocity&Kalman filter&Radial and angular velocity estimation, estimated trajectory accuracy, SNR, and Kalman filter performance&Tracking errors were within the range of the observation noise variance, that was a standard deviation of approximately $3.16$ for the measured velocities&\cite{track_new1}\\
\hline
Multi-rotor UAV&Networked radar systems&Detection from multiple radars&Recursive random sample consensus algorithm, tuned Kalman filter&Detection range, tracking accuracy, false alarm rate, latency, and coverage area&Tracking accuracy was within the range of $0.5$~m to $2.5$~m &\cite{track_new3}\\
\hline
Multi-rotor UAVs&A dynamic radar network onboard UAVs&State space model&Local Bayesian estimator, extended Kalman filter& Localization and tracking accuracy, and control and navigation performance&RMSE value of $0.15$~m for position tracking was achieved with $10$ UAVs in precise ranging-only configuration &\cite{track_new4}\\
\hline
DJI Phantom 3, Parrot Bebop 2, Parrot Disco, DJI MAVIC Pro.&Velodyne VLP-16 lidar&Speed, overall motion, laser scanned data&Kalman filter&DCT-U accuracy using lidar, maximum range, and corresponding range resolution &NA&\cite{track_new6}\\
\hline
UAVs&Mobile radar&3D position coordinates and velocity of the UAV&Kalman filter variants with east-north-up improvements&Type of Kalman filtering algorithm used and their deviation and divergence, and computational complexity of Kalman filters&For static and slow-moving radars, RMSE was $44.6$~m to $45.2$~m. For moderately moving radars, the RMSE was $44.75$~m to $44.77$~m. For fast-moving radars, the RMSE value was $50.8$~m&\cite{white_noise_track}\\
\hline
\end{tabular}
		\end{center}
			\end{table*}
   
   \subsubsection{Interferometric Radars} Interferometric radars use the phase shift of the received radar signals at two different points to determine the direction of arrival. In \cite{track_new1}, an interferometric radar was used for the detection and tracking of UAVs. The range and angular velocity information were obtained through interferometric radar measurements. The UAV was also tracked by predicting future positions using range and angular velocity measurements. In \cite{UAV_interferometric_new}, a phase interferometric Doppler radar was used for the detection and tracking of UAVs. The measurements were obtained using a dual-channel Doppler radar. Range, azimuth position, and Doppler processing were carried out. The joint range-Doppler-azimuth processing was used for the detection and tracking of micro-UAVs.
 
    \subsubsection{MIMO Radars} 
\begin{figure}[!t]
	\centering	\includegraphics[width=\columnwidth]{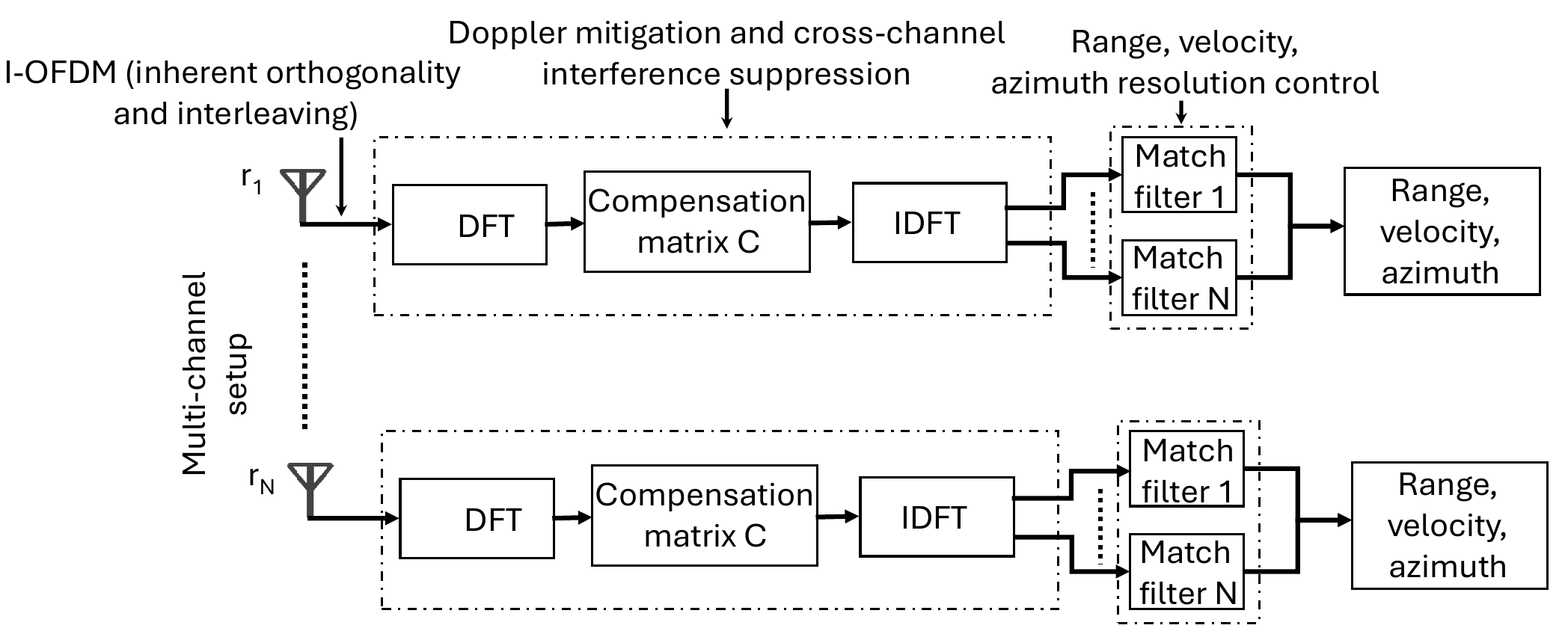}
	\caption{Processing at the receiver of interleaved OFDM MIMO radar~\cite{radar_centric5}, where $r_1,\cdots,r_{N}$ are the received signals at $N$ antenna elements, and DFT and IDFT represent the discrete Fourier transform and inverse discrete Fourier transform}. \label{Fig:Fig_modern_radar_MIMO}
\end{figure}

Multiple-input-multiple-output~(MIMO) radars operate similarly to multistatic radars. However, instead of spatially distributed nodes as in multistatic radars, the antenna elements are placed closely. MIMO radars provide high spatial and Doppler resolution and dynamic range. In \cite{UAV_MIMO1,MIMO_radar1}, MIMO radars were used for the detection and tracking of UAVs. In \cite{UAV_MIMO_new}, small UAV detection was carried out using a MIMO radar and sparse array optimization and calibration method. In \cite{digital_beamforming}, an S-band MIMO radar was used for the detection of small UAVs. A target-based calibration of the MIMO radar array was carried out to improve SNR. The MIMO radar was able to detect and track Phantom~$4$ UAV at a distance of $5$~km. In \cite{radar_centric7}, a novel frequency hopping waveform was used for MIMO radar that helped to estimate the hopping frequency sequence~(without degrading radar performance), timing offset, and channel. In \cite{radar_centric5}, a novel method for generating mutually orthogonal inherent-OFDM~(I-OFDM) signals for MIMO radar was introduced. This approach enhances Doppler immunity and suppresses inter-channel interference. The receiver processing, illustrated in Fig.~\ref{Fig:Fig_modern_radar_MIMO}, includes Doppler mitigation, cross-channel interference reduction, and adjustment of range, velocity, and azimuth resolution for I-OFDM-based signals.
    \subsubsection{Passive Radars} Passive radar systems can also be used for the detection and tracking of UAVs. A major advantage of passive radar systems is that a dedicated emitter is not required. The target illuminator can be any RF source. The illuminator sources for passive radars can be GSM, LTE, $5$G, WiFi, radio, or TV signal transmissions. Moreover, the multiple passive listener nodes at different locations can help to collect weak scattered reflections from a UAV. The majority of communication-based radars discussed in Section~\ref{Section:Comm_Sys} are passive. 
    \subsubsection{Compressed Sensing-based Radars} Compressed sensing helps to achieve better range-Doppler resolution compared to traditional radar systems in cluttered and sparse radar signal reception scenarios~\cite{compressed1,compressed2}.  In \cite{clutter_compress_sen}, a UWB radar with a compressed sensing-based adaptive filter was used for UAV detection. Compressed sensing helped to suppress the clutter and achieved a low false alarm rate. In \cite{compressed_uav}, simulations were carried out using passive radars for the detection of UAVs using compressed sensing and the Monte Carlo method. The effect of different factors on the detection of UAVs, e.g., antenna placement, and directivity was also covered.
    \subsubsection{Microwave Photonic Radars} Microwave photonic radars use large bandwidth, provide resistance to microwave interference at long ranges, and have the ability to simultaneously operate multiple coherent signals. In \cite{microwave_photonics, microwave_photonics2}, latest trends and advances using photonic radar technology were discussed. In \cite{photonic_uav}, a photonic radar operating in the X-band was used for the detection of small RCS UAVs. The radar used photonic frequency for transmission. The received signal was detected using a delay interferometer and photo-detector. The photonic radar was able to detect small RCS UAVs at a distance of $2.7$~km. In \cite{photonic_uav2}, a photonic radar using $11$~GHz bandwidth and a simple signal generation and processing setup was used for high-resolution identification of moving propellers of a UAV. 
    \subsubsection{Monopulse Radars} Monopulse radars can be used to obtain the accurate position of a UAV as well as the detection and tracking of a UAV. Such radars use additional coding of the RF signal to obtain accurate directional information. In~\cite{monopulse_uav1}, a multi-channel phased array radar operating in the C-band was used for the detection of multiple UAVs. The UAVs were closely spaced in the angular domain and were not resolvable in the range-Doppler domain. A monopulse algorithm was used to detect and find the azimuth position of multiple UAVs.  In \cite{monopulse_uav2}, a coordinated radar scenario for the tracking of a UAV was presented. The UAV was localized with the help of monopulse trackers that coordinated through a backhaul. The monopulse trackers achieved high localization performance at a low feedback rate. 
    \subsubsection{Polarimetric Radars} Polarimetric radars can be used to classify UAVs based on polarimetric parameters. In \cite{class_new8}, polarimetric parameters of S-band radar were used to distinguish between small UAVs and birds. In the time-frequency domain, the analysis of polarimetric parameters was carried out to differentiate different scatterers based on micro-motion that is resolved in time and velocity. The orientation of scatterers and polarimetric signature helped in the target classification. In \cite{polarimetric_radar2}, polarimetric analysis of UAVs was carried out from different aspect angles, using micro-Doppler signature that helps in the identification of UAV micromotions. 

 \begin{table*}[htbp]
	\begin{center}
     \footnotesize
		\caption{Popular counter-UAV radar systems~\cite{CSD_UAV,CSD_UAV2}. } \label{Table:CSD_UAV}
\begin{tabular}{@{}|P{ 2.4cm}|P{2.8cm}|P{1.6cm}|P{8.4cm}|P{0.6cm}|@{}}
 \hline
\textbf{Product Name}&\textbf{Manufacturer}&\textbf{Platform}&Key Specifications &\textbf{Ref.}\\
\hline
NM1-8A drone radar system&Accipter, Canada&Ground-based&Operates in X-band and S-band, typical radar output power is between $10$ to $25$~kW, uses antenna arrays, detection range for UAVs is approximately $4$~km, and a range resolution of less than $5$~m.&\cite{CSD1}\\
\hline
Spartiath&ALX Systems, Belgium&UAV/Ground-based&The detection range is up to $2$~km, the maximum range resolution is $20$~cm, and the modular antenna system operates effectively in all weather.&\cite{CSD2}\\
\hline
Gamekeeper 16U&Aveillant, United Kingdom&Ground-based&The detection range is up to $7.5$~km, the radar system can detect targets as small as $0.01$~m$^2$ RCS at a range of up to $5$~km, and uses a $2$D array of RXs.&\cite{CSD3}\\
\hline
Harrier Drone
Surveillance Radar&DeTect, Inc, United States/United Kingdom&Ground-based&The radar operates in S and X-bands, the maximum detection range is $30$~km for UAVs, and the maximum transmit power is $200$~W.&\cite{CSD5}\\
\hline
RadarOne&DroneShield, Australia&Ground-based&For small UAVs, the detection range is $1.5$~km, the angular accuracy is less than $0.5^{\circ}$ in both azimuth and elevation, the field of view is $90^{\circ}$ in azimuth and elevation, and multiple targets~(around $100$) can be tracked. &\cite{CSD6}\\
\hline
RadarZero&DroneShield, Australia&Ground-based&The maximum detection range for UAVs is $750$~m, and three antenna units with an azimuth coverage of $120^{\circ}$ each are used.&\cite{CSD7}\\
\hline
SABRE&DRS/Moog, United States&Ground-based&The maximum detection range for UAVs is $5$~km, the system uses $4$D AESA pulse-Doppler radar, and a combination of radar and EO sensor is used for UAV classification.&\cite{CSD8}\\
\hline
GroundAware&Dynetics, United States&Ground-based&The detection range is $1$~km to $15$~km and uses $2$D and $3$D digital beamforming for precise detection and tracking of multiple UAVs.&\cite{CSD9}\\
\hline
Red Sky 2 Drone Defender System&IMI Systems, Israel&Ground-based&The detection range is between $3$~km to $5$~km for UAVs, and an IR Scanners and EO thermal cameras are also used to provide a comprehensive $360^{\circ}$ surveillance capability.&\cite{CSD10}\\
\hline
SharpEye&Kelvin Hughes, United Kingdom&Ground-based&I/X/E/F/S-Bands can be used, and transmit power can vary between $80$~W to $300$~W.&\cite{CSD11}\\
\hline
SR-9000S&Meritis, Switzerland&Ground-based&The radar system operates on an X-band, the maximum detection range for UAVs is $5$~km, and the system is capable of tracking $400$ UAVs simultaneously.&\cite{CSD12}\\
\hline
Drone Detection Radar&Miltronix, United Kingdom&Ground-based&The radar operates in X-band with antenna gain $>41$~dBi, UAV detection range is between $4$~km and $10$~km, the range resolution is $50$~m, $360^{\circ}$ azimuth coverage and $42^{\circ}$elevation coverage, and capable of detecting and tracking more than $100$ UAVs simultaneously.&\cite{CSD13}\\
\hline
EAGLE&MyDefence Commun. ApS, Denmark&Ground-based&Detection range up to $1$~km for small UAVs, and simultaneous detection and tracking of multiple UAVs.&\cite{CSD14}\\
\hline
1L121-E&NNIIRT, Russia&Ground-based&The detection range is between $20$~km to $90$~km, position accuracy is $100$~m, angular accuracy is $1^{\circ}$, and simultaneous tracking of $64$ UAVs is possible.&\cite{CSD15}\\
\hline
3D Air Surveillance UAV Detection Radar&OIS-AT, India&Ground-based&Radar operates at Ku-band, UAV detection range is between $3$~km to $20$~km, radar operates at a nominal power of $4$~W, and $700$ targets can be simultaneously tracked. &\cite{CSD16}\\
\hline
OBSIDIAN&QinetiQ, United Kingdom&Ground-based&Radar uses phased array antennas and provides simultaneous detection and tracking of multiple targets.&\cite{CSD17}\\
\hline
Multi-Mission Hemispheric Radars&RADA Electronic
Industries, Israel&Ground-based&The detection range for UAVs is between $5$~km to $25$~km, AESA technology is used, and $90^{\circ}$ coverage in azimuth and elevation.&\cite{CSD18}\\
\hline
Elvira&Robin Radar Systems, Netherlands&Ground-based&The detection range is between $9$~km and $12$~km, a rotating antenna provides $360^{\circ}$ azimuth coverage, capable of tracking multiple UAVs simultaneously, and classification based on microDoppler signatures.&\cite{CSD20}\\
\hline
Drone Sentinel&Advanced Radar Technologies, Spain&Ground-based&The system uses a $3$D multibeam antenna, able to detect UAVs with a RCS of less than $0.01$~m$^2$, provides $360^{\circ}$ azimuth coverage, and capable of tracking multiple UAVs simultaneously.&\cite{CSD21}\\
\hline
RR Drone/radar detection system&Aaronia, Germany&Ground-based&The maximum detection range is $50$~km under optimal conditions, the angular accuracy is $0.1^{\circ}$, up to $16$ sector antennas provide $360^{\circ}$ coverage, and simultaneous tracking of multiple UAVs is possible.&\cite{CSD22}\\
\hline
Fortem TrueVIEW radar&Fortem Technologies&Ground-based/airborne&The detection range is $5$~km for smal UAVs, and AESA is used with R$30$ model provides $256$ received elements and $16$ digital channels.&\cite{Fortem}\\
\hline
EchoFlight&Echodyne corp.&Airborne&The operating frequency range is $24.45$ to $24.65$~GHz, metamaterials electronically scanned array is used, detection range is up to $1$~km for small UAVs, tracking accuracy is $1^{\circ}$ in azimuth and $1.5^{\circ}$ in elevation, and maximum transmit power is $45$~W.&\cite{echodyne}\\
\hline
LSTAR (V)2&SRC Inc.&Ground-based&The maximum detection range is $100$~km for different targets, and a $3$D, $360^{\circ}$ electronically scanned antenna is used.&\cite{lstar}\\
\hline
$60$~GHz FMCW MIMO&Custom built&Ground-based&The frequency range is $57$~GHz - $64$~GHz, $32$ channels are available, bandwidth is $4$~GHz, supports $2$D and $3$D antenna arrays, and $3$D array offers a horizontal resolution of $13^{\circ}$ and a vertical resolution of $25^{\circ}$.&\cite{MIMOradar_vasilii}\\
\hline
\end{tabular}
		\end{center}
			\end{table*}

    \subsubsection{Airborne Radars} Airborne radar systems can be used for DCT-U. Airborne radar systems have the advantage of longer detection ranges, better tracking with the help of maneuvering, and reduced clutter compared to ground-based radar systems. In \cite{Wahab_AESA}, MMPAR radar was used for the detection, tracking, and classification of multiple malicious UAVs in a swarm. The multiple radar beams were controlled autonomously through reinforcement learning. Reinforcement learning helped to overcome radar anomalies. In \cite{UAV_stealth_airborne}, active and passive radars onboard UAVs in a swarm were used for the detection of stealth aerial vehicles. The swarm of UAVs used a combination of active and passive radars that helped in the detection of stealth aerial vehicles. An airborne aerial dragnet system by DARPA is used for the detection and tracking of UAVs~\cite{DARPA_dragnet}. The aerial dragnet system used either tethered UAVs or long-endurance platforms as airborne surveillance platforms in urban environments.

The different algorithms available in the literature for tracking UAVs using radar systems are provided in Table~\ref{Table:Tracking_algos}. In Table~\ref{Table:Tracking_algos}, the type of UAV, radar type, tracking features, and tracking algorithm from different literature references are summarized. Table~\ref{Table:Tracking_algos} also provides performance metrics and performance results for different tracking algorithms and setups. From Table~\ref{Table:Tracking_algos}, we observe that different derivations of Kalman filters are frequently used to estimate the future position of the target based on the current measurements~\cite{track_new1,track_new2,track_new3,track_new4,track_new6,white_noise_track}. The major advantage of the Kalman filter includes short convergence time and accuracy in real-time tracking scenarios. The tracking features used mainly by Kalman filters for tracking are position and velocity. In some scenarios e.g., in \cite{track_new2}, angle information was not available, however, using different measurements from different bi-static radar pairs, the Kalman filter was able to track the UAV. In \cite{track_new10}, a Hough transform and linearly distributed features of micro-Doppler are used for the detection and tracking of a UAV.

\begin{table*}[htbp]
	\begin{center}
     \footnotesize
		\caption{Classification of UAVs by radar systems using different types of classification algorithms. Key performance metrics and performance results for the UAV classification are also provided.} \label{Table:Classification_algos}
\begin{tabular}{@{}|P{ 2.2cm}|P{ 1.7cm}|P{2.0cm}|P{2.0cm}|P{3.2cm}|P{3.2cm}|P{0.65cm}|@{}}
 \hline
\textbf{UAV type}&\textbf{Radar/sensor type}&\textbf{Classification features}&\textbf{Classification algorithm}&\textbf{Key performance metrics}&\textbf{Performance results}&\textbf{Ref.}\\
\hline
DJI Matrice 600, DJI Matrice 100, Trimble zx5, DJI Mavic Pro 1, DJI Inspire 1 Pro, and DJI Phantom 4 Pro&Surveillance radar operating at 15~GHz and 25~GHz&RCS of UAVs&15 different classifiers  provided in Table~IV of ~\cite{class_new9}&RCS measurement accuracy, measurement environment quality, operational frequency, polarization, and ML classifier performance&At $-5$~dB SNR a classification accuracy of $53.67$\% was obtained. At SNR $\geq 3$~dB) a classification accuracy of $100$\% was achieved&\cite{class_new9}\\
\hline
Multi-rotor UAVs and birds&S-band BirdRad system&Polarimetric parameters provided in Table~I of \cite{class_new8}&Nearest neighbor classifier &Accuracy and precision of results obtained, recall, confusion matrix, miscalculation rate, feature ranking, and cross-validation score &A UAV classification accuracy of $100$\% was achieved using an optimal set of features and averaging time&\cite{class_new8}\\
\hline
Metafly, MavicAir2, Parrot Disco&FMCW radar&Micro-Doppler signatures from radar spectrogram dataset&Deep learning&Classification accuracy of 
 time series images, denoising and image pre-processing, CNN model training and testing, and CNN model optimization&An overall UAV classification accuracy was $98.8$\% and was achieved across all SNR levels&\cite{ozturk_journal}\\
\hline
See Table~7.1 in \cite{class_new2}&CW radar, automotive radar&Diameter of the rotor& SVM, Naive Bayes&Separation in feature space, efficiency of feature extraction method, and frequency band invariance&The classification accuracy varied from $95.69$\% to $95.97$\% dependent on the classification algorithm used&\cite{class_new2}\\
\hline
Mobile objects in air&Mobile autonomous radar stations&Image of the environment&Neural network algorithm&Feature space extraction, phase coupling content utilization, and robustness to interference and noise&UAV classification accuracy of $92$\%, was achieved using neural network-based classification system&\cite{class_new4}\\
\hline
Planes, quadrocopter, helicopters, stationary rotors, and birds&X-band CW radar&Features based on micro-Doppler signatures of aerial vehicles&SVM, Naive Bayes &ML classifier type, noise suppression and signal processing efficiency, and handling variations in target characteristics&A UAV classification accuracy of $95.39$\% was achieved using a nonlinear SVM classifier, $94.91$\% using a linear SVM classifier, and $93.6$\% using a Naive Bayes classifier&\cite{class_new6}\\
\hline
Rotary UAVs&FMCW radar and acoustic sensor array&Spectral power bins&Neural network algorithms&Detection range and resolution, frequency band selection, integration and synchronization between radar and acoustic sensor, and robustness to environmental factors &UAV classification accuracy close to $100$\% was achieved with no false negatives and only a few transient false positives&\cite{class_new10}\\
\hline
Multi-rotor UAVs&X-band CW radar&Features based on micro-Doppler signatures of the UAVs&Spectrographic pattern analysis of UAVs &Detection accuracy, misclassification rate, SNR, spectrogram clarity, and propeller length and rotation speed estimation accuracy &NA&\cite{class_new5}\\
\hline
Fixed wing and multi-rotor UAVs&FMCW surveillance radar&Features based on micro-Doppler signatures&Total
error rate minimization based classification &Micro-Doppler signature analysis, features extraction and normalization, and total error rate&The classification accuracy varied from $90.59$\% to $94.39$~\% for three category and two category classification, respectively&\cite{class_new7}\\
\hline
\end{tabular}
		\end{center}
			\end{table*}

Table~\ref{Table:CSD_UAV} provides modern counter-UAV radar systems. These radar systems are specifically designed to counter different types of UAVs. The name of the radar system, installation platform, manufacturer information, and key specifications of the radar systems are also provided in Table~\ref{Table:CSD_UAV}. The majority of the counter-UAV radar systems are ground-based and provide small to medium-range detection of UAVs.

\subsection{Classification of UAVs Using Modern Radar Systems}
There are many methods available in the literature for the classification of UAVs using modern radar systems~\cite{classification1,classification2}. The main idea is to compare multiple features of the UAVs obtained from measurements with the database of stored features of different types of UAVs. Some of the important classification features are size, shape, velocity, maneuverability, and propeller motion patterns at different altitudes of a UAV flight. The features are analyzed in both time and frequency domains. In order to accurately classify a UAV, the following conditions should be met: 1) the scattered signal should be independent of the operating frequency, aspect angle, and polarization; 2) the noise, clutter, and interference from different sources should be minimum; and 3) there should be sufficient features to classify a UAV. Furthermore, the classification accuracy using a radar system depends on the degree of range and angular resolution, and clutter rejection capability.

Major challenges for the classification of UAVs using radar systems include determining the type and intent of the UAV, discriminating micro-UAVs from flying birds, identifying the number and type of UAVs flying in a swarm, complexity due to a large amount of training and real-time data, and optimum feature extraction. The popular methods for the classification of UAVs using radar systems in the recent literature are based on RCS and micro-Doppler signatures of the UAVs. The RCS is either taken as number values plotted across azimuth angles or as images. The micro-Doppler signatures obtained from spectrogram and cepstrogram plots are commonly used to classify small UAVs~\cite{cepstrogram_plot}. The classification methods for UAVs using radar systems can be broadly divided into AI and non-AI-based classification methods. The AI-based classification methods are dominantly used in the recent literature compared to non-AI-based classification methods.

\begin{figure}[!t]
	\centering	\includegraphics[width=\columnwidth]{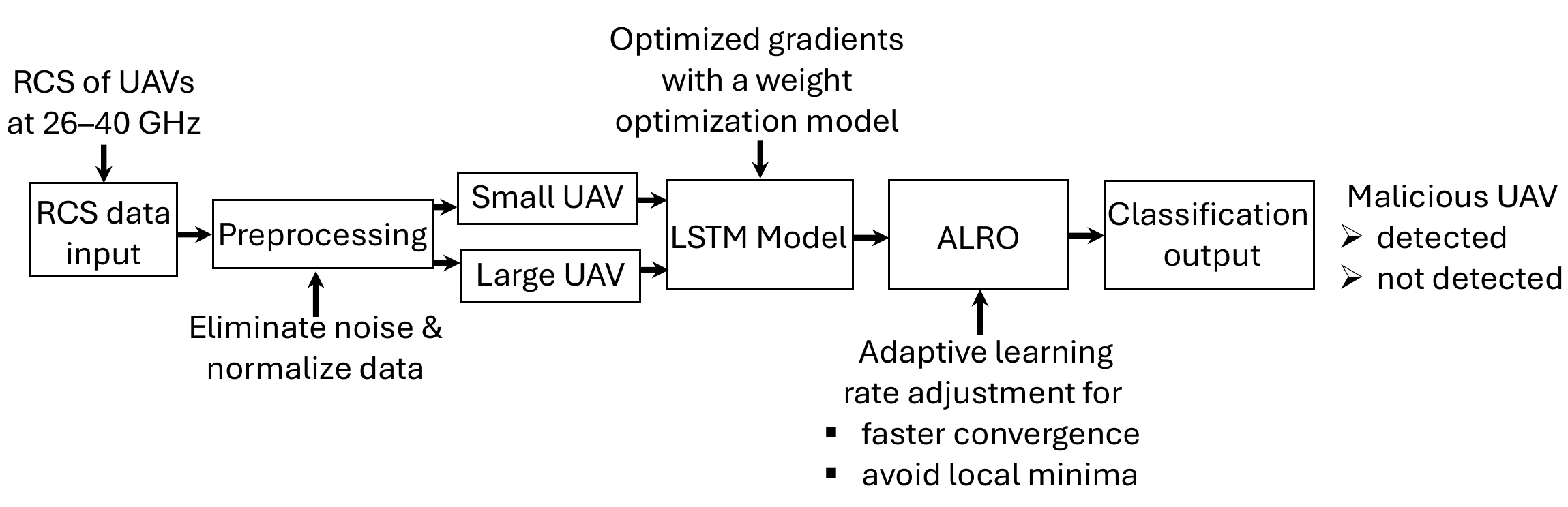}
	\caption{LSTM-ALRO model process for drone classification at mmWave frequencies~\cite{LSTM_classify}}. \label{Fig:Fig_classf_radar}
\end{figure}

AI algorithms learn the given features of a UAV using training data and use different algorithms for UAV classification~\cite{classification_AI,classification_AI2,new_radarAI}. Popular AI algorithms for the classification of UAVs using radar systems include neural networks. In \cite{RATR_HRRP1}, a radar automatic target recognition~(RATR) algorithm that used CNN and a high range resolution profile~(HRRP) algorithm were presented. The proposed method using RATR and HRRP was shown to work effectively even when the training data was sparse. In \cite{UAV_CNN2}, CNN was used for the classification of UAVs and birds. The radar signatures and CNN were used for distinguishing of UAVs from birds in \cite{UAV_CNN2}. Other literature using different neural network algorithms and radar data for the classification of UAV are available in~\cite{ozturk_journal,cognitive_deeplearning,UAV_ML_new2,AI_class1,LSTM_classify}. The long short-term memory adaptive learning rate optimization~(LSTM-ALRO) approach for UAV classification is provided in \cite{LSTM_classify}. Fig.~\ref{Fig:Fig_classf_radar} shows the workflow of the LSTM-ALRO model for drone classification at mmWave frequencies, highlighting its use of adaptive learning rate optimization and gradient balancing to improve UAV classification performance.

In \cite{boosting_algo} a boosting classifier was used on the radar measurements of birds and small UAVs. The micro-Doppler radar measurements in the form of time-velocity diagrams and a boosting classifier were used to distinguish different types of UAVs and birds. The boosting classifier was shown to perform better compared to the support vector machine~(SVM). In~\cite{GBM1}, different AI classification algorithms including grading boosting method~(GBM) were applied to the features obtained from range fast Fourier transform~(FFT) of mmWave FMCW radar measurements. A classification accuracy of $95.6~\%$ was achieved using GBM classifier. In~\cite{new_mmwaveradar}, a low-cost mmWave radar and CNN were used to simultaneously track and classify UAVs. The approach used advanced signal processing, cloud point clustering, and CNN for classification. In \cite{NB1}, the Naive Bayes classifier and other AI algorithms were used for the classification of UAVs based on RCS measurements. The classification results of different classifiers were also compared. Similarly, SVM and logistic regression~\cite{SVM_classify1}, and hybrid CNN-memetic algorithm~\cite{hyb_neural_net} were used for the classification of UAVs using radar data. 
 
Non-AI algorithms are also used for the classification of UAVs using radar data. In \cite{RCS_clutter1} a statistical-based analysis of the RCS of different types of UAVs was carried out for classification. The RCS measurements of different UAVs were carried out in an anechoic chamber at $15$~GHz and $25$~GHz. The best statistical model fitting on the measurement data was carried out. The class conditional probability density functions~(PDFs) based on the best match for a given UAV form the basis of classification. In \cite{UAV_classification_new} a symmetrical feature based on the motion of the UAV propellers during the approaching and receding phases was used to distinguish UAVs from birds. The symmetry feature was extracted from the micro-Doppler signatures of the radar returns. A peak extraction method based on the micro-Doppler returns provided the symmetry feature. In \cite{class_new6}, a CW radar operating at $9.5$~GHz was used to extract classification features from the micro-Doppler signature of a UAV. The eigenpairs obtained from the correlation matrix of the micro-Doppler signatures were used as features for classification. Different types of UAVs and birds were classified using this approach. 

Another non-AI-based classification method using FMCW radar was provided in \cite{class_new7}, where radar returns from different UAVs and non-UAV objects were collected. A classification method based on the pattern of micro-Doppler returns was developed. The proposed method showed higher classification accuracy compared to other classification methods. In \cite{nonAI_classification2}, a two-step UAV classification method was used. In the first step, a cadence velocity diagram was used to obtain the micro-Doppler features and in the second step, the data from the current and previous steps were jointly arranged for classification. The proposed method was tested at distributed locations for the classification of UAVs. In \cite{track_new10}, Hough Transform was used for the detection and classification of a multi-rotor UAV from micro-Doppler returns. 

Different algorithms used for the classification of UAVs using radar systems are provided in Table~\ref{Table:Classification_algos}. It can be observed that the majority of the current research relies on AI algorithms for the classification of UAVs. In Table~\ref{Table:Classification_algos}, deep learning, SVM, Naive Bayes, neural network, k-nearest neighbors~(KNN), and discriminant analysis were used for UAV classification and provided in \cite{ozturk_journal,class_new2,class_new4,class_new6,class_new8,class_new10,class_new11}. Non-AI classification methods are also provided in Table~\ref{Table:Classification_algos}. In~\cite{class_new9}, non-AI classification based on statistical learning was used in addition to machine and deep learning. It was observed that even though statistical learning provided good classification results, the computational time was large compared to machine and deep learning. Furthermore, non-AI UAV classification methods in Table~\ref{Table:Classification_algos} used micro-Doppler signatures and spectrographic pattern analysis and total error rate minimization for classification~\cite{class_new5,class_new7}. Additionally, key performance metrics and performance results are summarized in Table~\ref{Table:Classification_algos} for different classification methods.

In some studies, both AI and non-AI classification methods are used for comparison. For example, in \cite{class_new9}, different classification algorithms from statistical learning, ML, and deep learning were used and a comparative analysis of UAV classification techniques based on the RCS was provided. The RCS of six different types of UAVs was measured at $15$~GHz and $25$~GHz in an anechoic chamber. For example, assuming equal apriori probabilities for each of the UAV type or class, the statistical rule for UAV classification is given in \cite{class_new9} as
\begin{eqnarray}
\begin{aligned}
\widehat{C}  &=  \arg\max_{C=1,2,\cdots, M}{\ln P({C}=j|\boldsymbol{\sigma})},\\
   &=  \arg\max_{C=1,2,\cdots, M} {\ln P(\boldsymbol{\sigma}|C=j)},
   \label{decision_rule}
\end{aligned}
\end{eqnarray}
where $P(C = j|\sigma)$ is the posterior probability for a $j^{\rm th}$ class, and $P(\sigma|C=j)$ is the conditional class density, and there are $M$ different UAV classes. 

\subsection{How Modern Radar Systems Overcome Typical Limitations}
In contrast to conventional radar systems, modern radar systems leverage advancements in both hardware and software technologies~\cite{modern_radars}. Modern radar systems utilize cutting-edge digital signal processing, advanced RF front-ends and antenna technology, and ML. Modern radar systems can provide detection at long ranges, maintain a low probability of intercept, adaptively select transmission parameters based on the scenario, and process sparse and weak signals. Moreover, modern radar systems exhibit superior clutter rejection capabilities, high range and angular resolution, enabling the detection and tracking of multiple targets simultaneously, enhanced classification capabilities, a low false alarm rate, and resistance to ECM. A one-to-one comparison of the limitations of typical radar systems provided in Section~\ref{Section:Limitations_radars} and their mitigation using modern radar systems is given as follows:
\begin{itemize}
    \item Modern radar systems use adaptive resource scheduling. Limited resources are used for the initial search, whereas once a target is detected, large resources are allocated, e.g., using high-resolution and high-powered beams. Furthermore, modern passive radar systems can detect targets without active RF emissions. These passive radar systems operate anonymously and do not require transmit power resources. 
    \item Modern radar systems, including bistatic and multistatic, cognitive, and compressed sensing-based radars, can be utilized for the detection and tracking of small RCS UAVs.  
    \item Airborne and compressed sensing-based radar systems can detect low-flying UAVs in cluttered environments. Furthermore, advanced clutter rejection algorithms~(e.g., the space-time adaptive processing algorithm) and AI can be used to detect and track low-flying UAVs.
    \item Modern radar systems use different methods to overcome the range and Doppler ambiguity problem. For example, adaptively selecting the PRF~\cite{new_PRF_ambg}, coded pulses\cite{new_coding_ambg}, and interpulse modulation~\cite{new_ipm_ambg}. 
    \item Modern radar systems use multiple narrow beams to detect and track multiple targets simultaneously in a swarm, e.g., using AESA radar. Each beam can be assigned parameters independent of the other beams and based on the scenario.
    \item The adaptive adjustment of radar thresholds in a cluttered environment can be handled using cognitive and software-defined radar systems.
    \item Modern radar systems can use phased arrays, adaptive beamforming, frequency selection, and advanced signal processing algorithms to detect and track high-altitude, highly maneuverable, and high-speed UAVs.  
    \item Modern radar systems use different ECCM to overcome ECM. The ECCM includes robust frequency hopping over broadband, multilayered authentication and data encryption, detection and reporting of jammed strobes, and dedicated RX for monitoring the ECM~\cite{ECCM,ECCM2}.
    \item Modern radar systems use advanced and adaptive signal processing algorithms to overcome noise, RF interference, and clutter, such as pulse compression~\cite{new_interf_pc}, adaptive filtering~\cite{new_interf_af}, and AI algorithms~\cite{new_interf_ai}.     
\end{itemize}


\section{DCT-U using Communication Systems} \label{Section:Comm_Sys}
In this section, we highlight the similarities between radar and communication systems, and the advantages of communication systems for DCT-U. Furthermore, JC\&S in active mode and the use of communication systems in passive mode for DCT-U are provided. The limitations of communication systems for the DCT-U are also discussed. 

\subsection{Similarities and Differences of Communication and Radar Systems}
There are many similarities between communication and radar systems. Therefore, communication systems can be used for DCT-U. The similarities and differences between the communication systems and radar systems are summarized as follows:
\begin{itemize}
    \item Both communication systems and radar systems use radio waves at different frequencies.
    \item The radio wave propagation through free space takes place similarly for communication systems and radar systems. 
    \item Both radar systems and communication systems use a signal generation stage and an RF stage. The signal generation stage typically involves waveform generation, shaping, coding and modulation, and encryption. The RF stage includes amplification and beam steering~(by changing the signal phase across antenna elements) from the antenna array. 
    \item The RX of a communication system and a radar system generally includes a low noise amplifier, matched filter, decoder, decrypter, demodulator, and detector.
    \item Both communication systems and radar systems are time-critical systems.
    \item Majority of communication systems and radar systems use encryption and authentication.
    \item The major difference between the communication systems and the radar systems is the transmit power. The pulse-based radar systems transmit radio waves at much higher power compared to common RF communications. The high transmit power is obtained by using amplifiers and high-gain antennas.
    \item Complex frequency hopping patterns are frequently used in radar systems compared to communication systems to avoid jamming.
    \item The selection of frequencies for the communication systems takes into account the transmission~(or penetration losses) through the building materials. However, for radar systems, the transmission through materials is not mainly considered during frequency selection and the main focus is on the reflection properties. 
    \item Complex encoding and signal processing techniques are used for communication systems. The techniques help to attain a high data rate and serve multiple users simultaneously. On the other hand, for radar systems, simple pulse signals are commonly used because the main focus is reliability. 
    \item The radar systems operate mainly in the line-of-sight~(LOS), whereas, the communication systems can operate in the LOS and non-LOS~(NLOS).
    \item The RX sensitivity is higher for radar systems compared to common communication systems. 
    \item The TX and RX are generally co-located in a radar system~(monostatic radar), whereas TX and RX are apart in the communication systems. 
\end{itemize}

\subsection{Advantages of Communication Systems for DCT-U}
The major advantages of communications systems for DCT-U can be listed as follows: 
\begin{itemize}
    \item Communication systems offer a thorough coverage in the majority of urban areas compared to radar systems.
    \item Communication systems can operate and subsequently detect and track UAVs in NLOS conditions. On the other hand, the majority of radar systems require a LOS path with the UAV for precise detection and tracking.
    \item Multiple types of communication systems are available in an urban area, e.g., FM and amplitude modulation~(AM) radio broadcasts, television broadcasts, and mobile and satellite communications. Each communication system uses a particular frequency band and communication methodology. Moreover, different types of communication systems offer long, medium, and short-range connectivity. Overall, the diverse types of communication systems available in an urban area can help DCT-U in different scenarios. 
    \item The presence of multiple communication nodes along the path of a non-cooperative UAV can help to update the localization estimates. 
    \item The height of the antenna(s) on the communication towers is generally selected to maximize coverage in a given area. The height of the majority of the communication tower antennas is such that small UAVs flying at typical heights~(e.g., lower than $400$~feet in the United States under FAA Part $107$ Rules) can be detected and tracked through a LOS path. 
    \item Communication systems can be used to obtain the Doppler signature of the UAV, which can help to classify a UAV. 
    \item The communications using satellites offer ubiquitous and LOS coverage on the ground mainly due to the height of the satellites above the ground. The ubiquitous LOS coverage by satellites can help in the detection and tracking of UAVs.
    \item Modern communication systems, e.g., $5$G and beyond use phased arrays and can electronically steer antenna beams similar to phased array radar systems.  
    \item Different communication systems update the channel state information~(CSI) at regular intervals of time, e.g., massive MIMO systems. These periodic CSI updates can help in UAV detection.
    \item Both active and passive DCT-U can be achieved using communication systems. 
    \item Popular communication waveforms, e.g., orthogonal frequency-division multiplexing~(OFDM) can be used for JC\&S. Moreover, both uplink and downlink channels can be used for sensing in a JC\&S system using a mobile network. 
    \item A significant research literature and standardization details are available for communication systems in the public domain that can help research for UAV-based detection, tracking, and classification. 
\end{itemize}
 
\subsection{DCT-U Using JC\&S}
JC\&S also known as joint communication and radar systems has gained significant importance in the recent decade~\cite{JCS_book}. The term sensing generally includes DCT. The majority of JC\&S literature uses active radar and mobile communication networks. The major goal of JC\&S is to integrate communications and sensing of the environment through a single device and preferably using a single transmitted signal. The JC\&S can offer improved spectral efficiency, and reduce cost, size, and power consumption. Advanced hardware and software including complex signal processing techniques are required to make the integration of the two domains efficient. JC\&S can also be used for DCT-U. There are studies available in the literature that provide JC\&S techniques for DCT-U~\cite{UAV_comm_type2_5, JCS_DCTU2}.

In \cite{comm_jcas_survey1}, a survey on JC\&S in mobile networks was provided. In the survey, studies on coexisting radar and communication systems and their limitation were discussed. Mainly three types of JC\&S systems were covered namely, radar centric and communication centric designs, and joint design and optimization. Furthermore, modifications were discussed to integrate sensing operations into the current communication-only architecture. UAV detection was provided as an application of the JC\&S systems. In \cite{jcs_6G_2}, JC\&S using MIMO systems was discussed. Three novel JC\&S models using MIMO and cloud random access networks, UAVs, and reconfigurable intelligent surfaces were discussed. The survey provided different Internet of things~(IoT) scenarios for JC\&S. Different waveform designs for MIMO-based JC\&S were also provided. Other survey studies on JC\&S are available in \cite{UAV_comm_survey_other1,UAV_comm_survey_other2,UAV_comm_survey_other3,JCS_same_sig2}. We focus on active JC\&S systems in this survey. 

\begin{table*}[htbp]
	\begin{center}
    \footnotesize
		\caption{Comparison of JC\&S systems based on resources and waveform. } \label{Table:JCS}
\begin{tabular}{@{}|P{ 2cm}|P{2cm}|P{ 2.5cm}|P{3.75cm}|P{4.5cm}|P{0.7cm}|@{}}
\hline
\textbf{System}&\textbf{Resources}&\textbf{Waveform}&\textbf{Pros}&\textbf{Cons}&\textbf{Ref.}\\
\hline
JC\&S with different transmit signals&Resources are generally shared&Different types of waveform& 1) The design of waveform for communication and sensing is approximately independent, and 2) small mutual interference & 1) Complex and expensive hardware, and 2) low spectral efficiency&\cite{jcs_diff_sig_new1,jcs_diff_sig_new2,jcs_diff_sig_new3,jcs_diff_sig_new4,Wahab_AESA} \\
\hline
JC\&S with same transmit signals&Resources can be shared or dedicated&Same waveform&1) Shared TX and RX, and 2) high spectral efficiency& Requires clock synchronization between TX and RX to avoid sensing ambiguity& \cite{UAV_comm_type2_2,UAV_comm_type2_3,UAV_comm_type2_4,UAV_comm_type2_5}\\
\hline
Distributed JC\&S&Resources are shared&Different waveform optimized for sensing and communications, respectively&1) Improved spectral efficiency, and 2) less complex TX and RX compared to different waveform JC\&S system&1) Coordination among the sensing and communication nodes require synchronization that is challenging if far apart, and 2) interference between communication and sensing nodes&\cite{JCS_resource_same1, JCS_resource_same2}\\
\hline
Cognitive radio&Resource shared cognitively&Different waveform&Efficient spectral usage&1) Interference can affect the primary users, and 2) performance of secondary users depends on the cognitive algorithm and surroundings&\cite{cognitive_comm_sense,cognitive_comm_sense2}\\
\hline
\end{tabular}
		\end{center}
			\end{table*}

\subsubsection{Active JC\&S Using Different Signals for Communications and Sensing}
In this type of JC\&S, the communication and sensing signals are separated in frequency, time, code, and polarization. The communication and sensing hardware is often partially shared. In \cite{jcs_diff_sig_new1} a JC\&S system comprising of MIMO radar and a configuration of sparse antenna array was used. Independent transmit waveforms were associated with antennas resulting in different pulse repetition periods. Moreover, by antenna selection and the ordering of antenna-waveform pairs a high communication data rate was achieved. The sensed target was an aerial vehicle. In \cite{jcs_diff_sig_new2}, an intelligent transportation system operating at mmWave was used for simultaneous communication and radar functions. Different waveforms were used for communications and radar functions controlled by a voltage-controlled oscillator switch. 

In~\cite{jcs_diff_sig_new3} a software-defined automotive system providing JC\&S was provided. In the proposed approach, FMCW was used to obtain the range of the target, whereas pseudorandom~(PN) code pulse was used for inter-vehicle communication. The switching between FMCW and PN was controlled through software. In \cite{Wahab_AESA}, a phased array radar with six simultaneous beams was used for DCT-U. Two beams were used for communications and the other four beams were used for radar operation. The radar and communication beams used independent waveforms. The system resources were cognitively scheduled among the beams dependent on the scenario through reinforcement learning.

\subsubsection{Active JC\&S Using Same Signals for Communications and Sensing}
A major goal of JC\&S is to design a common signal that optimizes communications and sensing. There are many research efforts where a common signal is used and optimized for communications and sensing. In \cite{UAV_comm_type2_2}, multiple phased array beams originating from a MIMO base station~(BS) were steered in different directions to serve mobile users on the ground and detect UAVs in space. OFDM signals with a standard cyclic prefix were used for both communications and sensing. The variations in the channel impulse response~(CIR) were used to detect the presence of a target. The detection of a target was considered a binary hypothesis set. The example target sensed was a UAV. In \cite{UAV_comm_type2_3}, UAV-based machine type communications are considered for JC\&S. Code-division OFDM JC\&S system was used through unified spectrum and transceivers. A novel code-division OFDM signal was used for both communications and sensing. The successive interference cancellations were used to provide code-division gain. The code division gain was used to reduce the multiple access interference. In \cite{UAV_comm_type2_4}, JC\&S using orthogonal time frequency space~(OTFS) waveform was used for industrial IoT including ground vehicles and UAVs. The problem of lack of sensing by OFTS was overcome using different pre-processing methods to remove the effect of communication symbols in the time and frequency domains. 

\begin{figure}[!t] 
    \centering
	\begin{subfigure}{\columnwidth}
    \centering
	\includegraphics[width=\columnwidth]{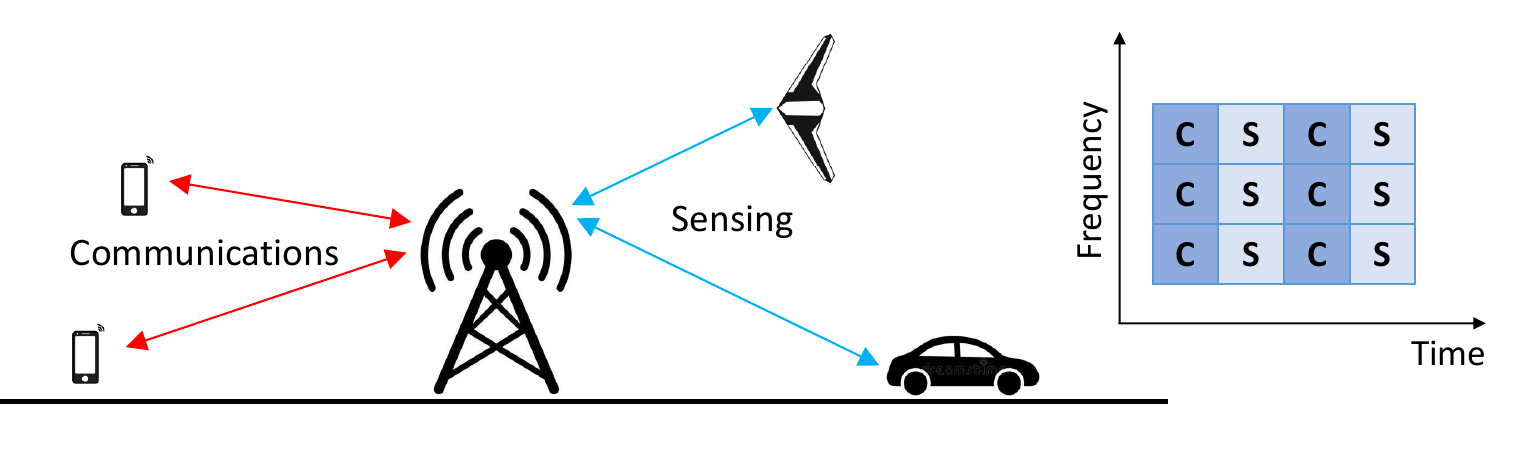}
	  \caption{}  
    \end{subfigure}    
    \begin{subfigure}{\columnwidth}
    \centering
	\includegraphics[width=\columnwidth]{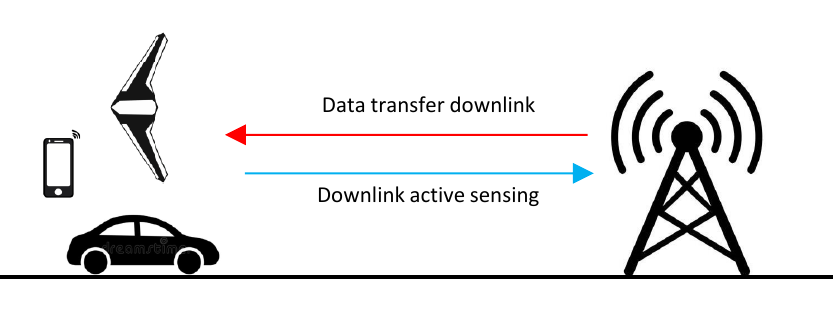}
	  \caption{}  
    \end{subfigure}
    \begin{subfigure}{\columnwidth}
    \centering
	\includegraphics[width=\columnwidth]{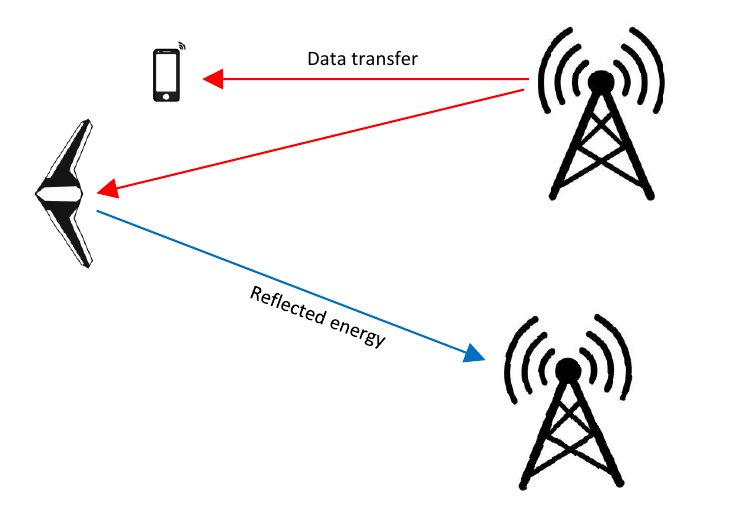}
	  \caption{}  
    \end{subfigure}
    \caption{(a) A JC\&S system example in which time-frequency resources are equally distributed between the communication and sensing operations~(C represents communications and S represents sensing). The signals for communications and sensing operation can be the same or different; (b) a JC\&S system example with simultaneous downlink communications and active sensing. The signals for communications and sensing are preferred to be the same in this scenario; (c) passive sensing of a target scenario using a communication system. } \label{Fig:JCR}
\end{figure}

\subsubsection{Active Distributed JC\&S Systems}
In a distributed JC\&S system the communication and sensing subsystems may not be co-located. In a distributed JC\&S system the signals are different but the resources are shared. Moreover, the communication and sensing subsystems in a distributed JC\&S can be cooperative or non-cooperative. The cooperative architecture requires information sharing and is complex compared to the non-cooperative architecture. Fig.~\ref{Fig:JCR}(a) shows a cooperative resource-sharing example for JC\&S where the communication and sensing subsystems are co-located. In \cite{JCS_resource_same1} a  joint MIMO matrix completion radar and communication system was proposed that operated concurrently by sharing the spectrum. The sharing of spectrum helped to reduce the interference between the two systems. Both cooperative and non-cooperative approaches were used. The target considered for JC\&S was an aerial vehicle. In \cite{JCS_resource_same2}, coexistence and spectrum sharing between a MIMO radar and a downlink multiuser-MIMO communication system was provided. The transmit beamforming of the MIMO downlink communications was designed to optimize the communications and probability of detection of MIMO radar. The beamformer proposed was robust to imperfect channel state estimates. 

\subsubsection{Active JC\&S Using Cognitive Radios}
In \cite{cognitive_comm_sense}, a rotating radar was used as a primary device and a communication system as a secondary device that cognitively accessed the available spectrum. The secondary communication device was allowed to transmit as long as the interference levels were bounded within the range allowed by the primary radar device. The secondary communication device was orthogonal frequency-division multiple access based operating in non-contiguous cells that used a given dedicated spectrum and a shared spectrum with the radar. Similarly, in \cite{cognitive_comm_sense2} the effect of opportunistic access to the radar spectrum by a communication link was investigated. The spatial and temporal opportunities were explored for the co-existence of radar and communication system keeping the interference level low to the primary radar device while maintaining the communication throughput. Table~\ref{Table:JCS} provides a comparison of JC\&S systems based on waveform and resources.   

\begin{figure*}[!t]
	\centering
	\includegraphics[width=\textwidth]{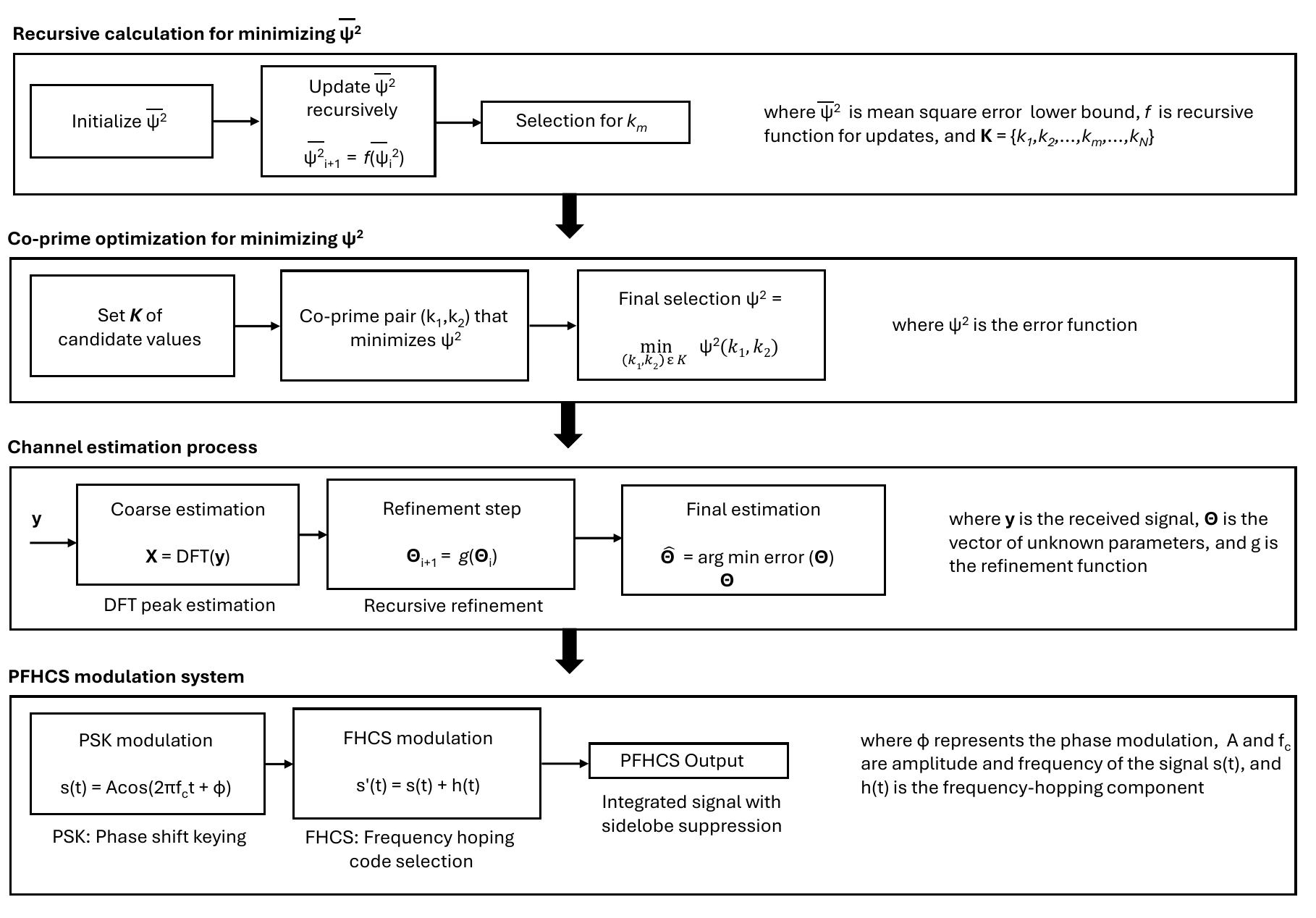}
	\caption{Frequency-hopping MIMO-based dual function
radar communication system~\cite{radar_centric7}.}  \label{Fig:Fig_JCS1}
\end{figure*}

\subsubsection{Active Radar Centric JC\&S Design}
In a radar-centric JC\&S design, the communications are realized in the primary radar system. The radar systems have generally long ranges and communication systems can take advantage of the long-range, however, the throughput is limited due to the radar-specific waveform. In a radar-centric system, multiple access channels for JC\&S were considered in \cite{radar_centric1}. Spectrum-time-space convergence was considered for the co-existence of radar and communication systems. In \cite{radar_centric} a MIMO radar with quasi-orthogonal multi-carrier linear frequency modulation continuous phase modulation waveform was used for sensing and multi-user communications. In \cite{radar_centric4} a radar was implemented using SDR that is able to support communications. The system used wideband communications. The wideband communications supported radar operation and allowed communication of range-Doppler maps with the BS. There are other related studies also~\cite{radar_centric5,radar_centric6,radar_centric7} where different types of waveform were used for radar that support communications. The multi-stage processing for frequency-hopping MIMO radar-based
communications in \cite{radar_centric7} is shown in Fig.~\ref{Fig:Fig_JCS1}. The multi-stage process includes initializing candidate parameters, recursively refining them to minimize error functions, selecting optimized co-prime values, performing channel estimation through DFT peak identification and refinement, and integrating PSK with FHCS modulation for efficient signal design.

\subsubsection{Active Communication Centric JC\&S Design}
In a communication-centric design, sensing capability is introduced to a primary communication system. Communication systems that support sensing can be further categorized into small-scale point-to-point communication systems, e.g., vehicular communications or large-scale mobile networks. In \cite{comm_centric1} radar-based sensing was implemented for the mmWave 802.11ad vehicular communication system. A full duplex radar operation was considered due to limited self-interference and sufficient isolation. In \cite{comm_centric2} velocity and range estimation of multiple scatterers in an OFDM vehicular communication system was carried out by different algorithms. JC\&S in $5$G mobile networks using compressed sensing was introduced in \cite{jcs_5G_1}. The compressed sensing helped in the target detection using $5$G communication signals.  

The sensing in a communication-centric mobile network can take place at the downlink or uplink. Fig.~\ref{Fig:JCR}(b) shows a downlink JC\&S scenario using a mobile network. The downlink JC\&S can be active or passive, whereas, the uplink JC\&S is generally passive. A passive sensing scenario using a mobile network is shown in Fig.~\ref{Fig:JCR}(c). In an active downlink JC\&S, the TX and RX are generally co-located. As the TX and RX are co-located, the synchronization between the TX and RX can be easily achieved. Also, the active downlink JC\&S requires full-duplex capability. In \cite{JCS_downlink_active1}, the JC\&S system using MIMO was used. A single TX communicated with the downlink mobile user and simultaneously detected targets in the surroundings. Different design criteria were introduced to mitigate the multiuser interference including omnidirectional and directional antennas, and optimized waveforms. In \cite{UAV_comm_type1_3}, both uplink and downlink sensing using a mobile network was demonstrated and the same signal was used for communications and sensing.  

\subsubsection{Active Non-centric JS\&S Design}
There is a third category where the design of JC\&S is not centric around existing radar or communication systems. These offer better freedom and a balanced approach in the design of waveforms, resource distributions, and corresponding algorithms for communications and sensing. In \cite{UAV_comm_survey_other3}, waveform designs and signal processing aspects for JC\&S at mmWave were covered without a centric design. Both UAV and ground vehicle targets were considered for JC\&S. In \cite{JSC4}, ultra-massive JC\&S was provided at the terahertz band. In the proposed JC\&S system both model-based and model-free hybrid beamforming was used. The BS consisted of ultra-massive MIMO and multiple beams were generated toward the multi-antenna mobile users and radar targets. Another non-centric JC\&S system was proposed in \cite{JSC3}. The proposed system used analog antenna arrays and two sub-beams were superimposed to generate an optimal combined beam. The optimal beam was used for JC\&S.

\subsection{DCT-U Using Communication Systems in Passive Mode} \label{Section:Section_V_D}
Communication signals reflected from UAVs can be passively monitored for DCT-U. Commonly used communication systems for DCT-U are GSM, LTE, $5$G, WiFi, digital audio and video broadcasting, and satellites. Table~\ref{Table:Comm_radars} lists some examples where different communication systems are used for DCT-U in passive mode. Specifications and major findings from the relevant literature are also provided in Table~\ref{Table:Comm_radars} with further details as follows.

\begin{table*}[htbp]
	\begin{center}
    \footnotesize
		\caption{Example passive radar systems that passively capture communication signals reflected from UAVs for DCT-U.}   \label{Table:Comm_radars}
\begin{tabular}{@{}|P{ 2cm}|P{10.5cm}|P{1.5cm}|P{0.7cm}|@{}}
\hline
\textbf{Communication System}& \textbf{Specifications and Findings}& \textbf{Bistatic or multistatic radar}&\textbf{Ref.}\\
\hline
GSM & A passive radar with multibeam and multiband capability is used for the detection of low RCS UAVs using GSM signals. Field experiments were carried out in a remote area. Time-Doppler map is used for detection of UAV. The ranging distance was approximately $1$~km.  The radar system detected and tracked multiple targets based on their Doppler characteristics.& Bistatic&\cite{UAV_passive_bistatic}\\
\hline 
LTE&A SDR-based LTE passive radar is used for the detection of a UAV in an urban environment. A frequency division duplex downlink LTE signal is used as an illuminator of opportunity. The maximum Doppler shift calculated was around $250$~Hz. The SNR was approximately $25$~dB after clutter cancellation. The high SNR helped to achieve high detection and tracking accuracy.&Bistatic&\cite{LTE5}\\
\hline
$5$G&A passive radar with Rényi entropy using $5$G network is used for the detection of UAVs. The Rényi entropy helps to select time frames at the downlink channel in both frequency and time. The PFA varied between $10^{-4}$ and $10^{-8}$ and improved with resource allocation and high SNR. Rényi entropy was $> 25.5$, and stable Rényi entropy helped to reliably analyze signals for accurate UAV detection. &Bistatic&\cite{passive_5G_3}\\
\hline
WiFi&A WiFi-based passive radar is used for the detection of UAVs. In addition to detection, $3$D localization of the UAV can be obtained using a quad-channel RX. Small UAVs were detected and tracked up to $40$~m with an overall PFA of $10^{-3}$. &Multistatic&\cite{passive_Wifi1}\\
\hline
DAB&DAB signals are used as illuminators of opportunity for a passive multistatic radar for the detection of low-flying UAVs. The detection ranges varied between $30$~km to $120$~km for manned aircraft, and the average bistatic RCS increased by $10$~dB from the L-band to the VHF band.&Multistatic&\cite{DAB4}\\
\hline
DVB-T&A DVB-T-based passive radar is used for the detection and tracking of UAVs. The range-Doppler matrix is used for the identification of UAVs. UAVs were detected at ranges of $3$~km for stationary targets and $3.5$~km for moving targets. The detection was dependent mainly on the presence and characteristics of rotating propellers.&Bistatic&\cite{UAV_DVBT}\\ 
\hline 
Satellite&Passive FSR and DVB-S signals are used for UAV detection. The proposed system is similar to an air traffic monitoring system. The maximum Doppler frequency was $88$~Hz, mainly due to the UAV's rotating propellers. The SNR was $14$~dB for a detection range of $100$~m. &Bistatic&\cite{Satellite1}\\
\hline
\end{tabular}
		\end{center}
			\end{table*}
   
\subsubsection{GSM}
Major advantages of using a GSM network for the detection and tracking of UAVs include the collection of detection measurements from different aspect angles using BSs, and combining measurements from different BSs to overcome weak measurements from a single BS. In~\cite{UAV_passive_radar}, a GSM-based radar in the passive mode was used to collect weak reflections from a UAV. Experiments were conducted to detect and track a quadcopter using passive GSM-based radar achieving a high detection rate. A track-before-detect method was used to detect and track small UAVs using GSM signals. A GSM-based detection system for small UAVs operating in the passive mode was introduced in \cite{GSM_new1}. Passive bistatic radar measurements were carried out using GSM signals in \cite{UAV_passive_bistatic} for the detection of small RCS UAVs. Experiments were carried out in \cite{GSM_new2} using multiple mobile telecommunication illuminators including GSM in the passive mode for the detection of aerial targets.  

\subsubsection{LTE}
Major benefits of LTE include variable bandwidth available from $1.4$~MHz to $20$~MHz and high range and velocity resolution compared to GSM. Also, the ambiguity function in LTE results in lower sidelobes, and the global coverage allows coordinated passive sensing. Similar to GSM, an LTE network can be used for the detection and tracking of UAVs in the passive mode. A USRP-based setup in \cite{LTE1} used LTE signals from multiple TXs and a single RX was used for the detection of UAV in the passive mode. In \cite{LTE4}, a synthetic aperture radar~(SAR) image of the UAV was obtained using OFDM signals of an LTE system. A passive LTE-based radar system was used in \cite{LTE5} for the detection of UAVs. Experiments were conducted to demonstrate the capability of the passive LTE radar to detect low-flying UAVs. In \cite{LTE6}, an LTE-based multistatic radar was used for the detection of UAVs. A new target detection-based feature was added to the existing downlink LTE signal.

\subsubsection{$5$G and Beyond}
\begin{figure}[!t]
	\centering
	\includegraphics[width=\columnwidth]{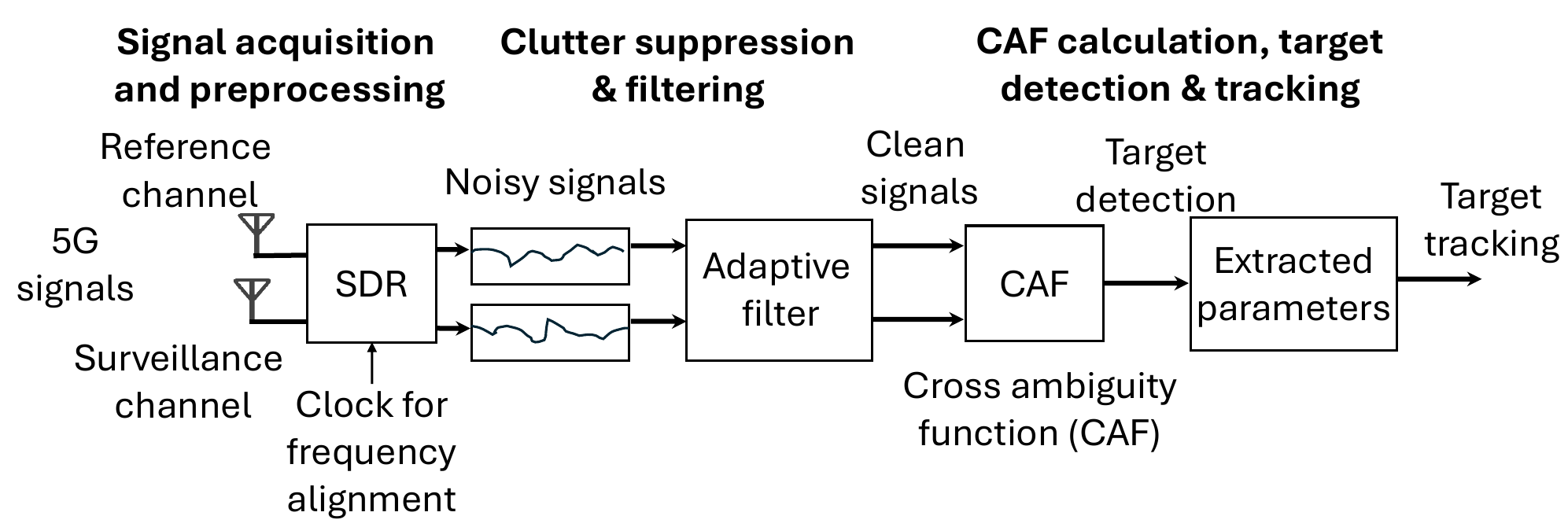}
	\caption{A passive radar for target detection and tracking using $5$G signals~\cite{passive_5G_1}.}  \label{Fig:Fig_JCS2}
\end{figure}

To fulfill high data rate requirements, mmWave, and higher frequency bands are envisioned for $5$G and beyond communications. Salient features of $5$G communications include dense network coverage through a large number of access points and connected devices, highly directional antenna beams, and beam steering capabilities. In \cite{passive_5G_2}, a passive $5$G radar was used for the detection of UAVs. The system took advantage of variable resource allocation in a $5$G system for the detection of low RCS UAVs. In \cite{passive_5G_1}, a fully cooperative $5$G cellular system was used as a source of illumination for a passive radar for target detection. A moving target was illuminated by the $5$G BS in a bistatic geometry. The processing of the $5$G signals for target detection and tracking in \cite{passive_5G_1} are shown in Fig.~\ref{Fig:Fig_JCS2}. In \cite{passive_5G_3}, Rényi entropy was used for the selection of time frames in a downlink channel of $5$G system. The resource allocation was found to have a significant effect on the detection and range resolution of a target. 

\subsubsection{WiFi}
The WiFi signals reflected from targets in the surroundings can be captured passively and subsequently used for detection. In \cite{passive_Wifi1}, WiFi transmissions were used for the detection and $3$D localization of UAVs and small planes using a passive radar. A passive bistatic radar based on WiFi transmissions is also provided in \cite{passive_Wifi2, passive_Wifi3}. In \cite{passive_Wifi3}, a WiFi-based passive bistatic radar using least square adaptive filtering and moving target indicator was used. The proposed approach provided a cost-effective solution that can be implemented using SDR for the detection of low-flying aerial vehicles.

\subsubsection{Digital Audio Broadcast}
Digital Audio Broadcast (DAB) has been a popular source of communication to large populations. DAB can also be used for the detection of UAVs in the passive mode. A passive radar in the VHF band and using DAB transmissions were used for the detection of a fixed-wing UAV in~\cite{DAB1}. The detection range of the fixed-wing UAV reported was $1.2$~km. In~\cite{DAB2} FM~(using analog signals) radio-based passive radar sensor was used. The sensor can be used for air surveillance. Similarly, passive multi-static aerial surveillance using DAB signals was carried out in~\cite{DAB3}. The Doppler and bi-static time difference of arrival measurements were used and a multi-hypothesis tracking algorithm was developed to help at different stages of tracking. 

In~\cite{DAB4}, DAB-based passive multistatic radar was used to detect stealth UAVs. Moreover, a passive multistatic radar based on the FM radio signal transmissions in~\cite{DAB6} was used to detect aerial vehicles. The tracking was carried out using three passive radars placed at different locations. In~\cite{DAB5}, a passive coherent location~(PCL) system using DAB, FM, and cellular communication signals was used. The PCL was used to detect UAVs. 

\subsubsection{Digital Video Broadcast Terrestrial}
Similar to DAB, digital video broadcast terrestrial~(DVB-T) signals can be used for the detection and tracking of UAVs. The potential of a passive radar using DVB-T was shown in \cite{UAV_DVBT}. Experimental data for the detection and identification of multi-rotor and fixed-wing UAVs was provided. It was claimed that passive radars can play an important role in the detection of different types of UAVs at long ranges using DVB-T transmissions. The potential uses of DVB passive radar for the detection and identification of different small UAVs were provided in \cite{UAV_DVBT}. The detection and identification using DVB passive radar helped to counter illegal UAV flights. The detection of UAVs using DVB-T2 passive coherent radar was provided in \cite{UAV_DVBT2}. 

A passive radar using DVB-T illuminators of opportunity was used for $4$D~(range, Doppler, azimuth, and elevation) detection and tracking of small UAVs in~\cite{UAV_4D}. The approach used a uniform linear array and two separate antennas for the estimation of range, Doppler, azimuth, and elevation angles of the UAVs. A passive multistatic digital TV-based radar for UAV detection was provided in \cite{UAV_SDPR}. Passive digital TV-based radar was used for the detection and tracking of UAVs in a highly cluttered scenario near airports in~\cite{UAV_SDPR}. In \cite{UAV_DVBT2_Clean} it was observed that the strong reflections from large RCS of commercial planes made software-based detection and tracking of small RCS UAVs difficult. Long coherent processing intervals and multi-stage CLEAN algorithms were, therefore, used to filter strong reflections from commercial planes.

 \subsubsection{Satellites}
 Satellite communications can be used for the detection and tracking of UAVs near the ground. However, the long-distance of the satellites from the ground, low data rates, and small RCS of common UAVs make detection and tracking challenging. In \cite{Satellite1}, a DVB satellite~(DVB-S) was used for the detection of UAVs. The micro-Doppler signature from rotating blades was used for UAV identification. To overcome the challenges of UAV detection by satellite networks, forward scattering radar~(FSR) configuration was used in the passive mode. The passive DVB-S and FSR were also used in \cite{Satellite2} for the detection of multi-rotor UAVs and performing micro-Doppler analysis of the copter UAVs. In \cite{Satellite4}, DVB-S was used for the detection of multi-rotor UAVs and micro-Doppler signature extraction. The micro-Doppler signature helped in the UAV classification. A reference channel and a surveillance channel were used. The reference channel pointed to the satellite for reference signaling, whereas the surveillance channel pointed to the UAV. The two channels were used to calculate the cross-ambiguity function which assisted in UAV detection. 

\subsection{Limitations of Communication Systems for DCT-U}
There are many limitations of communication systems when used for the DCT-U~(in active or passive modes). Further research efforts are required to address these limitations. The limitations are summarized as follows:
\begin{itemize}
    \item A common signal that optimizes both communications and sensing in JC\&S is desired. However, designing a signal that optimizes communications and sensing simultaneously is challenging. 
    \item In JC\&S systems where different signals are employed for communication and sensing, sharing the same resources can lead to interference between the communication and sensing signals.
    \item In a JC\&S system that utilizes a common signal for both communications and sensing, a necessity for full duplex operation arises.
    \item Synchronization between the TX and the RX is a major challenge for JC\&S systems when TX and RX are not co-located. 
    \item The effectiveness of sensing using passive radar when the target illuminator is a communication system depends on the characteristics of the communication system. 
    \item The accuracy of a passive radar based on the communication system is dependent on many factors including communication signal type, interference levels, and propagation environment. The accuracy is generally low compared to active radar systems.
    \item The range and coverage of target detection in passive mode rely on the strength of the communication signals. Communication systems operating in high-frequency bands, such as $5$G, may exhibit limited range for passive mode target detection. 
    \item The communication system-based passive radar systems are complex and often require multiple RX nodes and significant training data for correct detection. 
    \item The tracking of a UAV requires multiple communication nodes that coordinate with each other. The coordination for tracking UAVs requires additional resources and infrastructure. 
    \item The classification of UAVs requires training data. The training data obtained through a given communication system may not be applicable directly to another type of communication system.  
\end{itemize}


\section{DCT-U using Passive monitoring of RF SOO Radiated from UAVs} \label{Section:SOO}
In this section, the passive monitoring of RF signals radiated from UAVs for DCT-U is discussed. Note that this is different from what is covered in Section~\ref{Section:Section_V_D}, since here we consider signals explicitly transmitted by the UAVs. The advantages and limitations of DCT using passive monitoring of SOO radiated from UAVs are also provided. 


\subsection{RF Signals Radiated from UAV for DCT-U}
The RF signals radiated from UAVs for remote control and data transfer can be used for DCT-U. The RF signals are captured either between two communicating UAVs or between a UAV and a ground station~(GS). Different methods are available in the literature for capturing RF signals radiated from UAVs and distinguishing these RF signals from other communication signals for the detection and tracking of UAVs. Furthermore, the classification of UAVs is performed based on the features of the detected RF signals. In \cite{RF_analysis_clas}, a passive radio surveillance system was used to capture and analyze RF fingerprints of the radiated signals between UAV and GS in the presence of WiFi and Bluetooth interference. After the signal was confirmed to be from a UAV, the UAV was classified using different ML methods.

The effect of interference from WiFi and Bluetooth signals on the detection and classification of UAVs through the RF analysis of captured signals was provided in \cite{interference_Wifi_bluetooth2}. It was shown that by using CNN and ML classification, a high percentage of accuracy for the detection and classification of different types of UAVs was achieved. There are also similar studies where RF fingerprints of the radiated signal between UAV and GS were captured and analyzed passively to detect UAVs~\cite{class_new11,Wifi1,ozturk_journal}. 

In \cite{SOO1}, the detection and localization of UAVs relied on the utilization of multidimensional features extracted from the signals between the UAV and the GS. Features included the frequency transform of the signal, power spectral entropy, and wavelet energy entropy. ML algorithms were employed to detect UAVs based on these selected features. \cite{SOO2} involved the creation of a database comprising RF signals emitted by various UAVs in diverse flight scenarios, including being on the ground and connected to the GS, flying, hovering, and during video recording. Using the collected data and deep neural networks, UAVs were detected and classified. 

\begin{table*}[htbp]
	\begin{center}
    \footnotesize
		\caption{Unique aspects of DCT-U that are based on the passive detection of RF signals emitted by UAVs are summarized.}  \label{Table:SOO_radiated_UAVs}
\begin{tabular}{@{}|P{ 0.5cm}|P{ 2.0cm}|P{ 1.1cm}|P{ 1.9cm}|P{2.3cm}|P{3.6cm}|P{3.3cm}|@{}}
\hline
\textbf{Ref.} & \textbf{RF SOO considered} \big($f_{\rm c}$ is center frequency, BW is bandwidth\big) & \textbf{\# of UAVs/ controllers} & \textbf{AI/ML techniques used} & \textbf{Hardware for RF signal detection} & \textbf{Unique aspects}&\textbf{Computational complexity}\\
\hline
\cite{RF_analysis_clas}& $f_{\rm c} =2.4$~GHz, $5$~M samples&$17$ UAV controllers&kNN, discriminant analysis, SVM, neural network, RandF& High-frequency oscilloscope, directional antenna&Detection and classification of UAVs performed using RF energy detection between UAVs and GS in the presence of WiFi and Bluetooth interferences& Computational time provided in Table~IV of \cite{RF_analysis_clas}\\
\hline
\cite{interference_Wifi_bluetooth2}&$f_{\rm c}=2.4375$~GHz, BW = $28$~MHz&$7$ UAVs&kNN, logistic regression, CNN&SDR and directional antenna& RF signals between UAV and GS were collected and analyzed in the presence of WiFi and Bluetooth interferences& $O\big(N_{\rm I}+N_{\rm T}^2\big)$, where $N_{\rm I}$ is number of images, and $N_{\rm T}$ is number of training samples for kNN\\
\hline
\cite{SOO1}&$f_{\rm c}=2.4$~GHz, $20$~Mbit/s sampling rate&$2$ UAVs&SVM, random forest, kNN, Naive Bayes, ensemble learning&X$310$ USRP, UBX-160 daughter board, and circular antenna array& Detection and tracking of UAVs was carried out based on multidimensional signal features&$O\big(N_{\rm L}logN_{\rm L}\big)$, where $N_{\rm L}$ is the signal length\\
\hline
\cite{SOO3}&$f_{\rm c}=2.44$~GHz, BW = $20$~MHz&$1$ UAV&$-$&RF sensor and omnidirectional antenna&Multiple RF sensors were used to localize and track a UAV using extended Kalman filter&$O\big(N_{\rm k}n_{\rm d}^3\big)$, where $N_{\rm k}$ is number of time steps and $n_{\rm d}$ is state vector dimension\\
\hline
\cite{SOO8}&$f_{\rm c}=2.437$~GHz, $100$~kHz sample rate&$1$ UAV&$-$&SDR and directional antenna&Frequency spectrum analysis of signals between UAV and GS was carried out&$O\big(N_{\rm L}logN_{\rm L}\big)$\\
\hline
\cite{Wifi1}&$f_{\rm c}=2.4$~GHz&$2$ UAVs&Random tree, random forest, Logit boost&Guardian WiFi sniffer, and laptop& WiFi traffic generated by different UAVs was captured and analyzed& $O\big(FlogF\big)$, where $F$ is ratio of packet count by partitioning window\\
\hline
\cite{Wifi3}&$f_{\rm c}=2.4$~GHz&$2$ UAVs&kNN, SVM, RandF&SDR, directional and omnidirectional antennas& Wireless signals emitted by UAV were captured and fingerprinting of wireless signals was carried out&$O\big(N_{\rm s}N_{\rm L}logN_{\rm L}\big)$, where $N_{\rm s}$ is the number of segments\\
\hline
\cite{SOO2}&$f_{\rm c}=2.4$~GHz&$3$ UAVs&Deep neural network&USRP-$2943$&RF signals emitted from UAVs were collected and analyzed&$O\big(EK\sum_{l=1}^LH_{l-1}.H_l\big)$, where $E$ is epoch and $K$ is folds of cross-validation, and $H$ is neuron count \\
\hline
\cite{SOO4}&$f_{\rm c}=2.422$~GHz&$3$ UAVs&CNN&USRP-$2943$R&RF communications between UAV and GS were captured and analyzed&$O\big(N_{\rm l}DN_{\rm c}\big)$, where $N_{\rm l}$ is number of convolutional layers, $D$ is data dimension, $N_{\rm c}$ is number of channels\\
\hline
\cite{SOO5}&$f_{\rm c}=2.441$~GHz&$3$ UAVs&CNN&USRP-
$2943$&UAV RF signal database was used for detection and classification of UAVs&$O\big(N_{\rm s}logN_{\rm s}\big)$, where $N_{\rm s}$ is the number of samples in RF segment\\
\hline
\cite{SOO6}&$f_{\rm c}=2.4$~GHz, BW = $9.8$~GHz&$1$ UAV&Artificial neural network&Passive receiver and computer&Clutter suppression was applied on UAV communication signals&$O\big(N_{\rm f}^2N_{\rm a}\big)$, where $N_{\rm f}$ is number of frequency samples, and $N_{\rm a}$ is number of angles\\
\hline
\cite{SOO9}&$f_{\rm c}=900$~MHz&$2$ TX and $5$ RX UAVs&Deep CNN&USRP B$200$mini&Customized RF fingerprinting was applied on RF signals captured between UAV and GS&$O\big(p^4E+N_{\rm e}N_{\rm R}N_{\rm T}\big)$, where $p$ is CNN parameter size, $N_{\rm e}$, $N_{\rm R}$, $N_{\rm T}$ are number of channel estimates, and RX and TX UAVs \\
\hline
\cite{SOO10}&$f_{\rm c}=5.775$~GHz, and $2.441$~GHz&$10$ UAVs&Single and multi-neural network&Spectrum analyzer, and USRP X$310$ with UBX~$160$ daughterboard&RF signals from multiple UAVs in a swarm were collected and analyzed&$O\big(N_{\rm e}N_{\rm p}\big)$, where $N_{\rm e}$ and $N_{\rm p}$ are number of examples and neuron network parameters\\
\hline
\cite{SOO11}&$f_{\rm c}=2.4$~GHz&$10$ UAV controllers&kNN, SVM, SqueezeNet &Digital oscilloscope and directional antenna&Signals emitted from different UAV controllers were collected and analyzed& Computational time provided in Table~8 of \cite{SOO11}\\
\hline
\cite{wifi4}&$f_{\rm c}=2.4$~GHz&$8$ UAVs&$-$&$8260$ WiFi adapter on a personal computer&Encrypted WiFi traffic of different types of UAVs were collected and analyzed&$O\big(N_{\rm p}N_{\rm F}\big)$, where $N_{\rm p}$ is number of packets, and $N_{\rm F}$ is number of features\\
\hline
\end{tabular}
		\end{center}
			\end{table*}
   
The detection and localization of a UAV was carried out by passively monitoring the RF signals emitted from the UAV in an open area using RF sensors in \cite{SOO3}. An envelope detection algorithm and RF database of different UAVs were compared with the measurement data for UAV detection. The time difference of arrival was used to localize the UAV. There are other studies available that detect UAVs based on RF signals radiated from UAVs and use different ML methods~\cite{SOO4,SOO5,SOO6,SOO7}. In Table~\ref{Table:SOO_radiated_UAVs} RF SOO radiated from UAV/controller, number of UAVs/controllers used for classification, AI/ML techniques used, hardware used and unique aspects of DCT-U based on passively capturing RF signals radiated from UAVs are summarized. Computational complexities of different methods is also provided in Table~\ref{Table:SOO_radiated_UAVs}.

Observations from Table~\ref{Table:SOO_radiated_UAVs} highlight that most RF SOO-based passive UAV detection operate at a center frequency of $2.4$~GHz. Additionally, the hardware used in the majority of studies listed in Table~\ref{Table:SOO_radiated_UAVs} is simple and off-the-shelf. While deep learning offers highly accurate UAV classification, its accuracy might diminish in scenarios divergent from the training data. Table~\ref{Table:SOO_example} presents a comparative performance analysis of various ML algorithms employed for detecting and classifying different UAVs/controllers using RF fingerprinting techniques in \cite{RF_analysis_clas}. From Table~\ref{Table:SOO_example}, it can be observed that RandF offers the highest accuracy for the classification of UAVs.

\begin{table}[htbp]
\centering
\begin{tabular}{@{}|P{ 1cm}|P{1cm}|P{2.5cm}|P{2.5cm}|@{}}
\hline
\textbf{\# of controllers}& Classifier& Accuracy using all features $(\%)^2$& Computational time using all features $(\rm{s})^2$  \\ \hline
\multirow{ 5}{*}{15} & kNN & $97.30$ & $24.85$  \\
& DA & $96.30$ & $19.42$    \\ 
& SVM & $96.47$ & $119.22$   \\  
& NN & $96.73$ & $38.73$    \\ 
& RandF & $98.53$ & $21.37$    \\ \hline
\multirow{ 5}{*}{17} & kNN & $95.62$ & $26.16$  \\
& DA & $92.77$ & $19.36$    \\ 
& SVM & $93.82$ & $139.94$   \\  
& NN & $92.88$ & $46.04$    \\ 
& RandF & $9.32$ & $24.71$    \\ \hline
\end{tabular}
\caption{Performance comparison of different ML algorithms for detecting and classifying different types of UAV controllers using RF fingerprinting~(regenerated from \cite{RF_analysis_clas})}.
\label{Table:SOO_example}
\end{table}

\subsection{Advantages of DCT-U using RF SOO Radiated from UAVs}
DCT-U using passive RF SOO captured between the UAV and the GS offers many advantages given as follows:
\begin{itemize}
    \item Passive monitoring of RF SOO between UAV and GS provides DCT-U that is undetectable by the UAV operators. 
    \item No RF interference is introduced to the monitored UAV communications or other communication systems.
    \item Passive RF analysis of SOO offers longer detection and tracking ranges compared to visual and acoustic sensors.
    \item Simultaneous DCT of multiple UAVs is possible using RF analysis of signals radiated by different UAVs.
    \item The classification of a UAV using RF analysis of captured SOO is based on the specific frequencies and patterns of the signals used between the UAV and the GS. Compared to radar systems, where RCS is mainly used for classification, the classification is simpler and more detailed.
    \item Analyzing the control signals between UAV and GS can provide information regarding the UAV's mission, e.g., streaming video or reconnaissance.
    \item Passive RF analysis of SOO offers all-weather, day, and night DCT-U compared to EO/IR sensors.
    \item The cost for DCT-U using passive RF analysis of SOO is significantly less than that of radar systems. 
    \item The RF SOO-based passive monitoring systems for DCT-U have longer operational lifespans and require reduced maintenance due to fewer active components. 
    \end{itemize}

\subsection{Communication Systems for Capturing RF Signals Radiated from UAVs for DCT-U}
The communication protocols used by UAVs commonly use the same frequency bands as used by WiFi, e.g., $2.400–2.483$~GHz and $5.725–5.825$~GHz. Therefore, RF signals emitted by UAVs can be captured using WiFi and other communication RXs operating in the same frequency band. In \cite{Wifi1}, statistical fingerprint analysis of the WiFi signals was carried out to detect the presence of unauthorized UAVs. The characteristics of the UAV control signals and first-person view transmissions, features of WiFi traffic generated by UAVs, and ML algorithms helped to identify a UAV in a WiFi coverage area. Similarly, in \cite{Wifi2}, a WiFi-based statistical fingerprinting method was used for the detection and identification of UAVs. The proposed method claimed to achieve an identification rate of $96$\%. In \cite{new_rfanalysis}, a three-stage low-cost UAV detection system through WiFi traffic monitoring and ML algorithms was provided. The three stages included WiFi data monitoring, UAV classification, and UAV database management. The proposed approach has low computational complexity and can provide detection over encrypted WiFi traffic. 

\begin{figure}[!t]
	\centering
	\includegraphics[width=\columnwidth]{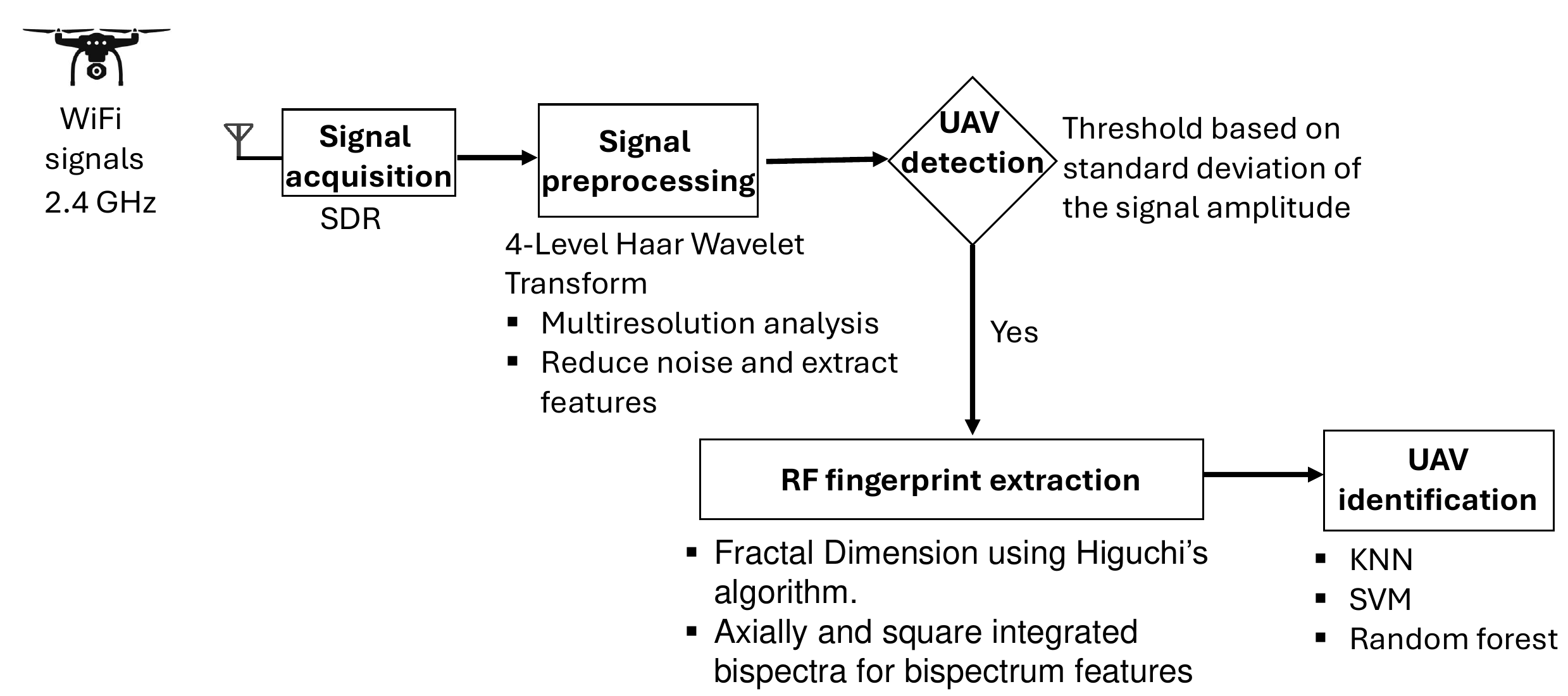}
	\caption{WiFi signals emitted by the UAV are captured, preprocessed, and RF fingerprinting is applied for UAV classification~\cite{Wifi3}. }\label{Fig:Fig_SOO1}
\end{figure}

In \cite{Wifi3}, the detection and identification of UAVs were carried out using captured WiFi signals from UAVs and RF fingerprinting. Firstly, a UAV was detected based on the captured RF signals by the WiFi RX, then ML algorithms were used to identify the category of the UAV. The identification accuracy was higher outdoors compared to indoors. The UAV detection and identification process using WiFi signals in \cite{Wifi3} is provided in Fig.~\ref{Fig:Fig_SOO1}. In \cite{wifi4}, ML algorithms were used to analyze the WiFi signals originating from UAVs for control and video streaming. The analysis helped to detect and identify UAVs. The proposed approach worked on the encrypted WiFi traffic also by using the packet size and WiFi traffic inter-arrival time features. In \cite{bluetooth_SOO}, long-range bluetooth was used to identify and track UAVs. The project provided a beacon capability for UAVs that helped in their identification when in the range of the bluetooth RX.

\subsection{Limitations of Passively Capturing RF SOO Radiated from UAVs}
There are limitations to passively detecting SOO radiated from UAVs for DCT-U. These limitations include, 1) the range of the RF analysis methods using different sensors is dependent on the strength of the radio link, and propagation channel between the UAV and the controller. Therefore, a sensor needs to be in the vicinity of the UAV and ground controller and the propagation channel should be favorable, 2) if a UAV is flying autonomously without any active radio link, or UAV is mechanically controlled, then the RF analysis method may not be effective, 3) RF analysis method operates in a passive mode and the precise location of the UAV cannot be precisely determined using a single antenna, 4) UAVs equipped with ECM can be used to deceive, hack, and jam RF analysis methods.


\section{Future Directions}    \label{Section:Future_directions}
In this section, future directions for DCT-U based on the lessons learned are provided. 

\subsection{Communication Systems}
Exploration of new methods for passive UAV detection through communication systems is a promising research area. Moreover, there is significant potential for the future of JC\&S with the utilization of $5$G and beyond mobile networks~\cite{sungjoon_journal,jcs_5G_1,jcs_6G_2}. The progress in JC\&S can directly benefit DCT-U. The $5$G and upcoming $6$G network deployments will enable high-throughput, ultra-low latency communications essential for real-time UAV monitoring and control. The massive MIMO capabilities of these networks and beamforming techniques can improve the accuracy and reliability of DCT-U. Additionally, using ML and AI to analyze communication signal patterns can significantly enhance the classification of UAVs. These technologies can enable the development of cognitive systems that can forecast UAV trajectories~\cite{trajectory_wahab} and identify potential threats in real-time.

The integration of satellite communication systems with terrestrial networks~\cite{space_terrestrial_integrate} is a promising direction for DCT-U. This hybrid approach can ensure uninterrupted tracking of UAVs, especially in remote areas, and provide comprehensive coverage. By combining data from various sources, including satellites, GSs, and UAVs, a robust and resilient DCT-U system can be established. There are additional possibilities in fields such as microwave communication connections between BSs, maritime radio transmissions, public safety communications, and combining data from different communication systems which can all be utilized for DCT-U.

\subsection{Single Long-range and High-resolution Radar}
Long-range radar with wide area coverage and high resolution can be used for DCT-U. Typically, long-range radars cover large areas and operate at low frequencies, but they have limited resolution. However, if we can develop a single long-range radar with wide coverage and high resolution, it can serve as a single point for DCT-U. A single long-range, high-resolution radar can have a shorter processing delay compared to distributed sensors. Achieving higher resolution without sacrificing range requires advancements in radar technology, which may be facilitated by using higher frequency bands, such as mmWave frequencies, that offer better resolution while maintaining sufficient coverage. Additionally, adaptive beamforming techniques~\cite{adaptive_beamforming} can focus the radar's energy on specific areas of interest, thereby enhancing DCT-U performance.

Integrating advanced signal processing algorithms, including ML and AI, can enhance the capabilities of long-range high-resolution radars. These algorithms can be used to identify patterns, filter out noise, and classify different types of UAVs with high accuracy. Another promising direction is the development of long-range, high-resolution, multifunctional cognitive radars that can perform multiple tasks simultaneously for DCT-U by adaptively scheduling available resources based on task priority~\cite{Wahab_AESA}.

Additionally, the effectiveness of long-range, high-resolution radars can be maximized by placing them in strategic locations. For instance, installing these radars on high-altitude platforms or satellites can broaden the coverage area and minimize scenarios where there is NLOS. This approach would be advantageous for monitoring extensive mountainous regions.

\subsection{Quantum Radars}
Quantum radars are considered to be the future of radar technology and can be utilized for the detection and tracking of UAVs. There is limited literature available on quantum radars at present~\cite{Quantum_review,UAV_quantum}. However, the future prospects for quantum radar technology in the field of DCT-U are promising. One significant area of research is focused on enhancing quantum entanglement techniques. By improving the generation and manipulation of entangled photon pairs, quantum radars can achieve high levels of resolution and sensitivity. This approach will help in the precise detection of small UAVs even in cluttered or noisy environments.

Another important aspect involves integrating quantum radar systems with classical radar systems. This hybrid approach can harness the strengths of both technologies, combining the high resolution and noise resilience of quantum radars with the established range and operational capabilities of classical radars. This integration has the potential to lead to the development of robust and versatile systems for DCT-U that are expected to operate effectively in a wide range of scenarios.

Developing scalable and practical quantum radar systems is a crucial area of focus. Current quantum radar prototypes often have limited operational range and require complex and sensitive equipment. Research into more compact and resilient quantum radar designs will be essential for their deployment in real-world scenarios. Advances in quantum optics and photonics, such as developing efficient single photon detectors and sources, can help achieve this goal.

In addition, investigating the capabilities of quantum radar for multi-static and networked radar systems is an exciting opportunity. Coordinating multiple quantum radars in a network can provide extensive coverage and improved DCT-U capabilities. Progress in quantum computing can also have a positive impact on quantum radar technology. Creating quantum algorithms could result in more efficient processing of the large amount of data generated by quantum radars, enabling real-time analysis and decision-making.

\subsection{Hybrid Radars}
A hybrid radar system consists of two or more radar types that can benefit from the strengths of different radar systems. For example, in~\cite{hybrid_active_passive}, a hybrid active and passive radar system was utilized. The active and passive radars operated at the same frequency band to maximize spectral efficiency. To address timing uncertainty and mutual interference caused by the simultaneous operation of radars, min-max and weighted sum criteria were employed. Future directions for DCT-U using hybrid radar systems are promising and extensive. One significant direction involves enhancing signal processing techniques to mitigate interference and timing uncertainties in multi-mode radar systems. By employing advanced algorithms, such as adaptive filtering and ML-based interference cancellation, hybrid radars can achieve high spectral efficiency and detection performance even in congested environments.

Additionally, integrating hybrid radar systems with other sensor technologies can create a robust and comprehensive network for DCT-U. For example, combining radar data with information from optical sensors, infrared cameras, and acoustic sensors can enhance the accuracy of DCT-U. This multi-sensor fusion approach can be particularly effective in differentiating UAVs from other airborne objects and detecting small RCS UAVs.

One possible area of progress involves integrating dynamic mode-switching into hybrid radar systems. This capability allows the radar to switch between various modes, such as active-passive and FMCW-interferometry, based on the operational context and the identified UAV characteristics. These systems can improve their performance to accommodate different scenarios, providing valuable adaptability in dynamic threat environments, including UAV swarms.

Moreover, advancements in hardware technology, including the creation of smaller and more energy-efficient radar components, will facilitate the use of hybrid radars in various environments. To create flexible and scalable detection networks, portable and miniaturized hybrid radar systems can be deployed across different platforms, including ground vehicles, ships, and UAVs.

\subsection{Sparse Signal Processing}
A significant limitation in the detection of UAVs at long ranges using RF-based systems is sparse signal returns. To better detect UAVs, the sparse signals can be processed using modern sparse signal processing techniques available in the literature~\cite{compressed2}. These techniques can exploit the sparsity of the signal returns to enhance the accuracy of DCT-U and reduce false alarms. Compressed sensing, sparse Bayesian learning, and matrix completion can be used for long-range DCT-U.

Another direction is the integration of sparse signal processing with ML and AI techniques. By combining the strengths of both fields, it is possible to develop intelligent systems that can learn from sparse signal data and improve their detection capabilities over time. For example, deep learning models can be trained on sparse signal datasets to identify patterns and features specific to UAVs, leading to accurate classification and tracking. Similarly, adaptive sparse signal processing techniques are also an important area for future exploration. This adaptability is essential in scenarios where the SNR is low or the environment is highly dynamic, such as in urban settings or during adverse weather conditions.

Furthermore, the combination of sparse signal processing with multi-sensor data fusion offers significant opportunities. By integrating data from various sensors like radars, cameras, and acoustic sensors, we can create a comprehensive and accurate system for DCT-U. Sparse signal processing can effectively extract and combine relevant information from these different data sources, thus improving overall situational awareness and detection reliability.

The development of robust sparse signal processing techniques that can operate effectively in the presence of interference and jamming is crucial for the deployment of DCT-U systems in contested environments. Research into anti-jamming strategies and resilient signal-processing methods will be essential to ensure the reliability and effectiveness of these systems in real-world applications.

\subsection{Open-Ended, Modular, and Reconfigurable Design}
For future RF-based systems to detect UAVs, it is recommended to use an open-ended, modular, software-defined architecture. This type of architecture allows for updates and improvements to be implemented without requiring significant changes to the overall hardware. It enables seamless integration of new detection algorithms, signal-processing techniques, and communication protocols as they become available. By adopting this approach, cost savings, increased operational efficiency, extended system lifespan, and opportunities for collaborative efforts among the research community can be realized.

The modular nature of software-defined architectures facilitates the incorporation of multi-sensor fusion capabilities and the development of networked DCT-U systems. The software-defined architecture allows quick and simplified addition or removal of sensor modules, providing flexibility in response to evolving requirements and challenges.

Moreover, the combination of cloud computing and edge computing technologies can improve the processing capabilities of adaptable, software-defined systems. By transferring resource-intensive tasks to the cloud or spreading them across edge devices, the system can effectively manage large amounts of data and deliver real-time DCT-U. Furthermore, the security of these systems can be strengthened through modular and software-defined methods. Security updates and patches can be swiftly and efficiently deployed, addressing vulnerabilities without interrupting the overall system operation.

\subsection{Resilience to ECM}
RF-based systems need to withstand modern ECM for accurate UAV detection. Specifically, these systems should resist jamming, spoofing, hacking, and decoying (virtual and actual decoys). To achieve this resistance, RF-based systems can utilize secure RF communications with low-energy transmissions, frequency hopping over a wide frequency band, encryption, pulse coding, and multiple authentication layers. One key area for future research and enhancing the resilience of RF-based systems for DCT-U involves the implementation of adaptive and cognitive radio technologies using AI. This technology enables RF-based systems to dynamically sense and adapt to the RF environment, selecting the best frequencies and transmission parameters to avoid jamming and interference. These systems can predict and respond to ECM in real-time by employing AI algorithms.

Research into low-probability-of-intercept and low-probability-of-detection techniques can provide resilience against ECM. These techniques involve designing complex transmission signals for adversaries to detect and intercept, reducing the chances of being targeted by jamming or spoofing attacks. By minimizing the signal’s detectability, low-probability-of-intercept and low-probability-of-detection techniques can help ensure that UAV detection systems remain less susceptible to ECM.

The advancement of encryption and cybersecurity is essential for safeguarding RF-based systems from hacking and unauthorized access. The development and implementation of cutting-edge encryption protocols, such as quantum-resistant algorithms, can help secure communication channels and data transmitted between sensors and control systems. Additionally, the use of blockchain technology for secure and tamper-proof data logging and transmission can enhance the system's resistance to cyber-attacks.


\begin{figure}[!t]
	\centering
	\includegraphics[width=\columnwidth]{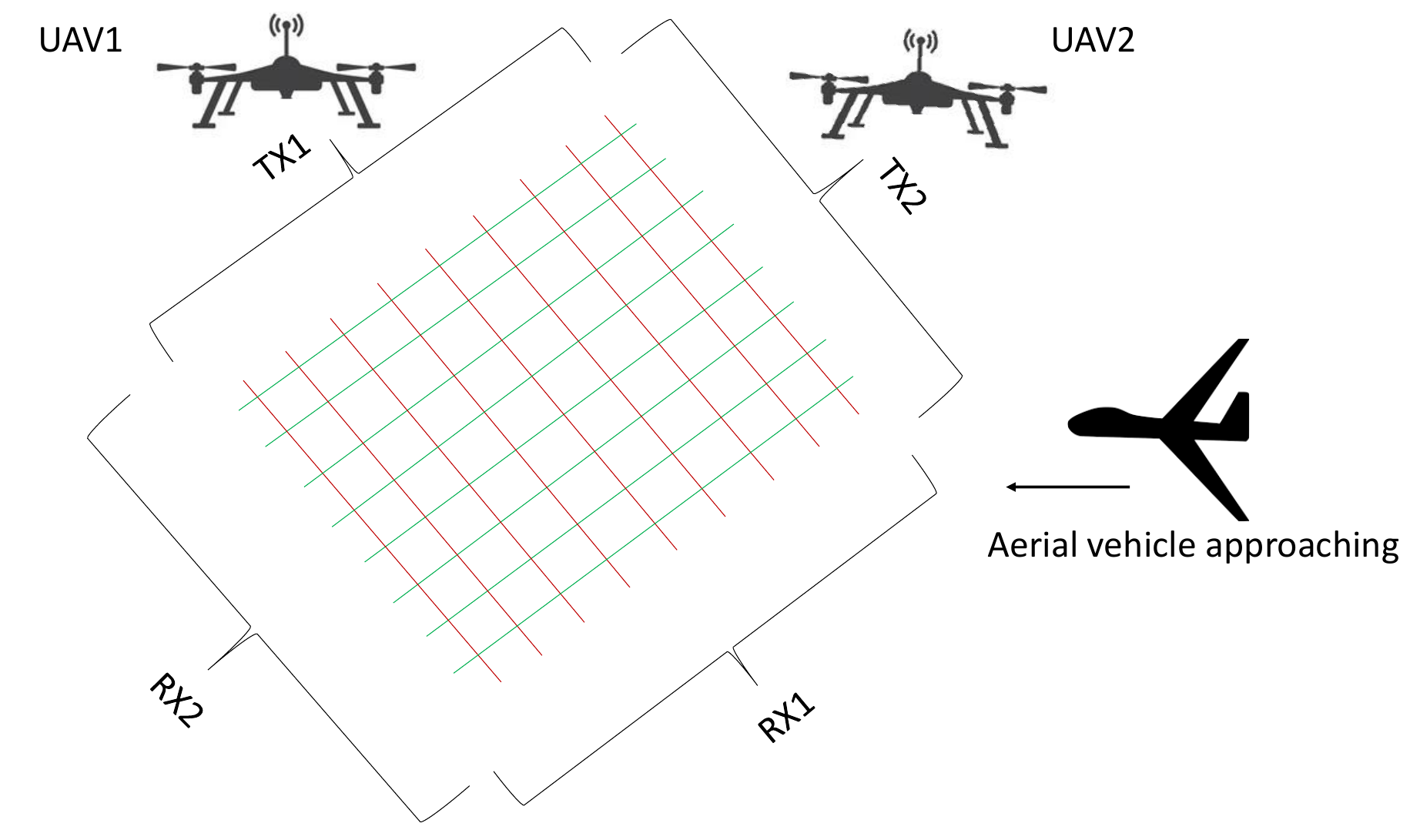}
	\caption{A laser mesh created in the air using two airborne UAVs for detection, tracking, and classification of aerial vehicles~(regenerated from \cite{UAV_wahab_laser}).}\label{Fig:laser_wahab}
\end{figure}

\begin{figure*}[!t]
	\centering
	\includegraphics[width=\textwidth]{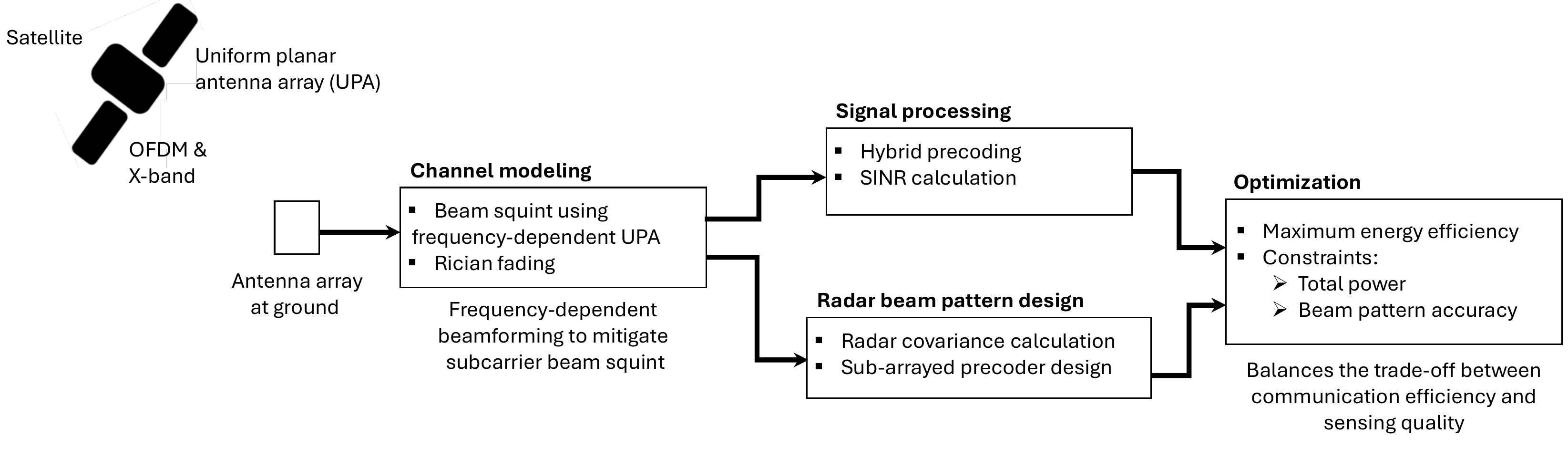}
	\caption{Beam squint-aware JC\&S for hybrid massive MIMO LEO satellite systems~\cite{sat_JCS}}.\label{Fig:sat_JCS}
\end{figure*}

\subsection{Laser-Based Techniques}
Laser is used for different civilian and defense applications. The laser can also be used for DCT-U.  For example, in \cite{track_new6}, a lidar sensor is used for the detection and tracking of small UAVs. Furthermore, a novel approach using a mesh of laser beams to detect, track, and classify small UAVs and stealth aerial vehicles was provided in \cite{UAV_wahab_laser}. At least two airborne platforms at different spatial positions were required. The two airborne platforms transmitted laser beams toward the ground RXs such that a mesh was formed as shown in Fig.~\ref{Fig:laser_wahab}. If an aerial vehicle blocked the path of the laser beams, it was detected and subsequently tracked and classified. Laser beam steering was also suggested to increase the coverage range and for better tracking and classification in \cite{UAV_wahab_laser}.  

\subsection{Distributed Sensors and Sensor Fusion}

A network comprising various off-the-shelf sensors such as imaging and acoustic sensors, as well as SDRs, can be deployed at different locations. These sensors may be situated on electricity and telecommunication towers, tall buildings, and other infrastructure. They are linked to a central network. Combining data from different sensors enhances situational awareness, surpassing the capabilities of a single sensor. The distributed sensors offer an initial layer of detection across a wide area. In \cite{new_sensorfusion}, radar, optical cameras, and microphones were utilized for gathering UAV detection measurements. Sensor fusion employing CNN and multinomial logistic regression was employed to enhance the accuracy of UAV detection and classification.

AI algorithms can be employed for advanced data fusion. Deep learning techniques like recurrent neural networks and transformer models can effectively handle temporal and spatial data correlations across multiple sensors. Additionally, advanced sensor fusion frameworks capable of real-time data processing and decision-making can be utilized. Techniques such as Bayesian inference, Kalman filtering, and particle filtering can enhance the robustness and reliability of the fused data. These methods can offer accurate position, velocity, and identity estimates of detected UAVs by optimally combining information from all available sensors.

\subsection{Regulating the UAV Air Traffic}
To minimize the threat posed by malicious and non-cooperative UAVs, regulating UAV air traffic is essential. This can be achieved by installing control towers similar to air traffic control towers, with the flexibility to be mounted on existing infrastructure such as telecommunication towers. These control towers should have the capability to autonomously manage UAV traffic in their vicinity by assigning tags to registered UAVs. The tags would be updated as the UAV approaches the next control tower limit, aiding in the identification of unregistered and malicious UAVs.

Blockchain technology can also be utilized to improve the security and traceability of UAV flight traffic. By creating a decentralized and immutable ledger of all UAV flights, blockchain ensures transparent and tamper-proof recording of all movements. This would aid in verifying the authenticity of UAV tags and tracking the history of each UAV, thereby making it challenging for malicious actors to go undetected.

Another solution is by using Automatic Dependent Surveillance-Broadcast Protocol~(ADS-B) protocol for the management of UAV traffic. ADS-B can provide improved situational awareness for UAV and civilian airline operators ensuring safe operations. Moreover, by adopting different security measures for ADS-B~\cite{ADS-B}, ECM can be minimized. Other future possibilities for UAV traffic management include using remote ID~\cite{remote_id}, unmanned aircraft system traffic management~(UTM)~\cite{utm}, advanced air mobility~(AAM)~\cite{aam}, and drone corridors~\cite{drone_corridor}. 

The cooperation among governments, industry stakeholders, and international organizations is crucial for creating standardized regulations and protocols for managing UAV traffic. Standard regulations can help facilitate UAV operations across borders and ensure consistent safety and security measures globally. Partnerships between the public and private sectors can encourage innovation and investment in the development and deployment of advanced UAV traffic management systems.

\subsection{Airborne UAVs}
Sensors on airborne vehicles, such as radars, can provide long-range detection against aerial threats~\cite{Wahab_AESA}. However, the operational and maintenance cost of manned aerial surveillance vehicles is high. A possible solution is to use sensors onboard UAVs for surveillance. The surveillance UAVs can be either remotely controlled by a human operator, work autonomously using AI, or both, and fly as a network in a swarm. These AI-driven UAVs can perform diverse tasks simultaneously with minimal human intervention.

Tethered UAVs and high-altitude platforms, such as hot air balloons and airships, can be further explored to extend the coverage and endurance of DCT-U operations. Tethered UAVs can provide continuous power and data connectivity, enabling prolonged missions without frequent recharging or maintenance. High-altitude platforms can serve as persistent surveillance nodes, offering a stable high-altitude point for monitoring large areas and relaying data to GS. Adopting advanced energy solutions, such as solar power and fuel cells, can extend the operational endurance of aerial surveillance platforms.

\subsection{Space Assets}
It is possible to monitor UAVs using sensors in space, such as radar, imaging, and infrared sensors~\cite{UAV_space1,UAV_space2}. Space-based sensors can provide early warnings for various aerial threats, including UAV swarms. These sensors have a wide coverage area and the added advantage of mobility. However, space sensors need to account for atmospheric and space effects as well as long delays. Advances in satellite technology, like high-resolution SAR and hyperspectral imaging, can greatly enhance the ability to detect UAVs from space. These high-resolution sensors can capture detailed images and signatures of UAVs, even in challenging conditions such as low visibility or adverse weather.

In addition, the satellite communications can be used for sensing also. In\cite{sat_JCS}, JC\&S in massive MIMO low-earth orbit~(LEO) satellite systems is discussed, addressing beam squint effects. A beam squint-aware hybrid precoding technique, signal-to-interference-plus-noise ratio~(SINR) calculation, using statistical channel state information is proposed, achieving efficient communication and sensing with mitigated squint effects. The implementation is summarized in Fig.~\ref{Fig:sat_JCS}. The deployment of constellations of small satellites, also known as CubeSats, can improve DCT-U capabilities. CubeSats can work together, offering continuous and overlapping surveillance over large areas. Their relatively low cost and quick deployment make CubeSats an appealing option for expanding the space-based surveillance network.

\subsection{Exploring Cost Effective Solutions}
To counter malicious UAVs, cost-effective solutions should be explored. For example, a cost-effective solution is to use simulators. Obtaining data from real-time measurements is time-consuming and expensive. Moreover, it is difficult to cover different possible scenarios. On the other hand, simulators imitating real-world scenarios can be used to generate large amounts of data in different scenarios at a low cost. The data obtained from simulators can be used to train the radar systems for DCT in real-time. 


\section{Concluding Remarks}     \label{Section:conclusions}
In this survey paper, we conduct an in-depth analysis of the capabilities, threats, and challenges associated with UAVs. We offer a comprehensive overview of three RF-based systems: radars, communication systems, and RF analyzers for DCT-U. Our survey encompasses fundamental radar processing and discusses both conventional and modern radar systems for DCT-U. Additionally, we explore JC\&S in active mode and passive radars based on communication systems for DCT-U. We also discuss various methods for capturing RF signals of interest emitted by UAVs for DCT-U. The paper addresses the limitations of RF-based systems for DCT-U and provides insights into future directions based on lessons learned.

\bibliographystyle{IEEEtran}
\bibliography{References}

\end{document}